\documentclass[aps,twocolumn,pre]{revtex4-2}

\usepackage{xcolor,hyperref}
\hypersetup{
   colorlinks,
   linkcolor={blue!50!black},
   citecolor={blue!50!black},
   urlcolor={blue!80!black}
}

\usepackage{epsfig, amsmath}
\usepackage{bm}
\usepackage{soul}
\usepackage{hyperref}
\usepackage{footmisc}
\DeclareMathAlphabet{\mathitbf}{OML}{cmm}{b}{it}

\newcommand{\C}[1]{{\mathcal{#1}}}
\newcommand{\zerovector}{\bm{0}}

\renewcommand{\=}{\!=\!}
\newcommand{\tripleCdot}{:\!\cdot\,}

\newcommand{\xv}{\mathitbf x}
\newcommand{\zv}{\mathitbf z}
\newcommand{\piv}{\bm{\pi}}
\newcommand{\Tx}{T_{\mbox{\tiny$\mathsf{X}$}}}
\newcommand{\calBold}[1]{\mbox{\boldmath${\cal #1}$}}
\newcommand{\dbar}{{\,\mathchar'26\mkern-12mu d}}
\newcommand{\Jx}{J_{\mbox{\tiny$\mathsf{X}$}}}
\newcommand{\wx}{\omega_{\mbox{\tiny$\mathsf{x}$}}}
\newcommand{\mathBold}[1]{\mbox{\boldmath$#1$}}

\begin{document}

\title{Low-energy quasilocalized excitations in structural glasses}
\author{Edan Lerner$^{1}$}
\email{Corresponding author: e.lerner@uva.nl}
\author{Eran Bouchbinder$^{2}$}
\email{Corresponding author: eran.bouchbinder@weizmann.ac.il}
\affiliation{$^{1}$Institute of Theoretical Physics, University of Amsterdam, Science Park 904, 1098 XH Amsterdam, the Netherlands\\
$^{2}$Chemical and Biological Physics Department, Weizmann Institute of Science, Rehovot 7610001, Israel}

\begin{abstract}
Glassy solids exhibit a wide variety of generic thermomechanical properties, ranging from universal anomalous specific heat at cryogenic temperatures to nonlinear plastic yielding and failure under external driving forces, which qualitatively differ from their crystalline counterparts. For a long time, it has been believed that many of these properties are intimately related to nonphononic, low-energy quasilocalized excitations (QLEs) in glasses. Indeed, recent computer simulations have conclusively revealed that the self-organization of glasses during vitrification upon cooling from a melt leads to the emergence of such QLEs. In this Perspective Article, we review developments over the past three decades towards understanding the emergence of QLEs in structural glasses, and the degree of universality in their statistical and structural properties. We discuss the challenges and difficulties that hindered progress in achieving these goals, and review the frameworks put forward to overcome them. We conclude with an outlook on future research directions and open questions.
\end{abstract}

\maketitle

\section{Introduction}
\label{sec:intro}

Structural glasses are formed by cooling liquids quickly enough so that they avoid crystallization. As liquids are supercooled below their melting temperature, their viscosity increases dramatically, by several orders of magnitude; at some point in this process, the viscosity becomes so large that the supercooled liquid falls out of equilibrium and is deemed to become a solid. This vitrification occurs at a temperature known as the glass transition temperature $T_{\rm g}$, which is operationally defined as the temperature at which the liquid's viscosity reaches $10^{13}$ Poise~\cite{Cavagna200951,Debenedetti2001}. The nature of the glass transition remains a highly debated topic under intense investigation~\cite{Cavagna200951,Debenedetti2001,MW_cates_length_discussion_prl_2017,tarjus_no_length}.

Glassy solids below their respective glass-transition temperature feature several intriguing properties associated with their disordered nature, absent in their crystalline counterparts that feature long-range order. The experimental work of Zeller and Pohl~\cite{Zeller_and_Pohl_prb_1971} revealed the anomalous thermal conductivity and specific heat of glasses below 10K. It is now well established that the specific heat of glasses grows from zero temperature approximately as $T$ and that the thermal conductivity of glasses grows from zero temperature approximately as $T^2$ for a very large variety of glasses~\cite{HUNKLINGER1976155,Pohl_review_2002}, instead of the phonon-mediated $T^3$ scaling predicted by Debye for both observables~\cite{kittel2005introduction}. These anomalies were addressed in the early 1970s by phenomenological tunneling models~\cite{phillips1972tunneling,anderson1972anomalous}. These models postulate the existence of localized excitations --- known as Two-Level Systems (TLSs) --- which are small, localized groups of particles that can tunnel between two mechanically-stable configurations.

Nonphononic low-energy excitations find another widely studied manifestation in glassy solids, known as the \emph{boson peak} (BP). The BP is observed when plotting the vibrational density of states (VDoS) of a glass, ${\cal D}(\omega)$, normalized by Debye's phononic VDoS (in three dimensions) ${\cal D}_{\rm D}(\omega)\!=\!{\cal A}_{\rm D}\,\omega^2$, where $\omega$ denotes the angular frequency and ${\cal A}_{\rm D}$ is a known frequency-independent  prefactor~\cite{kittel2005introduction}. The reduced VDoS ${\cal D}(\omega)/{\cal D}_{\rm D}(\omega)$ of glasses generically exceeds unity at low frequencies, indicating the existence of excess low-frequency vibrational modes on top of the low-frequency phononic excitations. Furthermore, the reduced VDoS of glasses generically features a peak, typically at a frequency $\omega_{\mbox{\tiny BP}}$ in the THz range. It is commonly believed that this BP is more pronounced in glassy states that feature a greater degree of structural disorder~\cite{Shintani2008,Ruocco2008,KIRILLOV1999279,Schirmacher_2013_boson_peak,Zaccone_prl_2019}. Despite decades of investigations, there is no consensus regarding the origin of the boson peak~\cite{MALINOVSKY1986757,schirmacher_1998,taraskin2001origin,Lubchenko1515,Shintani2008,Chumakov_2011_bosonPeak,eric_boson_peak_emt,Gurevich2003,Gurevich2005,Gurevich2007,Schirmacher_2013_boson_peak,Zaccone_prl_2019}.

The structural disorder of glasses manifests itself in their mechanical response to external forces as well. Different from ordered crystalline solids, the micro-scale disorder of glasses leads to correlated, non-affine motions of the constituent particles in response to external forces~\cite{barrat_prb_2002,lemaitre_sum_rules_2006,delgado_prl_2010}, even in the elastic/reversible regime of small deformation. At larger deformation levels, plastic/irreversible processes become abundant, taking the form of localized immobile rearrangements of a few tens of particles, yet again in sharp contrast to mobile dislocations in crystalline solids. These rearrangements --- coined `shear transformations' by Argon~\cite{argon_st,argon_bubble_raft} (occurring at regions that were later on coined `shear transformations zones' (STZs) by Falk and Langer~\cite{falk_langer_stz}) --- have been the subject of extensive theoretical, computational and experimental investigations~\cite{spaepen_1977,argon_st,argon_bubble_raft,argon_simulations,falk_langer_stz,Malandro_Lacks,lemaitre2004,Argon_prb_2005,lemaitre2006_avalanches,schall_stz_colloids,manning2011,carmel_pre_2013,falk_qlm_2014,falk_prl_2016,david_collaboration_2020}. Under a broad set of circumstances~\cite{Ozawa6656,mw_sudden_failure_pre_2018,Fielding_prl_2020,david_macroscopic_shear_band}, predominantly at low temperatures/high strain-rates, STZs are activated in a collective/correlative manner, resulting in plastic strain localization in the form of shear bands~\cite{argon_simulations,johnson_shear_bands_2003,falk_shi_prl_2005,Schuh_review_2007,Falk2011,itamar_yielding_pnas_2017}.

\begin{figure*}[ht!]
  \includegraphics[width = 0.85\textwidth]{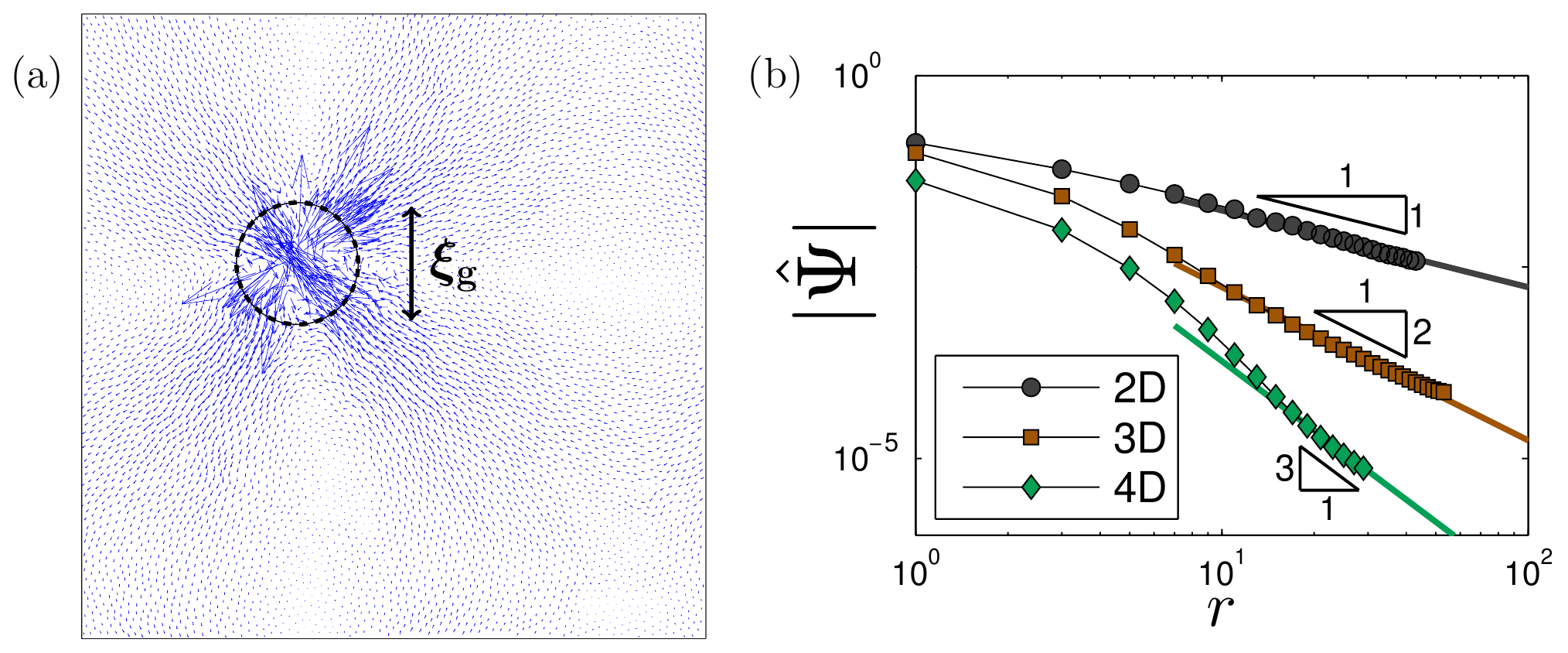}
  \caption{\footnotesize (a) A quasilocalized excitation (QLE) observed in a computer glass in two-dimensions (see~\cite{cge_paper} for details). The circle delineates the spatial extent of the QLE's core, of linear size $\xi_{\rm g}$. (b) The spatial decay of QLEs' amplitude, as observed in computer glasses in various spatial dimensions (see legend and~\cite{modes_prl_2018} for details). Agreement with the continuum elastic response to a local dipolar perturbation $\sim\!r^{-(\dbar-1)}$, in $\dbar$ spatial dimensions, is typically observed above $r\!\sim\!\xi_{\rm g}\!\approx\!10$ particle diameters. This Perspective Article reviews past and recent advances in understanding the properties of QLEs and the degree of universality of their emergent statistics in structural glasses.}
  \label{fig:fig1}
\end{figure*}

The generic nature of the aforementioned phenomena suggests that they share a common, universal origin, presumably associated with disorder-induced emergent excitations that differ from, and coexist with, phonons at low frequencies. Indeed, over the past few years it has been established that structural glasses quenched from a melt generically host a population of soft, quasilocalized nonphononic excitations, similar to that depicted in Fig.~\ref{fig:fig1}a. These soft excitations feature a disordered core of linear size $\xi_{\rm g}$, typically of the order of 10 particle diameters (with some exceptions to be discussed below). As shown in Fig.~\ref{fig:fig1}b, particle displacement a distance $r$ away from a QLE's core decay as $r^{-(\dbar-1)}$ in $\dbar$ spatial dimensions. As such, they echo the continuum-level elastic fields of Eshelby-like inclusions~\cite{avraham_core_properties_pre_2020}, e.g.~the far-field displacement response to local dipolar perturbations~\cite{cge_paper,pinching_pnas}. Under certain conditions --- to be discussed at length in what follows --- QLEs assume the form of harmonic vibrational modes whose frequencies $\omega$ follow a universal $\sim\!\omega^4$ distribution, apparently independent of spatial dimension~\cite{modes_prl_2018}, glass formation history~\cite{LB_modes_2019,pinching_pnas} or microscopic details~\cite{modes_prl_2016,ikeda_pnas,modes_prl_2020}. As described in detail below, this universal form of the nonphononic VDoS had been predicted theoretically since the late 1980s~\cite{soft_potential_model_1987,soft_potential_model_1989}, and has been more firmly established using computer simulations in recent years.

In this Perspective Article, we review past and recent developments in understanding the universal emergence of soft, quasilocalized excitations in structural glasses, and their connection to glass properties. While we build on extensive effort and progress accumulated in the literature over the last few decades, we are also quite strongly biased toward our own recent work and our understanding of the history of this scientific field. Consequently, this Perspective Article is necessarily somewhat subjective, and is not meant to provide a comprehensive and technical review of this topic. This Perspective Article is structured as follows; in Sect.~\ref{sec:history}, we review early theoretical, numerical and experimental developments in understanding the emergence of QLEs in structural glasses. In Sect.~\ref{sec:hybridization}, we explain how phononic modes dwelling in the low-frequency tail of the vibrational spectrum of structural glasses tend to hybridize with QLEs, and as such suppress the latter's realization as distinct harmonic vibrational modes. We further trace out the set of conditions in which QLEs can be cleanly and directly observed in the vibrational spectrum of finite-size computer glasses. In Sect.~\ref{sec:recent}, we review recent progress in resolving the degree of universality of the statistical properties of QLEs, and discuss what affects the characteristic length and frequency scales associated with them. In addition, we review recently proposed mean-field models of QLEs, and report recent progress in developing methods that allow to define non-hybridized QLEs by incorporating anharmonicities of the potential energy landscape. Finally, in Sect.~\ref{sec:exp} we discuss experimental evidence for QLEs and
in Sect.~\ref{sec:outlook}, we briefly discuss some open questions and future research directions.

\section{Early developments}
\label{sec:history}

In this Section, we provide a concise chronological perspective on what we view as the key observations, as well as the accompanying evolution of concepts, regarding the generic existence of QLEs in structural glasses, along with the roles they play in determining glass properties. As stressed above, we note that in the framework of such a Perspective Article, we cannot possibly offer an exhaustive account of the huge literature on this topic.

To the best of our knowledge, the first suggestion that structural glasses embed a population of low-energy localized excitations was put forward in 1962 by Rosenstock~\cite{rosenstock1962}. Rosenstock argued that ``non-elastic'' (i.e.~non-Debye or non-wave-like) soft localized excitations --- emanating from weakly-bounded groups of atoms in the glass structure --- should be expected to emerge in disordered solids. This claim was based on earlier observations~\cite{Anderson_1959,Flubacher_1959} of discrepancies between the measured specific heat of glasses, and that expected from elastic moduli measurements and Debye's theory~\cite{kittel2005introduction}. Later inelastic cold neutron scattering experiments by Leadbetter and Litchinsky~\cite{Leadbetter1970} suggested the existence of resonant modes associated with particular defects in the structure of Vitreous Germania. Indeed, some authors used the term ``resonant modes'' interchangeably with QLEs (cf.~\cite{Schober_Laird_numerics_PRL}), similarly to resonant modes associated with self-interstitials in metals~\cite{Resonant_Modes_1973}.

More well known are the works by Phillips~\cite{phillips1972tunneling} and by Anderson, Halperin and Varma~\cite{anderson1972anomalous}, who independently formulated phenomenological models that address the anomalous temperature-dependence of the thermal conductivity and specific heat of structural glasses at very low temperatures, as revealed earlier by the experimental work of Zeller and Pohl~\cite{Zeller_and_Pohl_prb_1971}. A key assumption in these models is the existence of localized excitations --- the `Two-Level-Systems' (TLSs) --- envisioned as small groups of atoms or molecules that can tunnel between two mechanically-stable states, typically at temperatures of $\approx$1K and below. In~\cite{Phillips1978}, Phillips proposed that the anomalous thermodynamic and transport properties of glasses reflect the behavior of intrinsic low-frequency vibrational modes of the structure --- a proposition that nicely corresponds to the subject of this Perspective Article. In Sect.~\ref{sec:qle_and_tls}, we discuss in more detail possible connections between tunneling TLSs --- more precisely double-well potentials --- and QLEs.

In the late 1970s, the fundamental mechanism of plastic deformation in externally driven structural glasses was studied by Spaepen~\cite{spaepen_1977} and Argon~\cite{argon_st}. Further reinforced by observations from mechanical experiments on bubble-rafts~\cite{argon_bubble_raft} and computer simulations~\cite{Maeda1978}, these studies argued that plastic flow in amorphous solids proceeds via immobile, localized shear-like rearrangements of a few tens of particles. Subsequent computer simulations~\cite{Egami_Maeda_Vitek_1980,Srolovitz1981,Srolovitz1983} defined various structural defects in computer glasses, and studied their statistical properties and the level of correlations between those defects and plastic-flow events.

\subsection{The Soft Potential Model}
\label{sec:spm}

Between the early 1980s and the early 1990s a series of papers by Klinger, Karpov, Ignatiev, Galperin, Il'in, Buchenau, Gurevich, Schober and others~\cite{karpov1982,klinger1983,karpov1983,soft_potential_model_1987,soft_potential_model_1989,soft_potential_model_1991,buchenau_prb_1992,buchenau_Phil_Mag_1992,soft_potential_model_prb_1993} proposed that glasses generically host local regions in which the stiffness associated with atomic motion is anomalously small. These frameworks were collectively termed the "Soft Potential Model" (SPM). The SPM assumes that localized groups of particles can be envisioned as {\em non-interacting} anharmonic oscillators, each of which is described by a smooth, random potential energy function $U(q)$ that admits a Taylor expansion in the form $U(q)\!=\!\sum_{n=1}^\infty a_n q^n/n!$ (for convenience we set $U(0)\=0$). Here the coefficients $\{a_n\}$ follow a regular (i.e., featuring no zeros or singularities) joint distribution function $p(\{a_n\})$. Assuming then that $U(q)$ attains a minimum at $q\=q_0$, one can obtain the following quartic expansion
\begin{equation}
\label{eq:SPM}
    U(s) \simeq U_0 + \frac{1}{2}b_2s^2 + \frac{1}{3!}b_3s^3 + \frac{1}{4!}b_4s^4 + {\cal O}(s^5)\,,
\end{equation}
where $s\!\equiv\!q-q_0$ and $U_0\!\equiv\!U(q_0)$. Transforming from $p(\{a_n\})$ to $p(\{b_n\})$, and integrating over all of the coefficients but $b_2$, one can show that $p(b_2)\!\sim\!|b_2|$ for $b_2\!\to\!0$~\cite{Chalker2003}.

In view of Eq.~\eqref{eq:SPM}, TLSs naturally emerge as double-well potentials in the framework of the SPM. Invoking then quantum tunneling (relevant at very low temperatures) and using $p(b_2)\!\sim\!|b_2|$ for $b_2\!\to\!0$, the SPM offers various predictions (not discussed here) for tunneling TLSs~\cite{soft_potential_model_1991}. More directly relevant for our discussion here is that $p(b_2)\!\sim\!|b_2|$, together with $\omega\!\sim\!\sqrt{b_2}\!\to\!0$, implies a density of vibrational modes that grows from zero frequency as $\omega^3$. As pointed out in~\cite{Chalker2003}, further demanding that the minimum at $q_0$ is a {\em global} minimum of $U(q)$~\cite{Chalker2003}, which for the quartic expansion of Eq.~\eqref{eq:SPM} is guaranteed for $b_3^2\!\le\!3b_2b_4$, the SPM predicts that the contribution of the local soft potentials to the density of vibrational modes grows from zero frequency as $\omega^4$~\cite{soft_potential_model_1987,soft_potential_model_1989,soft_potential_model_1991}. In Sect.~\ref{sec:nonlinear_qles}, we further discuss the condition $b_3^2\!\le\!3b_2b_4$ --- which may be viewed as a stability bound ---, as well as numerical evidence supporting it and its implications.

While the simplicity of the SPM picture is appealing, it leaves some key questions unanswered. For example, the SPM does not explain what determines the degree of localization of soft excitations, nor does it fully describe the physical factors that control their number density. In addition, the SPM appears to lack a physical description of the minimal set of conditions necessary for the model's key predictions to hold.

\subsection{The reconstruction picture}
\label{sec:reconstruction}

In the mid-2000s Gurevich, Parshin and Schober (GPS) ---  who were part of the group of workers that previously formulated the SPM --- put forward a phenomenological theory for the VDoS of glassy solids that goes beyond the SPM~\cite{Gurevich2003,Gurevich2005,Gurevich2007}. The theory envisions a glass as a collection of {\em interacting} anharmonic oscillators --- as opposed to the non-interacting SPM picture --- that are meant to represent mesoscopic material elements. According to the theory, anharmonic oscillators a distance $r$ from each other interact bilinearly, with an interaction strength that decays with distance $r$ as $r^{-\dbar}$ (in $\dbar$ spatial dimensions), mimicking elastic dipole-dipole interactions. The physical picture according to which a glass is represented by interacting anharmonic oscillators as described by GPS bears some similarities to earlier propositions by Grannan, Randeria, and Sethna~\cite{sethna_prb_1990_1,sethna_prb_1990_2} and by K\"uhn and Horstmann (KH)~\cite{Kuhn_Horstmann_prl_1997}.

The GPS approach involves two steps: the first step considers the effect of interactions on soft oscillators, which are a priori assumed to exist: the stiffnesses associated with soft oscillators are reduced due to interactions with stiff oscillators. This softening creates a `traffic' of oscillators' stiffnesses towards zero stiffness. This interaction-induced softening leads to the destabilization of some of the soft oscillators: they assume a \emph{negative} stiffness and --- together with the local anharmonicity --- become double-well potentials. Since the system must evolve towards a mechanically-stable state, those destabilized oscillators restore stability by assuming a shifted equilibrium position that corresponds to the minimum of one of the two potential wells formed. This reconstruction of soft oscillators leads to a stiffness distribution that is flat near zero stiffness, resulting in a generic ${\cal D}(\omega)\!\sim\!\omega$ reconstructed VDoS, independent of the initial distribution of oscillator-stiffnesses (as long as is has no hard gap, see further discussion in Sect.~\ref{sec:mean_field}).

In the second step of the GPS theory, the interactions between the \emph{reconstructed} soft oscillators are considered. Since those oscillators have assumed shifted equilibrium positions, they exert random static forces on each other. These may be viewed as frustration-induced internal stresses, which generically exist in glasses. In the presence of local anharmonicities, these static forces lead to a further \emph{stabilization} of the softest reconstructed oscillators, resulting in a universal VDoS ${\cal D}(\omega)\!\sim\!\omega^4$ for frequencies $\omega$ smaller than a characteristic crossover frequency $\omega_{\rm b}$~\cite{Gurevich2003,Gurevich2005,Gurevich2007}, above which the first reconstructed ${\cal D}(\omega)\!\sim\!\omega$ persists. These predictions imply that the reduced VDoS ${\cal D}(\omega)/{\cal D}_{\rm D}(\omega)$ features a boson peak in the vicinity of $\omega_{\rm b}$.

In addition to the phenomenological theory outlined above, GPS put forward a lattice model in three dimensions, and studied it numerically in order to validate their theoretical predictions. The lattice model assumes each lattice site is occupied by an anharmonic oscillator with a stiffness $\kappa_i\!\ge\!0$ drawn from a gapless parent distribution $g_0(\kappa)\!\sim\!\kappa^\beta$, and is described by the Hamiltonian
\begin{equation}
\label{eq:GPS}
    H_{\mbox{\tiny GPS}} = \frac{1}{2}\sum_i\kappa_ix_i^2 +\sum_{i<j}J_{ij}(r_{ij})x_ix_j + \frac{A}{4!}\sum_i x_i^4\,.
\end{equation}
Here $x_i$ denotes the (scalar) coordinate of the $i^{\mbox{\tiny th}}$ oscillator, and $J_{ij}(r_{ij})\!=\!Jg_{ij}/r_{ij}^3$ is a space-dependent random variable representing the elastic coupling between the $i^{\mbox{\tiny th}}$ and $j^{\mbox{\tiny th}}$ oscillators, where $g_{ij}\!\in\![-1/2,1/2]$ is a uniformly-distributed random variable, $J$ is an interaction-strength parameter and $r_{ij}$ is the distance between the oscillators. In~\cite{Gurevich2003}, GPS report on numerical simulations of this lattice model, verifying that it features a VDoS ${\cal D}(\omega)\!\sim\!\omega^4$ as $\omega\!\to\!0$, independent of the exponent $\beta$ that characterizes the initial gapless distribution $g_0(\kappa)$ of oscillator stiffnesses. In addition, at higher frequencies the model's VDoS was shown to follow a $\sim\!\omega$ scaling, resulting in a boson peak in the reduced VDoS as predicted by the phenomenological theory. Several additional predictions from the phenomenological theory were verified by numerical simulations of the lattice model in~\cite{Gurevich2003,Gurevich2007}. The GPS phenomenological theory and lattice model were critically discussed in~\cite{schirmacher2011comments,itamar_gps_prb2020}.

\begin{figure}[ht!]
  \includegraphics[width = 0.5\textwidth]{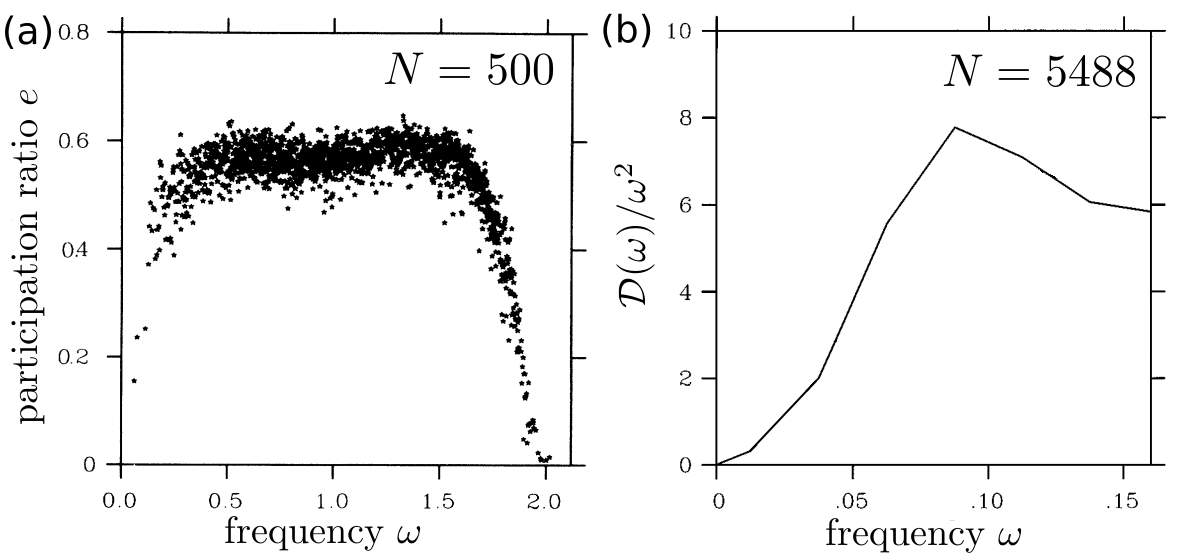}
  \caption{\footnotesize(a) Early numerical evidence for the existence of low-energy QLEs in computer glasses, by Schober and Laird~\cite{Schober_Laird_numerics_PRB}; shown is a scatter-plot of the participation ratio of vibrational modes vs.~their frequency, calculated in a single soft-sphere computer-glass of $N\!=\!500$ particles in three dimensions. Adapted with permission from H.~R.~Schober and B.~B.~Laird, Phys.~Rev.~B \textbf{44}, 6746 (1991). Copyright 1991 American Physical Society. (b) Early numerical observations by Schober and Oligschleger~\cite{Schober_Oligschleger_numerics_PRB} that are consistent with the universal $\sim\!\omega^4$ VDoS of QLEs. Plotted is the reduced low-frequency VDoS ${\cal D}(\omega)/\omega^2$ of the same computer-glass model~\cite{Schober_Oligschleger_numerics_PRB}, but with $N\!=\!5488$ particles, which appears to increase superlinearly from zero frequency. Adapted with permission from H.~R.~Schober and C.~Oligschleger, Phys.~Rev.~B \textbf{53}, 11469 (1996). Copyright 1996 American Physical Society.}
  \label{fig:schober_prb_1996}
\end{figure}

\subsection{Early atomistic simulations}

In the early 1990s, inspired by neutron scattering experiments on viteous silica~\cite{Buchenau_1988}, Schober and coworkers turned to atomistic simulations~\cite{Schober_Laird_numerics_PRL,Schober_Laird_numerics_PRB,schober1993_numerics,Schober_Oligschleger_se_1993_numerics,Schober_Oligschleger_numerics_PRB,Schober_numerics_prb_2000,schober_and_ruocco_2004,Schober_2004}, with the aim of testing and possibly validating the SPM predictions. To the best of our knowledge, the first robust numerical observations of low-frequency quasilocalized vibrations in simple computer glasses were put forward in 1991 by Schober and Laird~\cite{Schober_Laird_numerics_PRL,Schober_Laird_numerics_PRB} (however, see also~\cite{footnote5}). In these works, monodisperse soft-sphere computer glasses with $N\!=\!500$ and $N\!=\!1024$ particles were studied, and the (quasi-) localization of the lowest-frequency vibrational modes was established by measuring those modes' participation ratio
\begin{equation}
\label{eq:participation_ratio}
    e(\bm{\Psi}) \equiv \frac{\big(\sum_i \bm{\Psi}_i\cdot\bm{\Psi}_i\big)^2}{N\sum_i\big(\bm{\Psi}_i\cdot\bm{\Psi}_i\big)^2}\,,
\end{equation}
where $\bm{\Psi}_i$ denotes the $\dbar$-dimensional vector of Cartesian components pertaining to the $i^{\mbox{\tiny th}}$ particle of a vibrational mode $\bm{\Psi}$. The participation ratio $e(\bm{\Psi})$ of a given mode $\bm{\Psi}$ (i.e.~a displacement vector field defined on each particle in the system) is a quantifier of the degree of localization of that mode. If a mode is localized on a compact core of $N_{\rm c}$ particles, one expects $e\!\sim\!N_{\rm c}/N$, whereas if it is extended, then $e\!\sim\!1$. In Fig.~\ref{fig:schober_prb_1996}a, we show an example of a scatter plot of the participation ratio of vibrational modes of a simple computer glass, vs.~their frequency, adapted from the original work of Schober and Laird~\cite{Schober_Laird_numerics_PRB}. Those workers argued, based on their numerical results, that the low-frequency quasilocalized modes are centered typically on $\sim$20 particles. Later, in~\cite{Schober_Oligschleger_se_1993_numerics}, it was shown that localization of low-frequency vibrational modes also occurs in a computer model of Selenium, which includes directional bonds, reinforcing that QLEs generically emerge in glassy solids.

The statistical samples accessible in~\cite{Schober_Laird_numerics_PRL,Schober_Laird_numerics_PRB} were insufficient in order to robustly validate the predicted $\omega^4$ scaling of the VDoS. Later work~\cite{Schober_Oligschleger_numerics_PRB} employed the same computer glass model, but larger glass samples (up to $N\!=\!5488$). These larger computer-glass samples allowed the authors to establish that the reduced nonphononic VDoS grows at least as $\omega^3$ and possibly stronger (see numerical data of~\cite{Schober_Oligschleger_numerics_PRB} in Fig.~\ref{fig:schober_prb_1996}b), compatible with the predicted $\omega^4$ scaling of the SPM for quasilocalized vibrations' VDoS.

In that same work~\cite{Schober_Oligschleger_numerics_PRB}, the generic occurrence of \emph{hybridizations} between low-frequency extended, plane-wave (phononic) modes and low-frequency quasilocalized vibrations is discussed. These hybridizations were further elaborated upon in~\cite{schober_and_ruocco_2004}. Since phonons are always present in the low-frequency spectrum of solids, due to the breakdown of global continuous symmetries (Goldstone's theorem), their hybridization with QLEs in the harmonic spectrum of glasses has for a long time hindered progress in revealing the statistical and structural properties of QLEs, and in understanding the low-frequency spectrum of structural glasses using computational tools. In the following Section, we describe in more detail the hybridization of phononic modes and QLEs in the low-frequency spectrum of structural glasses.

\section{Phonon-QLE hybridization}
\label{sec:hybridization}

Within the harmonic approximation, one generically expects different classes of excitations to strongly hybridize and mix, if those different excitations interact and share similar frequencies. As described early on by several workers~\cite{Schober_Oligschleger_numerics_PRB,schober_and_ruocco_2004,klinger2002_hybridizations}, and more recently by others~\cite{parisi_spin_glass,SciPost2016,manning_defects,paoluzzi_pnas_2018_pinning,episode_1_2020,julia_arXiv}, low-energy QLEs and plane-wave-like phononic excitations are no exception to this rule. In Fig.~\ref{fig:hybridization_1} below, we show an example of a low-frequency vibrational mode in a computer glass in two dimensions (2D), which is comprised of a QLE hybridized with a plane-wave (phononic) mode.
\begin{figure}[ht!]
  \includegraphics[width = 0.45\textwidth]{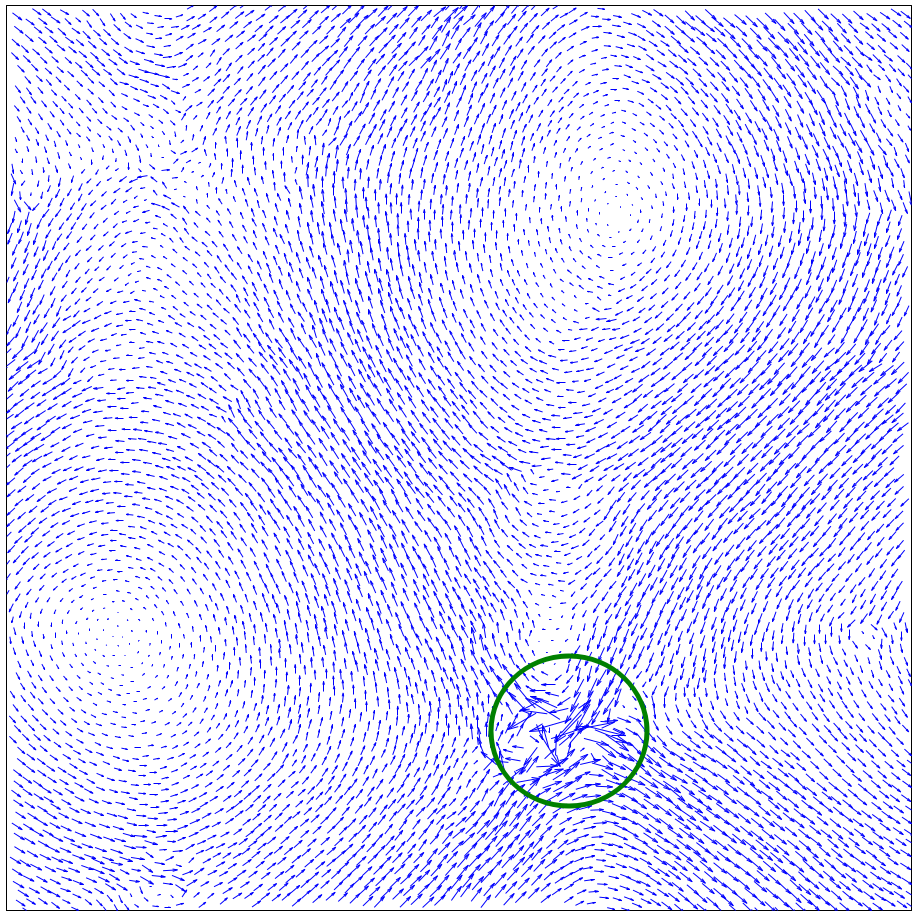}
  \caption{\footnotesize An example of a QLE (marked by the green circle) that is strongly hybridized with the lowest-frequency phononic vibrational mode in a 2D computer glass (see~\cite{cge_paper} for model details).}
  \label{fig:hybridization_1}
\end{figure}

\begin{figure}[ht!]
  \includegraphics[width = 0.51\textwidth]{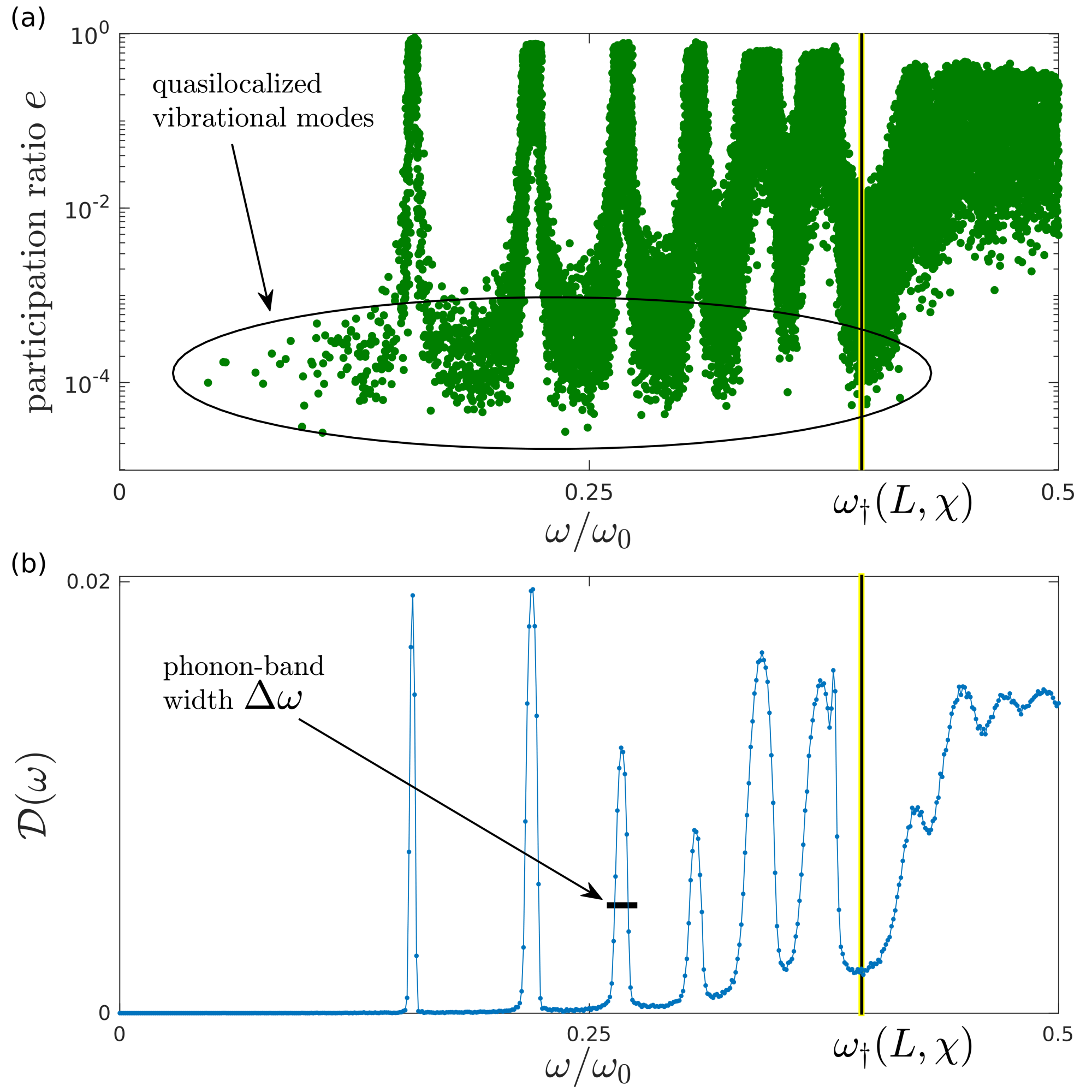}
  \caption{\footnotesize (a) Scatter-plot of the participation ratio $e$, cf.~Eq.~\eqref{eq:participation_ratio}, of vibrational modes of a 3D computer glass \emph{vs.}~their frequency $\omega$. Frequencies are expressed in terms of $\omega_0\!\equiv\!c_s/a_0$, where $c_s$ is the shear wave-speed, $a_0\!=\!(V/N)^{1/3}$ is an interparticle distance (with $V$ denoting the system's volume), and here $N\!=\!\mbox{64,000}$. The ellipse engulfs those quasilocalized modes that escaped hybridizations, see text for further discussion. (b) The VDoS of the same glasses analyzed in panel (a). Discrete phonon bands, which are apparent, acquire a finite width $\Delta\omega$ due to the glass's mechanical disorder~\cite{phonon_widths,phonon_widths2}. The vertical lines in both panels mark the crossover frequency scale $\omega_\dagger$, cf.~Eq.~\eqref{eq:omega_dagger}, above which phonon band widths become comparable to the gaps between them, such that QLEs can no longer be cleanly realized as harmonic vibrations. Note that the lower envelope of the presented VDoS in fact follows an $\omega^4$ scaling (not marked on the figure, but see Fig.~1 in~\cite{phonon_widths2}, where it is marked).}
  \label{fig:hybridization_2}
\end{figure}

The hybridization of QLEs and phonons in the harmonic spectrum of glasses affects these modes' localization properties. In particular, the participation ratio of hybridized phonon-QLE excitations can assume any value between $N_{\rm c}/N$ and 1, as demonstrated in Fig.~\ref{fig:hybridization_2}a. The data presented therein were obtained in a polydisperse soft-sphere computer glass of $N\!=\!\mbox{64,000}$ particles in 3D (see~\cite{phonon_widths2} for details). It is apparent from these data that QLEs can be realized as harmonic vibrations below the first phonon band and between phonon bands, which appear as peaks with $e\!\sim\!{\cal O}(1)$. In addition, it appears that phonon bands `burn' holes in the low-$e$ range corresponding to the participation ratio of this system's quasilocalized modes. In other words, although QLEs certainly exist at frequencies that match phonon bands' frequencies (as explicitly demonstrated in~\cite{phonon_widths}), their clean realization as quasilocalized harmonic vibrations is largely destroyed by hybridizations with phonons, if their frequencies lie in the close vicinity of phonon frequencies. It is crucial to stress, in this context and more generally, that QLEs manifest the existence of soft glassy structures embedded inside a glass, and as such their ontological status is independent of whether they can be realized as quasilocalized normal (harmonic) modes or not. Hybridizations with phonons, however, can have serious implications for one's ability to detect QLEs.

\begin{figure*}[ht!]
  \includegraphics[width = 1.0\textwidth]{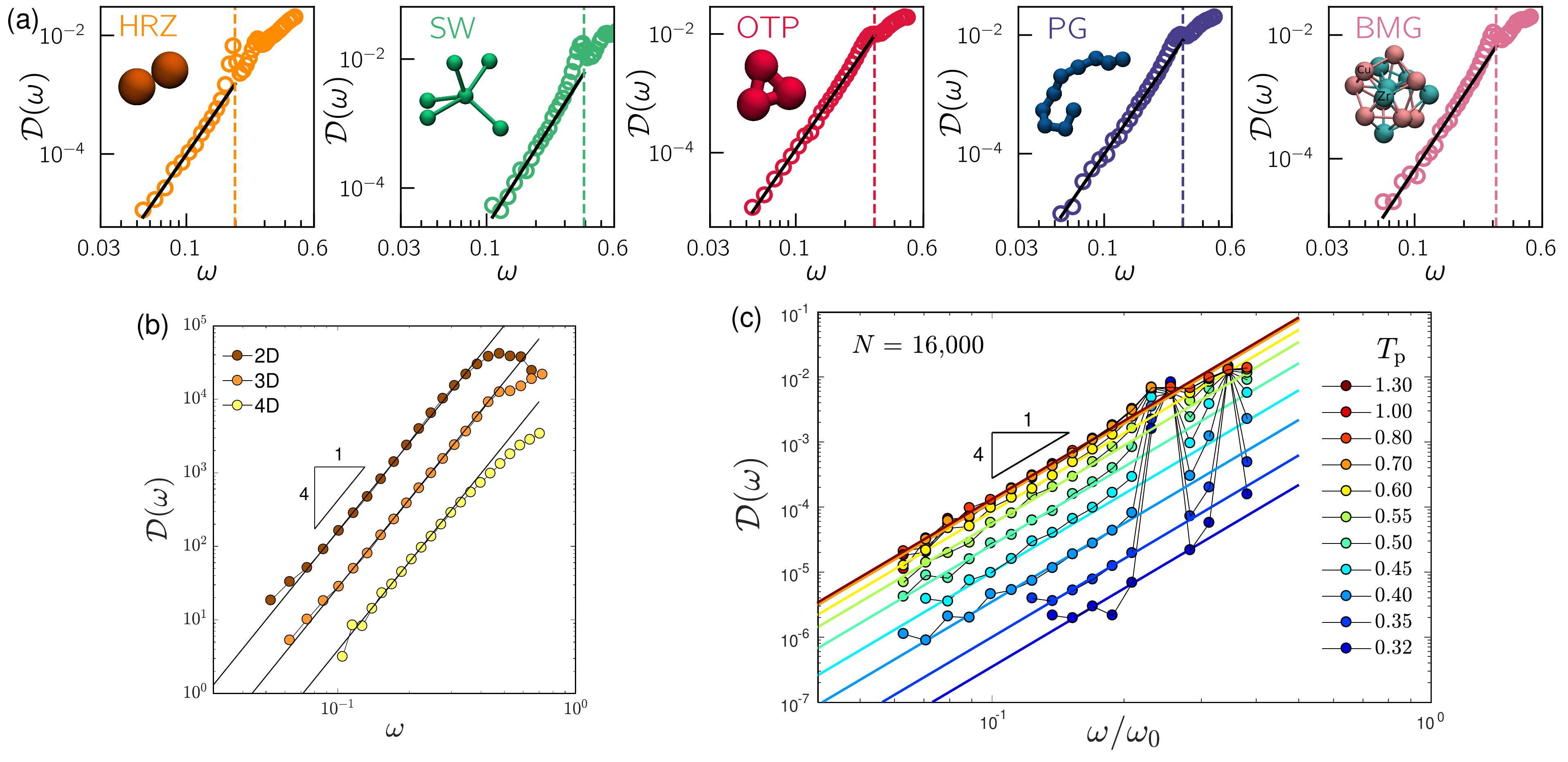}
  \caption{\footnotesize Numerical evidence supporting the universality of the $\omega^4$ nonphononic VDoS of structural glasses. (a) Low-frequency spectra of (from left to right) soft spheres interacting via the Hertz law, the Stillinger-Weber network glass-former, a triatomic OTP-like molecular glass former, a model polymeric glass, and a model CuZr bulk metallic glass. The vertical dashed lines mark the lowest phonon frequency, whereas the continuous lines correspond to the universal $\omega^4$ scaling. Details about the models and methods can be found in~\cite{modes_prl_2020}. (b) Low-frequency spectra of soft-sphere computer glasses in 2D, 3D and 4D (the spectra are shifted for visual clarity), see~\cite{modes_prl_2018} for details. (c) Low-frequency spectra of a polydisperse soft-sphere glass (see~\cite{phonon_widths2} for details) that can be subjected to extremely deep supercooling using the Swap-Monte-Carlo algorithm~\cite{LB_swap_prx}. Here, $T_{\rm p}$ is the equilibrium parent temperature from which glassy configurations were instantaneously quenched to zero temperature.}
  \label{fig:universal_modes_fig}
\end{figure*}

Also apparent from the data of Fig.~\ref{fig:hybridization_2} is that above a system-size- and mechanical-disorder-dependent frequency scale denoted by $\omega_\dagger(L,\chi)$, QLEs can no longer be realized as harmonic vibrations due to phonon-hybridizations. Here $L\!\sim\!N^{1/\dbar}$ denotes the linear size of the glass, and $\chi$ is a measure of mechanical disorder that is intimately related to the relative spatial fluctuations of the shear modulus field~\cite{phonon_widths2}. The frequency scale $\omega_\dagger(L,\chi)$ is understood as follows: the mechanical disorder intrinsic to structural glasses lifts the degeneracy of low-frequency phononic excitations that share the same wavelength. As a result, sets of low-frequency, iso-wavelength phonons form discrete \emph{bands} with finite widths $\Delta\omega$, as illustrated in Fig.~\ref{fig:hybridization_2}b. In~\cite{phonon_widths,phonon_widths2}, it was shown that
\begin{equation}
    \Delta\omega \propto \frac{\chi\, \omega \sqrt{n_z}}{\sqrt{N}}\,,
\end{equation}
where $n_z$ is the degeneracy level of the $z^{\mbox{\tiny th}}$ phonon band in an ideally isotropic homogeneous elastic medium~\cite{phonon_widths,phonon_widths2}. This relation implies that $\Delta\omega$ increases with increasing frequency. In~\cite{phonon_widths,phonon_widths2}, it was shown that at the crossover frequency
\begin{equation}\label{eq:omega_dagger}
    \omega_\dagger(L,\chi) \sim \big(\chi L)^{-\frac{2}{\dbar+2}}\,,
\end{equation}
$\Delta\omega$ becomes comparable to the gaps between consecutive phonon-bands. Consequently, above $\omega_\dagger$, phononic excitations are no longer clustered into discrete bands, but are instead distributed quasi-continuously over the frequency axis. Therefore, QLEs with frequencies $\omega\!>\!\omega_\dagger$ exclusively hybridize with phonons, as shown in Fig.~\ref{fig:hybridization_2}.

We finally note that in 2D, $\omega_\dagger\!\sim\!L^{-1/2}$, whereas the typical frequency of the \emph{softest} quasilocalized mode in a finite-size glassy sample, $\omega_{\mbox{\tiny min}}$, is of order $L^{-2/5}$ (with logarithmic corrections, see~\cite{modes_prl_2018,modes_prl_2016}). This implies that, in 2D, $\omega_\dagger\!<\!\omega_{\mbox{\tiny min}}$ and therefore 2D computer glasses of sizes of a few thousand particles and above typically do not feature many non-hybridized quasilocalized vibrations, as indeed observed numerically in several works~\cite{ikeda_pnas,grzegorz_arXiv_2021_2D_modes}.

\section{Recent developments}
\label{sec:recent}

\subsection{Evidence for universal nonphononic VDoS}

Since the numerical work of Schober in the 1990s, many theoretical, computational and experimental investigations of the vibrational spectra of structural glasses were put forward; some examples include~\cite{schirmacher_1998,ohern2003,Silbert_prl_2005,Shintani2008,Monaco_pnas_2009,Monaco_prl_2011,Schirmacher_2013_boson_peak,barrat_3d,Tanguy2015,eric_boson_peak_emt,eric_hard_sphere_vdos_pnas2014,silvio}, and see additional references therein. It was not until 2011, however, that the first numerical evidence of the universal $\omega^4$ nonphononic VDoS of quasilocalized excitations was indirectly revealed by Karmakar and coworkers~\cite{karmakar_lengthscale}, who studied the statistics of the lowest-frequency vibrational modes per computer glass. A few years later, in 2015, simulations by Baity-Jesi and coworkers~\cite{parisi_spin_glass} of the 3D Heisenberg spin glass revealed a quartic VDoS of QLEs. This was accomplished by applying a fluctuating external field that penalizes Goldstone modes that emerge due to the rotational invariance of the Heisenberg spin glass's Hamiltonian (analogous to phonons in structural glasses), hence overcoming the aforementioned hybridization issues and exposing the universal statistics of QLEs in a spin glass.

A more direct route to observe the form of the nonphononic spectrum of structural glasses was taken in 2016~\cite{modes_prl_2016}. The main idea of this work was that in finite systems of linear size $L$, the (finite) lowest-frequency phonons and QLEs follow a {\em different} scaling with $L$. Consequently, it was shown that the system size $L$ of computer glasses can be carefully tuned such that it is small enough so as to push phononic excitations to higher frequencies, cleanly exposing the VDoS of QLEs without hybridizations with phonons~\cite{Biswas_1988,Schober_2004,schober_and_ruocco_2004}, yet appreciably larger than QLEs' core size $\xi_{\rm g}$. Therefore, it turned out that small systems can in fact be beneficial in this context. Following this idea, extensive ensembles of glassy samples were generated, such that the nonphononic $\sim\!\omega^4$ VDoS of QLEs could be directly and robustly observed. In the very same work, the $\sim\!r^{-2}$ spatial decay of QLEs (in 3D) away from their respective cores was shown --- consistent with previous observations~\cite{lemaitre2006_avalanches,micromechanics2016}, and see also Fig.~\ref{fig:fig1}b. Furthermore, the core size $\xi_{\rm g}\!\approx\!10a_0$ was estimated, and the scaling $e\!\sim\!1/N$ of the participation ratio of QLEs was established. Finally, that work established that the softest QLEs per glassy sample follow Weibullian statistics, suggesting that QLEs are largely uncorrelated.

Subsequent work showed that the $\omega^4$ scaling of the nonphononic VDoS of structural glasses is robust to changes in spatial dimension~\cite{modes_prl_2018,Atsushi_high_d_pre_2020} and to extreme supercooling~\cite{LB_modes_2019,pinching_pnas} (made possible by applying the Swap-Monte-Carlo algorithm~\cite{tsai_swap,gazzillo_swap,grigera_swap} to the glass-forming model by Ninarello, Berthier and Coslovich~\cite{LB_swap_prx}). More recently, the robustness of the $\omega^4$ VDoS to changes in the interaction potential was established~\cite{modes_prl_2020,universal_VDoS_ip}. Some of these efforts to establish the universality of the nonphononic VDoS are presented in Fig.~\ref{fig:universal_modes_fig}.

\subsection{Effect of glass-formation history on the properties of QLEs}
\label{sec:qle_properties}

What are the salient properties of QLEs, apart from the universal $\sim\!\omega^4$ form of their distribution over frequency? As mentioned in the Introduction, QLEs are characterized by a core of linear size $\xi_{\rm g}$, typically on the order of 10 interparticle distances~\cite{modes_prl_2016,modes_prl_2018,pinching_pnas}, see e.g.~Fig.~\ref{fig:fig1}. Another important attribute of QLEs is their number density, as encapsulated in the prefactor ${\cal A}_{\rm g}$ of the nonphononic VDoS, which is written for small frequencies as
\begin{equation}
    {\cal D}(\omega) = {\cal A}_{\rm g}\omega^4\,.
\end{equation}
The prefactor ${\cal A}_{\rm g}$ has dimensions of [frequency]$^{-5}$; its physical significance was discussed at length in~\cite{cge_paper,atsushi_core_size_pre,pinching_pnas,david_fracture_2021}. It encompasses information both about the number density ${\cal N}$ of QLEs, and about their characteristic frequency $\omega_{\rm g}$~\cite{cge_paper,pinching_pnas}, in complete analogy to the corresponding prefactor ${\cal A}_{\rm D}$ (see Introduction) in Debye's VDoS of phonons. In~\cite{cge_paper,pinching_pnas}, it was suggested that the prefactor can be meaningfully decomposed into a product of the form ${\cal A}_{\rm g}\!=\!{\cal N}\omega_{\rm g}^{-5}$, since changes in ${\cal A}_{\rm g}$ may stem both from the stiffening or softening of QLEs (i.e.~variations in their characteristic frequency $\omega_{\rm g}$ as discussed in~\cite{pinching_pnas,cge2_jcp2020}) \emph{and} from QLEs' depletion or proliferation. If both $\omega_{\rm g}$ and ${\cal A}_{\rm g}$ can be measured independently, then the QLE number density ${\cal N}\!=\!{\cal A}_{\rm g}\omega_{\rm g}^5$ can be estimated, and its dependence on the formation history of a glass and on other factors can be studied.

How do the length $\xi_{\rm g}$, characteristic frequency $\omega_{\rm g}$, and nonphononic VDoS prefactor ${\cal A}_{\rm g}$ depend on the formation history of a glass? Here we discuss this question in the context of the `\emph{parent temperature}' simulational protocol. This protocol amounts to equilibrating a computer liquid at some parent temperature $T_{\rm p}$, and following it by an instantaneous quench to zero temperature to form a glassy solid. We note that at high $T_{\rm p}$s this protocol generates rather unrealistically unstable glasses compared to laboratory liquids, which cannot be driven through their respective glass transition temperatures at comparable cooling rates. Nevertheless, this protocol is useful as an investigative tool, as it allows to probe the full variety of glassy structures accessible to a single glass-forming liquid model.

\begin{figure}[ht!]
  \includegraphics[width = 0.50\textwidth]{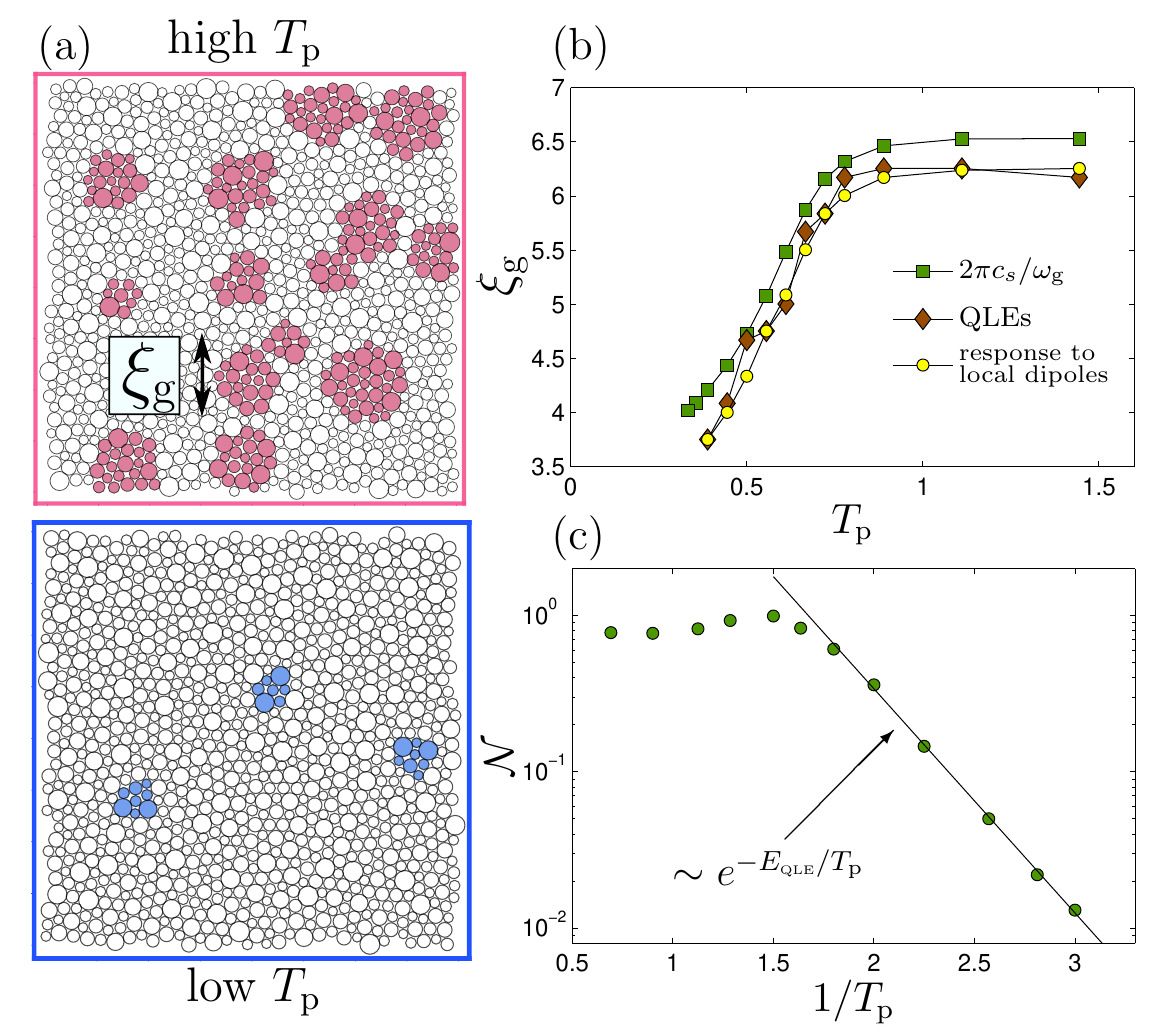}
  \caption{\footnotesize (a) Illustration of the $T_{\rm p}$ dependence of QLEs' properties, see text for discussion. (b) The lengthscale $\xi_{\rm g}$ that characterizes the core size of QLEs \emph{decreases} upon deeper supercooling of glasses' ancestral liquid configuration, see text and~\cite{pinching_pnas} for further details (and compare to panel (a)). (c) The number density ${\cal N}$ of QLEs follows a Boltzmann-like dependence on the parent temperature $T_{\rm p}$, see discussion in the text and in~\cite{pinching_pnas}.}
  \label{fig:Tp_dependence}
\end{figure}

In Fig.~\ref{fig:Tp_dependence}a, we illustrate the effect of $T_{\rm p}$ on QLEs' properties: upon deeper supercooling of glasses' ancestral liquid configurations, the core size of QLEs decreases, in parallel to their depletion, as discussed in~\cite{LB_modes_2019,pinching_pnas,experimental_inannealability_AM_2016} (note that TLSs have also been shown to undergo depletion upon deep supercooling~\cite{zamponi_tls_prl_2020}). Direct numerical evidence for the decrease in QLEs' core size is shown in Fig.~\ref{fig:Tp_dependence}b; these data were adapted from~\cite{pinching_pnas}, where in addition to directly probing the core size of QLEs (brown diamonds in Fig.~\ref{fig:Tp_dependence}b), it was further compared to (i) the length obtained via $2\pi c_s/\omega_{\rm g}$ (green squares), where $c_s$ is the shear wave-speed and $\omega_{\rm g}$ was estimated as the typical frequency associated with the response of the glass to local dipolar forces~\cite{cge_paper,pinching_pnas,sticky_spheres1_karina_pre2021}, and to (ii) the length obtained by analyzing the \emph{spatial response} to the same local force dipoles, where the distance from the imposed dipoles in which the expected $\sim\!r^{-(\dbar-1)}$ continuum-elastic scaling is observed was estimated~\cite{breakdown,lerner2019finite,sticky_spheres1_karina_pre2021} (yellow circles). The agreement between all of these lengths supports the relation $\xi_{\rm g}\!\sim\!c_s/\omega_{\rm g}$, and indicates that $\xi_{\rm g}$ can be accurately estimated via responses to local dipolar forces. The latter thus emerge as important physical quantities for probing QLEs' properties~\cite{cge_paper,pinching_pnas,sticky_spheres1_karina_pre2021}.

Finally, with an estimation of the characteristic frequency $\omega_{\rm g}$ of QLEs and the prefactors ${\cal A}_{\rm g}$ of their VDoS at hand, the number density of QLEs can be estimated as ${\cal N}\!\simeq\!{\cal A}_{\rm g}\omega_{\rm g}^5$; ${\cal N}$ is plotted against the inverse-parent-temperature in Fig.~\ref{fig:Tp_dependence}c, revealing a Boltzmann-like dependence~\cite{pinching_pnas}
\begin{equation}
\label{eq:Boltzmann}
{\cal N}(T_{\rm p})\sim\exp\!\left(-\frac{E_{\mbox{\tiny QLE}}}{T_{\rm p}}\!\right)\ ,
\end{equation}
below a crossover parent temperature $\Tx$ (discussed e.g.~in~\cite{karina_Tx_jcp_2020}), with $E_{\mbox{\tiny QLE}}$ representing the formation energy of a QLE. Equation~\eqref{eq:Boltzmann} suggests that $T_{\rm p}$ plays the role of a nonequilibrium thermodynamic temperature that carries memory of the equilibrium state at which a glass falls out of equilibrium, in particular of the configurational (as opposed to vibrational) degrees of freedom of the liquid, deep into the nonequilibrium glassy state~\cite{pinching_pnas}.

The existence of a nonequilibrium temperature in glasses, sometimes termed the fictive/effective/configurational temperature, and its relation to the number of QLEs (additional discussions regarding the number density of QLEs can be found in~\cite{uli_arXiv_stz_creation,mw_thermal_origin_of_qle_pre2020}), strongly echo the nonequilibrium thermodynamic Shear-Transformation-Zones (STZs) theory of glassy deformation~\cite{Bouchbinder2009b,Bouchbinder2009c,Falk2011}. This theory is based on a two-temperature nonequilibrium thermodynamic framework, where the number of STZs --- the ``flow defects'' in a glass --- follows a Boltzmann-like relation with the effective temperature as in Eq.~\eqref{eq:Boltzmann}, which satisfies its own field equation~\cite{Bouchbinder2009b,Bouchbinder2009c,Falk2011}. Consequently, if QLEs can be identified with STZs (or at least if they are strongly correlated with them, as shown in Subsects.~\ref{sec:stzs} and~\ref{sec:LHC}) and if $T_{\rm p}$ can be identified with the effective temperature, then Eq.~\eqref{eq:Boltzmann} and Fig.~\ref{fig:Tp_dependence}c provide interesting support to one of the main predictions of the STZ theory of glassy deformation.

The STZ theory provides various predictions regarding the elasto-plastic deformation of glasses in a wide variety of physical situations~\cite{Falk2011}. In particular, the strong depletion of STZs with decreasing effective temperature, consistent with the Boltzmann-like relation, has been shown to
give rise to a ductile-to-brittle transition in the fracture toughness
of glasses~\cite{Rycroft2012,Vasoya2016}. This prediction has been recently supported
by experiments on the fracture toughness of bulk metallic glasses, where
the effective temperature has been carefully controlled and varied~\cite{Eran_mechanical_glass_transition}. In addition to establishing a connection between individual QLEs and STZs, as discussed in Subsects.~\ref{sec:stzs} and~\ref{sec:LHC}, large-scale computer simulations can shed light on the collective effect of ${\cal A}_{\rm g}\!\sim\!{\cal N}$ (once the former is properly nondimensionalized) on the nonlinear and dissipative mechanics of glasses. Indeed, very recently it has been shown that the variation of the fracture toughness of various computer glasses with thermal history and the underlying interparticle interaction potential is largely controlled by the dimensionless ${\cal A}_{\rm g}$, proving strong support to the central role played by QLEs in the physics of glasses~\cite{david_fracture_2021}.

\subsection{Effect of interparticle interactions on the properties of QLEs}
\label{sec:effect_of_interactions}

In the previous Subsection, we have seen that while the quartic law of the nonphononic VDoS appears to universally hold, the number density, structural and energetic properties of QLEs' can be affected by the formation history of structural glasses. Recent work~\cite{sticky_spheres1_karina_pre2021,david_fracture_2021} has demonstrated that the nature of interparticle potentials may affect QLE-properties as well. Here, we concisely review the results of~\cite{sticky_spheres1_karina_pre2021} in which the effect of variations of a simple, pairwise interparticle potential on the properties of QLEs was studied.

\begin{figure}[ht!]
  \includegraphics[width = 0.50\textwidth]{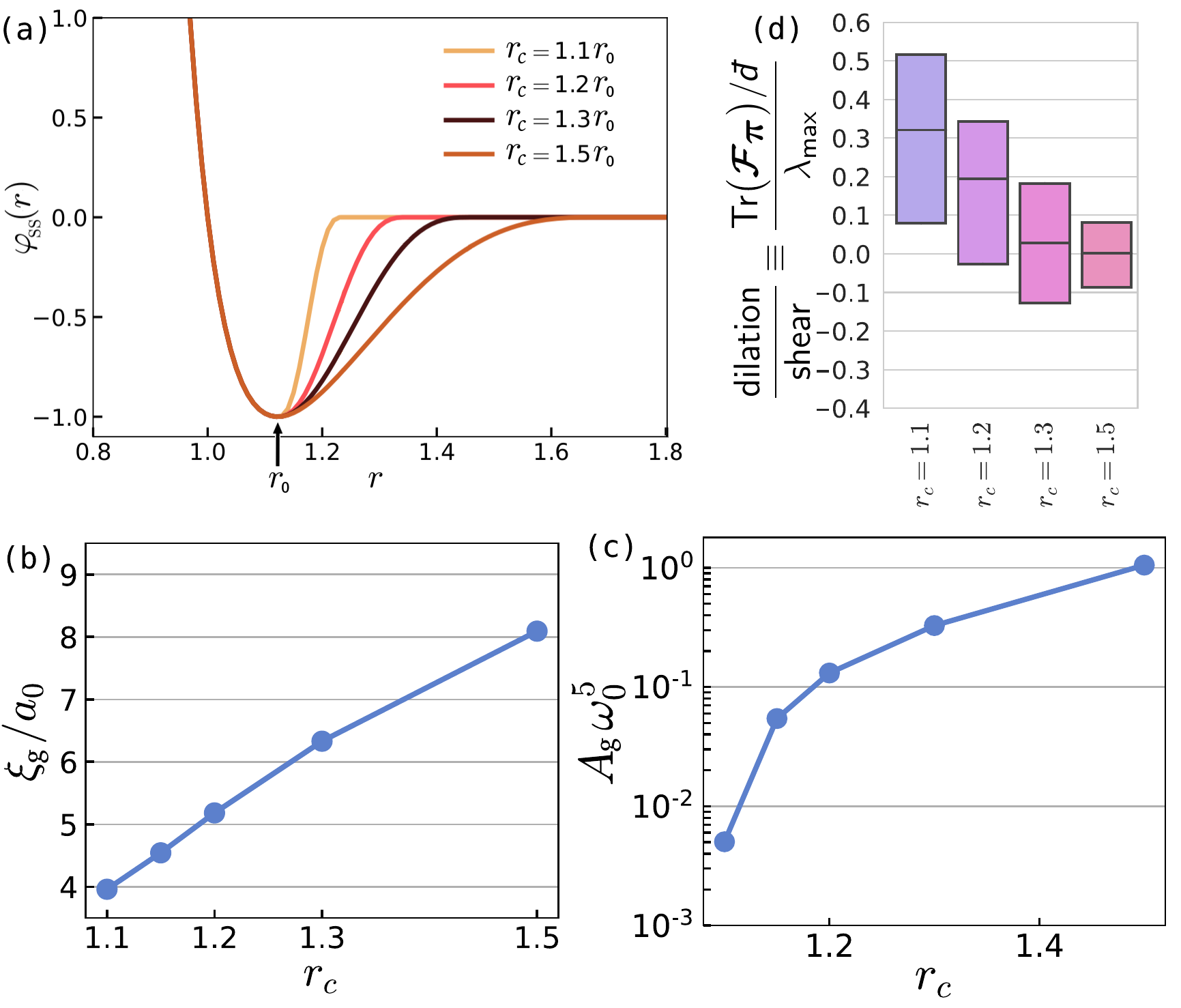}
  \caption{\footnotesize Effect of interparticle potential on QLE-properties, data from~\cite{sticky_spheres1_karina_pre2021}. (a) The sticky-spheres pairwise potential $\varphi_{_{\mbox{\tiny \rm SS}}}(r)$ put forward in~\cite{itamar_sticky_spheres_potential_pre_2011}. The interaction-range cutoff $r_{\rm c}$ serves as the key control parameter affecting QLEs' properties (note that $r_0$ denotes the location of the minimum of the potential). (b) The lengthscale $\xi_{\rm g}$ that characterizes the core size of QLEs \emph{decreases} with decreasing $r_{\rm c}$ by roughly a factor of two. (c) The dimensionless nonphononic VDoS prefactor ${\cal A}_{\rm g}\omega_0^5$ varies by over two decades under changes of $r_{\rm c}$. (d) The ratio of dilatant-to-shear strain associated with QLEs (see text for exact definitions) grows significantly with decreasing $r_{\rm c}$.}
  \label{fig:interaction_dependence}
\end{figure}

The main results of~\cite{sticky_spheres1_karina_pre2021} relevant to the current discussion are summarized in Fig.~\ref{fig:interaction_dependence}. Panel (a) shows the tunable pairwise interaction potential; the interaction range, denoted by $r_{\rm c}$, is the key control parameter that affects QLE-properties, along with other mechanical and elastic observables. Panels (b) and (c) show the dependence of the (linear) core size $\xi_{\rm g}$ of QLEs and the (dimensionless) prefactor ${\cal A}_{\rm g}\omega_0^5$ of the nonphononic VDoS, respectively, on the cutoff $r_{\rm c}$ (recall that $\omega_0\!\equiv\!c_s/a_0$, where $c_s$ is the shear wave-speed and $a_0$ a typical interparticle distance). The length $\xi_{\rm g}$ varies by over a factor of two, which exceeds the $T_{\rm p}$-induced variation of $\xi_{\rm g}$ as shown in Fig.~\ref{fig:Tp_dependence}b. The dimensionless prefactor ${\cal A}_{\rm g}\omega_0^5$ varies by over two decades; in~\cite{sticky_spheres1_karina_pre2021}, it is further demonstrated that the variability of ${\cal A}_{\rm g}\omega_0^5$ stems mostly from the stiffening of QLEs (i.e.~the increase in $\omega_{\rm g}$ or decrease in $\xi_{\rm g}$) with reducing $r_{\rm c}$, rather than from their depletion (with the exception of the $r_{\rm c}\!=\!1.1$ systems, see further discussions in~\cite{sticky_spheres1_karina_pre2021,karina_minimal_disorder}).

In the same work~\cite{sticky_spheres1_karina_pre2021}, a quantifier of the \emph{geometry} of QLEs was put forward, with the aim of assessing the ratio of shear vs.~dilatant strain that QLEs feature. The quantifier is constructed as follows; for a QLE given by a normalized displacement field $\piv$, we define the tensor
\begin{equation}\label{eq:cal_F_definition}
    \calBold{F}_{\piv}\equiv \frac{\partial^2U}{\partial \bm{\epsilon}\partial\xv}\cdot\piv\,,
\end{equation}
where $U(\xv)$ is the potential energy, which depends on coordinates $\xv$, and $\bm{\epsilon}$ is the strain tensor~\cite{landau_lifshitz_elasticity}. Next, $\calBold{F}_{\piv}$ is decomposed into its deviatoric and dilatational contributions as $\calBold{F}_{\piv}\!=\!\calBold{F}_{\piv}^{\mbox{\tiny iso}}\!+\calBold{F}_{\piv}^{\mbox{\tiny dev}}$, where $\calBold{F}_{\piv}^{\mbox{\tiny iso}}\!\equiv\!\calBold{I}\,\mbox{Tr}(\calBold{F}_{\piv})/\dbar$ ($\calBold{I}$ is the identity tensor) and $\calBold{F}_{\piv}^{\mbox{\tiny dev}}\!\equiv\!\calBold{F}_{\piv}\!-\!\calBold{F}_{\piv}^{\mbox{\tiny iso}}$. $\calBold{F}_{\piv}^{\mbox{\tiny dev}}$ is then diagonalized and its eigenvalue with the largest absolute magnitude $\lambda_{\mbox{\scriptsize max}}$ is recorded. The ratio of dilatational to shear strain of the QLE $\piv$ is finally defined as
\begin{equation}
    \frac{\mbox{dilation}}{\mbox{shear}} = \frac{\mbox{Tr}(\calBold{F}_{\piv})/\dbar}{\lambda_{\mbox{\scriptsize max}}}\,.
\end{equation}
The behavior of the dilation-to-shear ratio calculated over a few thousand QLEs observed in computer glasses of different $r_{\rm c}$'s is presented in Fig.~\ref{fig:interaction_dependence}d; the color bars cover the second and third quartiles of the dilation-to-shear ratio, and the middle horizontal line represents the mean ratio. Interestingly, reducing $r_{\rm c}$ leads to the development of a much larger dilatational component of the strain fields associated with QLEs.

We note that the same computer glasses whose QLEs feature large dilation-to-shear strain ratios were also shown to have relatively small Poisson's ratios~\cite{sticky_spheres1_karina_pre2021}, and to fail in a brittle fashion under uniaxial loading~\cite{david_fracture_2021}. A continuum analog of the geometric quantifier of QLEs discussed here was introduced and compared to the microscopic quantifier described above in~\cite{avraham_core_properties_pre_2020}. Similar approaches towards quantifying the geometry of plastic instabilities in computer glasses were discussed in~\cite{uli_arXiv_stz_creation,Rodney_pre_2016,Rottler_pre_2018}. We finally  note that while the introduction of strong attractive interactions {\em may} affect QLEs' properties --- as shown in Fig.~\ref{fig:interaction_dependence} --- they do not necessarily do so, as discussed at length in~\cite{sticky_spheres1_karina_pre2021}.

\subsection{QLEs near the unjamming transition}
\label{sec:unjamming}

The unjamming transition is an elasto-mechanical instability that occurs in gently compressed disordered packings of soft spheres upon reducing their pressure $p$ toward zero~\cite{ohern2003,liu2011jamming,liu_review,van_hecke_review}. Ikeda and coworkers~\cite{ikeda_pnas,atsushi_core_size_pre} have recently studied using computer simulations how QLEs' statistical and structural properties in harmonic soft-sphere packings are affected by the proximity of those packings to the unjamming transition.

The key microscopic observable in the context of the unjamming transition is the coordination difference $\delta Z\!\equiv\!Z\!-\!Z_c$ to the so-called Maxwell threshold $Z_c\!=\!2\dbar$, where $Z$ denotes the mean number of contacts per particle in a packing. For many canonical soft-sphere models near the unjamming point, $p/K\!\sim\!\delta Z^2$ (where $K$ is the bulk modulus)~\cite{ohern2003,liu2011jamming,liu_review,van_hecke_review}. In~\cite{atsushi_core_size_pre}, it was shown that $\xi_{\rm g}\!\sim\!\delta Z^{-1/2}$ using numerical simulations, consistent with previous observations of diverging lengthscales near the unjamming point~\cite{Silbert_prl_2005,breakdown,liu_transverse_length_2013,brian_prl_2017,quasilocalized_states_of_self_stress}. In~\cite{ikeda_pnas,atsushi_core_size_pre}, it was claimed based on scaling arguments and demonstrated numerically that the dimensionless prefactor of the quartic nonphononic VDoS near unjamming follows ${\cal A}_{\rm g}\omega_0^5\!\sim\!\delta Z^{-3/2}$.

Interestingly, it was shown in~\cite{atsushi_core_size_pre} that the product $Ne$ (with $e$ denoting the participation ratio, cf.~Eq.~(\ref{eq:participation_ratio})) --- which represents the effective volume of QLEs' core --- scales as $1/\delta Z\!\sim\!\xi_{\rm g}^2$ instead of the naive expectation $\sim\!\xi_{\rm g}^\dbar$ in $\dbar$ dimensions. This result is consistent with the $1/\delta Z$ scaling of the sum-of-squares of the displacement response to local dipole forces in disordered networks of relaxed Hookean springs, as spelled out in~\cite{breakdown}. According to a recent replica calculation of the overlap correlation function~\cite{mw_hopping_qle_arXiv_2021}, this scaling between the effective volume of a QLE and its core length $\xi_{\rm g}$ stems from the pre-asymptotic spatial decay $\sim\!r^{-(\dbar-2)/2}$ of QLE displacements at distances $r\!<\!\xi_{\rm g}$.

\subsection{Mean-field models of QLEs}
\label{sec:mean_field}

\begin{figure*}[ht!]
  \includegraphics[width = 0.95\textwidth]{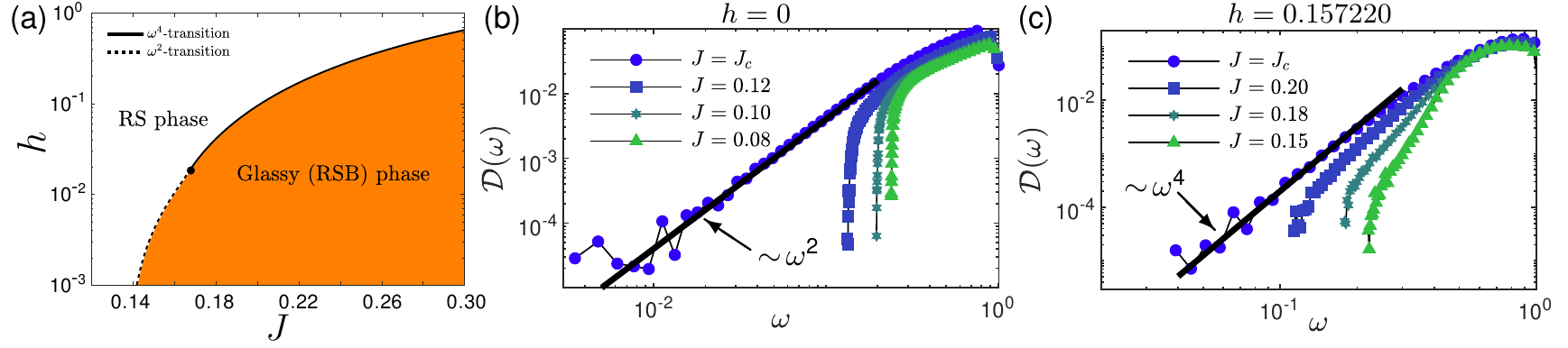}
  \caption{\footnotesize (a) The $h\!-\!J$ phase diagram of the KHGPS model, defined in Eq.~\eqref{eq:KHGPS}, with $\kappa_{\mbox{\tiny min}}\!=\!0.1$ and $\kappa_0\!=\!1$. The critical transition line $J_{\rm c}(h)$ separates the RS phase from the glassy (RSB) phase, see text and~\cite{meanfield_qle_pierfrancesco_prb_2021} for additional details. On the dotted line, one has ${\cal D}(\omega)\!\sim\!\omega^2$, while on the solid line, one has ${\cal D}(\omega)\!\sim\!\omega^4$. (b) ${\cal D}(\omega)$ upon approaching the $\omega^2$-transition line (dotted line in panel (a)). (c) ${\cal D}(\omega)$ upon approaching the $\omega^4$-transition line (solid line in panel (a)).}
  \label{fig:mean_field1}
\end{figure*}
Several efforts to understand the disorder-induced properties of the low-frequency spectra of structural glasses based on mean-field models have been put forward in previous literature; some notable examples include Fluctuating Elasticity Theory (FET)~\cite{Schirmacher_2006,schirmacher2011comments,Schirmacher_2013_boson_peak}, Effective Medium Theory~\cite{eric_boson_peak_emt,eric_hard_sphere_vdos_pnas2014}, the perceptron model~\cite{silvio}, and the mean-field theories for hard-sphere glasses~\cite{Zamponi_2014,parisi_fractal,zamponi_hard_spheres_review_2017} and jammed packings~\cite{parisi_mean_field_w4}. Some of these models, e.g.~\cite{eric_boson_peak_emt,silvio,Sharma_pre_2016}, predict that the nonphononic VDoS follows an $\omega^2$ scaling with frequency, independent of spatial dimension. FET predicts a dimension-dependent, Rayleigh-like scaling $\sim\!\omega^{\dbar+1}$ for excess modes that are spatially extended (i.e.~FET does not predict QLEs)~\cite{schirmacher2011comments}.

Recently, a mean-field model for QLEs in structural glasses was put forward~\cite{scipost_mean_field_qles_2021,meanfield_qle_pierfrancesco_prb_2021}. The model is a generalization of the 3D model for anharmonic interacting oscillators by Gurevich, Parshin and Schober (GPS)~\cite{Gurevich2003,Gurevich2005,Gurevich2007} discussed at length in Sect.~\ref{sec:reconstruction}, which is also somewhat reminiscent of the earlier model by K\"uhn and Horstmann (KH)~\cite{Kuhn_Horstmann_prl_1997}; as such, it was termed the KHGPS model in~\cite{scipost_mean_field_qles_2021,meanfield_qle_pierfrancesco_prb_2021}.

The KHGPS mean-field model, defined through the following Hamiltonian
\begin{equation}
\label{eq:KHGPS}
    H_{\mbox{\tiny KHGPS}} = \frac{1}{2}\!\sum_i\!\kappa_ix_i^2 +\!\sum_{i<j}\!J_{ij}x_ix_j + \frac{A}{4!}\!\sum_i x_i^4 -h\!\sum_ix_i\,,
\end{equation}
can be formally obtained from Eq.~\eqref{eq:GPS} by taking the space-dependent interaction coefficients $J_{ij}(r_{ij})$ to be {\em space independent} (corresponding to taking the infinite-dimensional limit, $\dbar\!\to\!\infty$, of the dipole-dipole elastic interaction $r^{-\dbar}$) and Gaussian, i.i.d. random variables of variance $J^2/N$; here $J$ represents the interaction strength and $N$ is the number of interacting oscillators described by the scalar coordinates $x_i$. In addition, the anharmonic oscillators are assumed to linearly interact with a constant field $h$, which breaks the $x_i\!\to\!-x_i$ symmetry of the Hamiltonian, a feature missing in Eq.~\eqref{eq:GPS}. Finally, the oscillator stiffnesses $\kappa_i$ are assumed to be drawn from a uniform parent distribution over the interval $[\kappa_{\mbox{\tiny min}},\kappa_0]$, where $\kappa_{\mbox{\tiny min}}\!<\!\kappa_0$ may be finite, which is yet another deviation from the GPS model that considered only gapless parent distributions for $\kappa_i$.

The KHGPS model defined in Eq.~\eqref{eq:KHGPS} is similar to the KH model considered in~\cite{Kuhn_Horstmann_prl_1997}, with the notable difference that in the latter the stiffnesses $\kappa_i$ were not taken to be random variables. Formally, in terms of the formulation above, the model in~\cite{Kuhn_Horstmann_prl_1997} corresponds to $\kappa_{\mbox{\tiny min}}\=\kappa_0\=1$. While the VDoS has not been studied in~\cite{Kuhn_Horstmann_prl_1997}, it was shown in~\cite{meanfield_qle_pierfrancesco_prb_2021} to give rise to an $\omega^2$ VDoS as $\omega\!\to\!0$ and to delocalized modes (i.e.~an ${\cal O}(N)$ oscillators feature sizable displacements $x_i$ at minima of the Hamiltonian that populate the $\omega^2$ regime). The KHGPS model defined in Eq.~\eqref{eq:KHGPS} also bears some similarity to the soft-spin version of the Sherrington-Kirkpatrick model~\cite{SK_model_1982}, though the latter has not been previously shown to be related to soft vibrational excitations. The KHGPS model has been recently analyzed in~\cite{scipost_mean_field_qles_2021,meanfield_qle_pierfrancesco_prb_2021} and the main results are briefly reviewed next.

In~\cite{meanfield_qle_pierfrancesco_prb_2021}, it was rigorously shown that if the parent stiffnesses distribution is gapped, i.e.~if $\kappa_{\mbox{\tiny min}}\!>\!0$, then for a fixed $h$ and sufficiently small interaction strength $J$, the VDoS corresponding to $H_{\mbox{\tiny KHGPS}}$ in Eq.~\eqref{eq:KHGPS} is also gapped. This phase is replica-symmetric (RS) and hence is denoted as the RS phase in Fig.~\ref{fig:mean_field1}a. With increasing $J$, there exists a critical line $J_{\rm c}(h)$ in the $h\!-\!J$ plane on which a gapless VDoS emerges. Interestingly, for small $h$ the gapless VDoS is populated by delocalized modes that follow a quadratic behavior $\sim\!\omega^2$, similarly to previous mean-field models~\cite{eric_boson_peak_emt,silvio,Sharma_pre_2016}. On the other hand, there exists a special point on the critical $J_{\rm c}(h)$ line above which (i.e.~for large enough $h$) the gapless VDoS is populated by localized modes that follow a quartic behavior $\sim\!\omega^4$. This result shows that in contrast to previous belief, mean-field models can in fact feature localized modes with an $\omega^4$ VDoS, similarly to direct observations in finite-dimensional computer glasses~\cite{modes_prl_2018}.

Finally, for $J\!>\!J_{\rm c}(h)$ the model features a replica-symmetry breaking (RSB), hence this regime is denoted as the glassy (RSB) phase in Fig.~\ref{fig:mean_field1}a. This phase has not been analyzed in~\cite{meanfield_qle_pierfrancesco_prb_2021}. The results described above are visually presented in Fig.~\ref{fig:mean_field1}, where the $h\!-\!J$ phase diagram is shown in panel (a), the VDoS upon approaching the critical $J_{\rm c}(h)$ line (from the RS phase) in the $\omega^2$ regime is shown in panel (b) and the VDoS upon approaching the critical $J_{\rm c}(h)$ line (from the RS phase) in the $\omega^4$ regime is shown in panel (c).

\begin{figure}[ht!]
  \includegraphics[width = 0.50\textwidth]{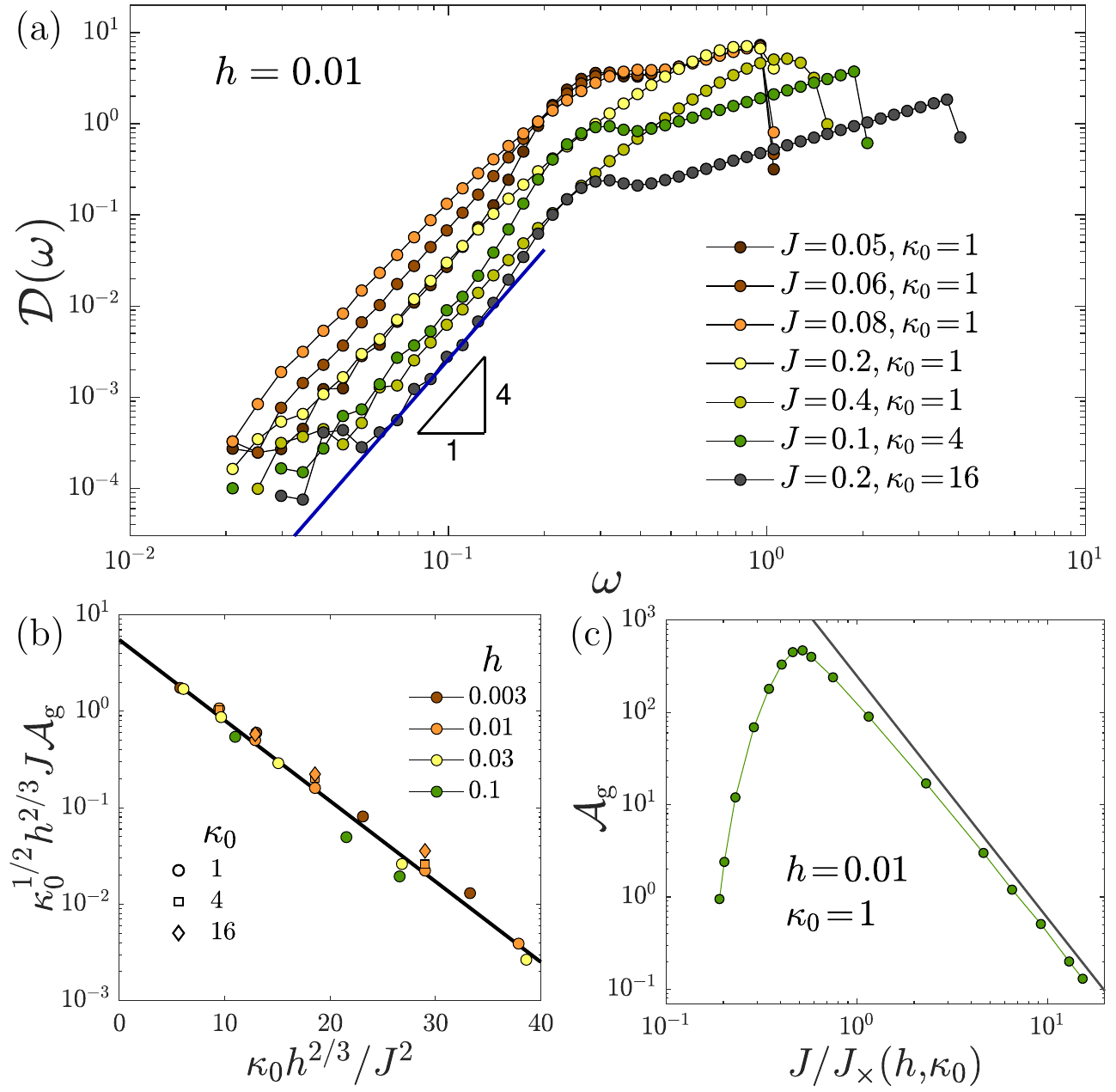}
  \caption{\footnotesize (a) ${\cal D}(\omega)$ of the KHGPS model in the glassy (RSB) phase, with $\kappa_{\mbox{\tiny min}}\!=\!0$, calculated numerically in systems of $N\!=\!\mbox{16,000}$ oscillators, for $h\!=\!0.01$ and various values of the parameters $J$ and $\kappa_0$ as indicated by the legend. The generic emergence of ${\cal D}(\omega)\!\sim\!\omega^4$ is demonstrated (see the $4\!:\!1$ triangle and note the double-logarithmic axes). (b) Numerical validation of the theoretical prediction for the predominantly exponential variation of ${\cal A}_{\rm g}(h, J,\kappa_0)$ with $-\kappa_0\,h^{2/3}J^{-2}$, see text and~\cite{scipost_mean_field_qles_2021} for details. (c) Numerical validation of the theoretical prediction for the predominantly power-law decay of ${\cal A}_{\rm g}(h, J,\kappa_0)$ with $J$, when the latter is sufficiently larger than the crossover interaction strength $\Jx(h,\kappa_0)$, see text and~\cite{scipost_mean_field_qles_2021} for details. The weak interactions regime, discussed in panel (b), is also presented such that the non-monotonic variation of ${\cal A}_{\rm g}(h, J,\kappa_0)$ with $J$ is evident.}
  \label{fig:mean_field2}
\end{figure}

When $\kappa_{\mbox{\tiny min}}$ is reduced toward zero, the $J_{\rm c}(h)$ line is pushed toward smaller values of $J$, and the $\omega^4$ regime increases; for $\kappa_{\mbox{\tiny min}}\=0$, the KHGPS model is in the RSB phase for all $J$ values~\cite{meanfield_qle_pierfrancesco_prb_2021}. While --- as stated above --- this glassy (RSB) regime has not been analyzed in~\cite{meanfield_qle_pierfrancesco_prb_2021}, it has been rather thoroughly studied numerically and through a scaling theory in~\cite{scipost_mean_field_qles_2021}. First, it was shown that the model in this regime gives rise to a gapless VDoS ${\cal D}(\omega)\= {\cal A}_{\rm g}\omega^4$ for a broad range of model parameters, as demonstrated in Fig.~\ref{fig:mean_field2}a. This result shows that the $\omega^4$ scaling of the VDoS in the KHGPS model persists deep inside the  glassy (RSB) regime. In addition, a complete understanding of the non-universal prefactor ${\cal A}_{\rm g}(h, J,\kappa_0)$ has been developed; recall that here $\kappa_{\mbox{\tiny min}}\=0$ and that the anharmonicity amplitude $A$ is fixed, hence the model is fully characterized by the parameters $h$, $J$ and $\kappa_0$.

It was theoretically predicted that in the weak interactions regime, i.e.~for $J$ smaller than a crossover level $\Jx(h,\kappa_0)$, the prefactor ${\cal A}_{\rm g}(h, J,\kappa_0)$ satisfies $\log\left[\kappa_0^{1/2}h^{2/3}J {\cal A}_{\rm g} \right]\!\sim\!-\kappa_0\,h^{2/3}J^{-2}$. The validity of this predominantly exponential variation
of ${\cal A}_{\rm g}(h, J,\kappa_0)$ with $-\kappa_0\,h^{2/3}J^{-2}$ is numerically demonstrated in Fig.~\ref{fig:mean_field2}b. Interestingly, this result is reminiscent of the predominantly exponential variation of the number density ${\cal N}$ of QLEs
with $-1/T_{\rm p}$ in computer glasses, shown above in Fig.~\ref{fig:Tp_dependence}c (see also Eq.~\eqref{eq:Boltzmann}). This similarity is suggestive, calling for a better understanding of the possible relations between the KHGPS model parameters $h$, $J$ and $\kappa_0$, and the parent
temperature $T_{\rm p}$ that characterizes the liquid state at which the glass falls out of equilibrium
during a quench.

Furthermore, it has been predicted in~\cite{scipost_mean_field_qles_2021} that for $J$ larger than a crossover interaction strength $\Jx(h,\kappa_0)$, ${\cal A}_{\rm g}$ predominantly decays with $J$ as a power-law, as is numerically demonstrated in Fig.~\ref{fig:mean_field2}c. In addition, the analysis in~\cite{scipost_mean_field_qles_2021} has revealed the existence of a characteristic frequency $\wx$ scale in the KHGPS model, which is reminiscent of the crossover frequency $\omega_{\rm b}$ mentioned in Sect.~\ref{sec:reconstruction} in the context of the reconstruction picture, and studied its properties. The importance of this frequency scale has been further highlighted in a very recent work~\cite{pierfrancesco_arXiv_2021}, where additional intriguing features of the KHGPS model are revealed using long time gradient descent dynamics in the glassy (RSB) phase (based on a dynamical mean field theory) and replica method calculations. All in all, the formulation of the KHGPS model and its revealed properties as of now show, contrary to previous belief, that mean-field models can share similar properties with finite-dimensional computer glasses in relation to QLEs, most notably modes localization and the $\sim\omega^4$ VDoS.

At the same time, the KHGPS model raises various questions and challenges. One class of questions concerns the relations between the model and finite-dimensional, realistic glasses. In particular, it remains a challenge to understand the relations between the model's inputs, such as the parent stiffnesses distribution and the parameters $J$ and $h$, and the self-organizational processes taking place during glass formation when a liquid is quenched. Establishing such connections, qualitative or even quantitative, appears essential for clarifying what mean-field models such as the KHGPS one may teach us about the physics of glasses at the fundamental level. Another class of challenges, more on the mathematical physics side, concerns the rigorous analysis of the glassy (RSB) phase, which has been so far mainly analyzed numerically and through a scaling theory (but see~\cite{pierfrancesco_arXiv_2021}). Finally, exploring whether such mean-field models can teach us something deep about glassy dynamics -- not just statistical properties of glassy structures --- appears to be an interesting direction for future investigation.

\subsection{Overcoming hybridizations: nonlinear frameworks for QLEs}
\label{sec:nonlinear}

In Sect.~\ref{sec:hybridization}, we described how finite-size scaling of QLEs and phononic excitations may be utilized to overcome their hybridization~\cite{modes_prl_2016}. Back in 1996, Schober and Oligschleger already proposed a demixing/dehybridization procedure to disentangle extended phonons and QLEs~\cite{Schober_Oligschleger_numerics_PRB}. In recent years, several other computational frameworks for overcoming phonon-QLE hybridizations have been put forward~\cite{parisi_spin_glass,SciPost2016,manning_defects,episode_1_2020,julia_arXiv}. In this Subsection, we describe one of these frameworks, which enables to robustly single out QLEs regardless of whether or not phononic excitations with similar frequencies exist in the glass.

Consider a zero-temperature glass, namely a configuration $\xv_0$ that constitutes a (local) minimum of the glass's potential energy $U(\xv)$. Consider next the energy variation $\delta U \!\equiv\! U(\xv)\!-\! U(\xv_0)$ that results from displacing the particles a distance $s$ in the configuration-space \emph{direction} prescribed by a given putative displacement field $\zv$, namely
\begin{equation}
\label{eq:cubic_expansion}
    \delta U(s) \simeq \frac{1}{2}b_2 s^2 + \frac{1}{3!}b_3 s^3\,,
\end{equation}
where
\begin{eqnarray}
b_2(\zv) & \equiv & \frac{\partial^2U}{\partial\xv\partial\xv}\bigg|_{\xv_0}\!\!:\zv\zv \,, \label{eq:b2_definition} \\
b_3(\zv) & \equiv & \frac{\partial^3U}{\partial\xv\partial\xv\partial\xv}\bigg|_{\xv_0}\!\!\tripleCdot\zv\zv\zv \,. \label{eq:tau_definition}
\end{eqnarray}
Here $:,\tripleCdot$ represent double and triple contractions, respectively,
and we assume that $\zv$ is normalized such that $\zv\!\cdot\!\zv\!=\!1$.
Notice that the absence of a first-order term in Eq.~(\ref{eq:cubic_expansion}) stems from the mechanical equilibrium condition $\frac{\partial U}{\partial\xv}\big|_{\xv_0}\!=\!\zerovector$ that applies to any minimum $\xv_0$ of $U(\xv)$. The cubic expansion as appears in Eq.~(\ref{eq:cubic_expansion}) implies that an energy barrier of height
\begin{equation}
\label{eq:Barrier}
    B(\zv) = \frac{2[b_2(\zv)]^3}{3[b_3(\zv)]^2}
\end{equation}
exists at a displacement of $s_\star\!=\!-2b_2/b_3$ away from the minimum at $s\!=\!0$ of the potential energy $U$.

Important to the present discussion is that the barrier $B(\zv)$ is fully expressed in terms of the configuration-space \emph{direction} $\zv$; all of its dependence on the details of the minimum $\xv_0$ and on the potential energy topography in the vicinity of $\xv_0$ are embodied in the tensors $\frac{\partial^2U}{\partial\xv\partial\xv}$ and $\frac{\partial^3U}{\partial\xv\partial\xv\partial\xv}$, evaluated at $\xv_0$. Consequently, one can ask: which configuration-space direction $\zv$ should be \emph{chosen}, such that the energy barrier on the way towards an adjacent, nearby minimum of the potential energy is small?

\begin{figure}[ht!]
  \includegraphics[width = 0.49\textwidth]{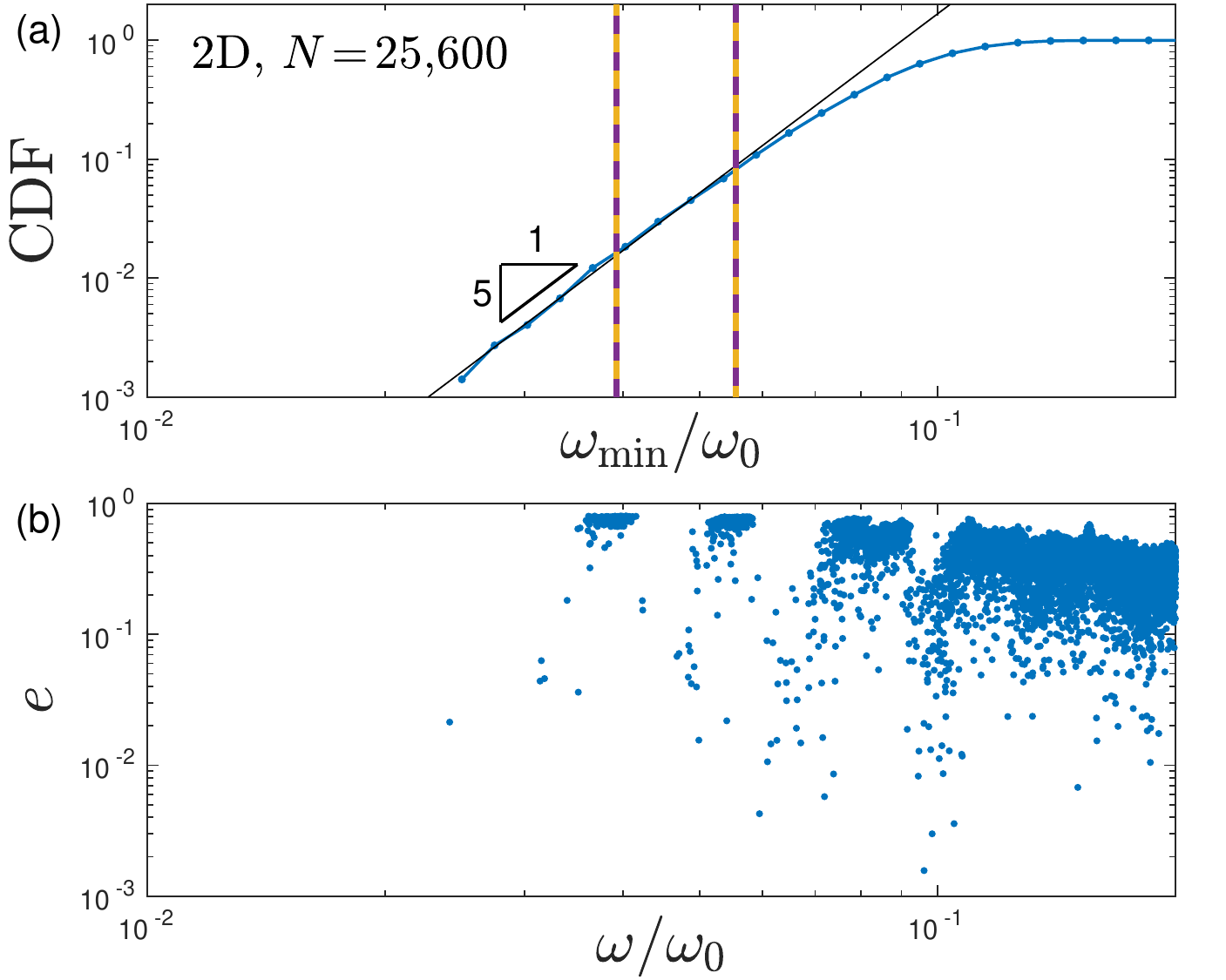}
  \caption{\footnotesize (a) Cumulative distribution function (CDF) of the sample-to-sample statistics of minimal-frequency QLEs (per sample), measured in $\approx\!\mbox{10,000}$ computer glasses in 2D using the nonlinear framework described here, see~\cite{cge_paper} for details about the model, and Appendix~B in~\cite{episode_1_2020} for details about the calculation of QLEs. The striped vertical lines mark the two lowest-phonon frequencies. (b) Participation ratio $e$ of vibrational modes measured in the same 2D computer glasses analyzed in (a),  scatter-plotted against frequency $\omega$. These data demonstrate the robustness of the nonlinear framework against phonon-hybridizations of QLEs seen in harmonic analyses. The results presented in the figure have not been published elsewhere.}
  \label{fig:omega_min_2D}
\end{figure}

Finding the aforementioned directions that minimize $B(\zv)$ is equivalent to finding fields $\piv$ that solve the equation
\begin{equation}\label{eq:NQLE_equation}
    \frac{\partial^2U}{\partial\xv\partial\xv}\cdot\piv = \frac{b_2}{b_3}\frac{\partial^3U}{\partial\xv\partial\xv\partial\xv}:\piv\piv\,,
\end{equation}
obtained by requiring $\frac{\partial B}{\partial \zv}\big|_{\piv}\!=\!\zerovector$. In~\cite{plastic_modes_prerc,micromechanics2016,SciPost2016,episode_1_2020}, it was established that solutions $\piv$ are QLEs; they can be obtained either by solving Eq.~(\ref{eq:NQLE_equation}) via an iterative scheme spelled out in~\cite{SciPost2016}, or by performing a numerical minimization of the barrier $B(\zv)$ with respect to the direction $\zv$~\cite{footnote2}. Solutions $\piv$ that pertain to minima of $B(\zv)$ are both low-energy --- due to the numerator $\sim\!b_2^3$ of $B(\zv)$ ---, and highly localized -- due to the denominator $\sim\!b_3^2$ of $B(\zv)$, which was shown in~\cite{plastic_modes_prerc,SciPost2016} to be larger for more localized modes. In addition, they feature the same properties as those described above for QLEs: their frequency distribution follows the same universal $\omega^4$ law, see Fig.~\ref{fig:omega_min_2D}a, where new results that have not been published earlier are presented, and their spatial structure features the same disordered core and algebraic far-field decay of QLEs~\cite{SciPost2016}. Importantly, in~\cite{episode_1_2020} it was shown that the energies associated with QLEs obtained via the nonlinear framework described here are comparable to the energies of non-hybridized QLEs obtained via harmonic analyses.

One key advantage of the nonlinear-QLE framework is its robustness against hybridization, primarily with phonons~\cite{SciPost2016}, but also between QLEs that are spatially adjacent and close in frequency~\cite{episode_1_2020}. These properties allow to establish the universal quartic law for QLE-frequencies in 2D, in system sizes for which hybridizations with phonons largely obscure QLEs within the harmonic analysis, as demonstrated in Fig.~\ref{fig:omega_min_2D}. Some useful generalizations of the nonlinear-QLE framework reviewed here are discussed in~\cite{SciPost2016,episode_1_2020,pseudo_harmonic_prl}.

\subsection{Properties of nonlinear QLEs}
\label{sec:nonlinear_qles}

QLEs obtained via the nonlinear framework discussed in the previous Subsection feature some interesting properties different from their harmonic counterparts. Consider again a soft potential given by a quartic expansion of the potential energy along a QLE $\piv$ (and recall Eq.~\eqref{eq:SPM}), namely
\begin{equation}
\label{eq:quartic_expansion}
    \delta U(s) \simeq \frac{1}{2}b_2s^2 + \frac{1}{3!}b_3s^3 +
    \frac{1}{4!}b_4s^4\,,
\end{equation}
where we have introduced the quartic coefficient associated with $\piv$,
\begin{equation}
\label{eq:b4_definition}
    b_4 \equiv \frac{\partial^4U}{\partial\xv\partial\xv\partial\xv\partial\xv}::\piv\piv\piv\piv\,,
\end{equation}
and the notation $::$ stands for a contraction over 4 fields. The nonlinear framework for QLEs $\piv$ and the accompanying definitions of the coefficients $b_2,b_3$ and $b_4$ (Eqs.~(\ref{eq:b2_definition}), (\ref{eq:tau_definition}) and (\ref{eq:b4_definition}), respectively) offer a concrete realization of the local soft-potentials as postulated within the Soft Potential Model (SPM) discussed at length in Sect.~\ref{sec:spm}.

\begin{figure}[ht!]
  \includegraphics[width = 0.49\textwidth]{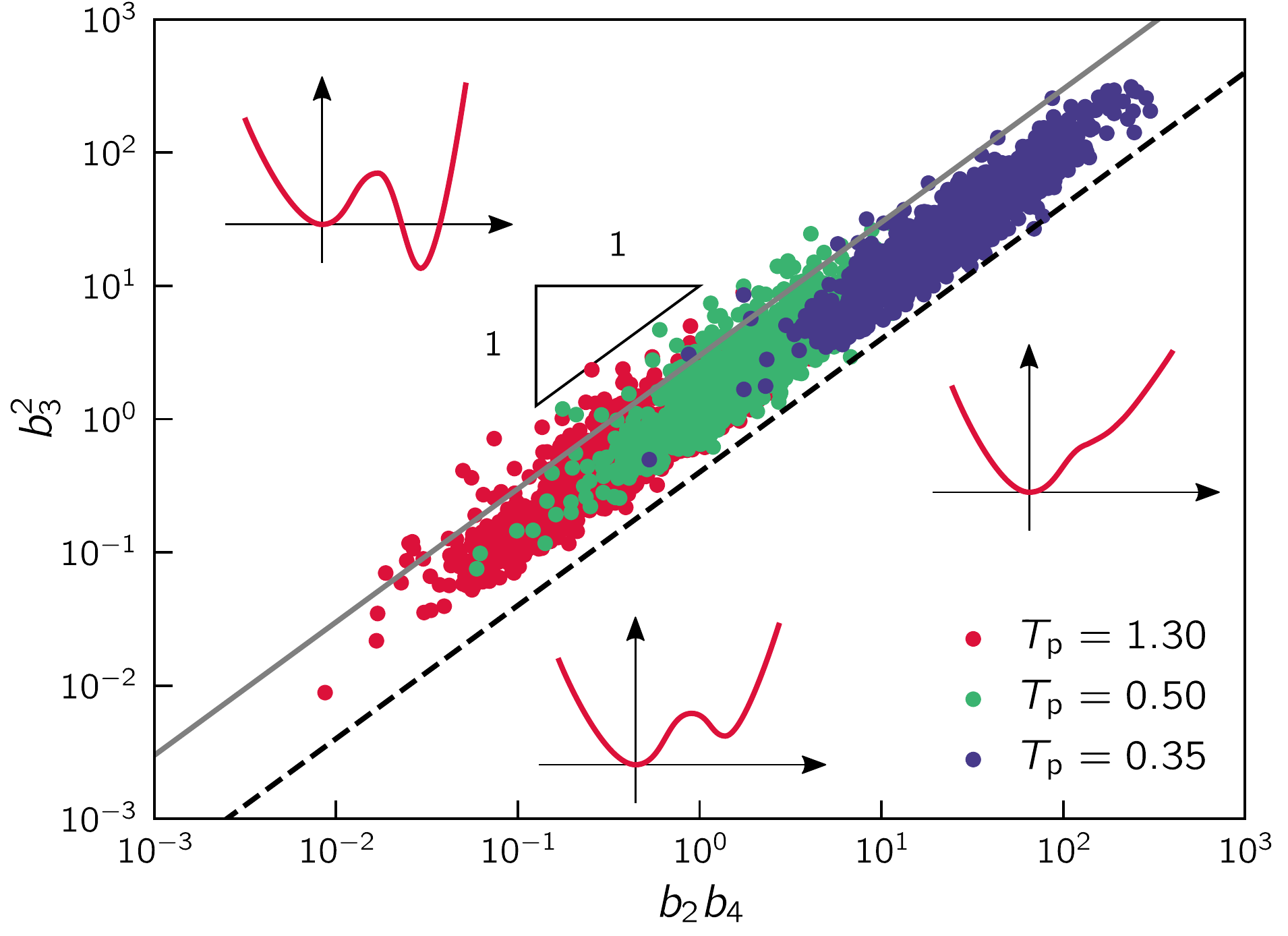}
  \caption{\footnotesize The square of the cubic expansion coefficient $b_3^2$ associated with QLEs is scatter-plotted against the product $b_2b_4$ of the quadratic and quartic expansion coefficients (cf.~Eq.~\eqref{eq:quartic_expansion}), all calculated using the nonlinear-QLE framework~\cite{plastic_modes_prerc,SciPost2016,episode_1_2020}. The continuous line represents the stability bound $b_3^2\!\le\!3b_2b4$ as spelled out in the SPM (cf.~Sect.~\ref{sec:spm}), whereas the dashed line marks an empirical lower-bound that still lacks a theoretical explanation.}
  \label{fig:b2b3b4}
\end{figure}

In Fig.~\ref{fig:b2b3b4}, we scatter-plot $b_3^2$ vs.~the product $b_2b_4$ measured for QLEs obtained via the nonlinear framework applied to glasses quenched from equilibrium liquid states at parent temperatures as seen in the legend, and see~\cite{episode_1_2020} for further details about the model and calculation. Also marked by the continuous line is the stability bound $b_3^2\!=\!3b_2b_4$ as spelled out in the SPM. In essence, this bound assumes that if soft potentials have a double-well form, the system will typically reside in the lower-energy well. Indeed, lower-$T_{\rm p}$ ensembles --- comprised of glasses with enhanced mechanical stability --- seem to better satisfy this bound.

Not predicted by the SPM is the curious observation that a lower bound $b_3^2\!\gtrsim\!0.4b_2b_4$ --- marked by the dashed line in Fig.~\ref{fig:b2b3b4} --- appears to exist. In~\cite{episode_1_2020}, it was shown that QLEs calculated via other definitions (e.g.~QLEs that assume the form of harmonic vibrations) do not satisfy any lower bound --- their associated cubic coefficient $b_3$ can be arbitrarily small. There is currently no theoretical explanation for the emergence of this lower bound.

In addition, since $b_4$ is largely independent of $b_2\!\sim\!\omega^2$~\cite{modes_prl_2016,episode_1_2020}, the lower and upper bounds on $b_3^2$ imply a tight correlation between a QLE's stiffness $b_2$ and its associated cubic coefficient $b_3$; in particular, we deduce that
\begin{equation}\label{eq:b3_b2_relation}
|b_3|\sim\sqrt{b_2}\sim\omega\,.
\end{equation}
Consider next a quartic soft potential in the form of Eq.~(\ref{eq:quartic_expansion}), and let us assume that $b_3^2\!>\!(8/3)b_2b_4$, i.e.~that the quartic expansion corresponds to a double-well potential. In such cases, and using the scaling relation Eq.~(\ref{eq:b3_b2_relation}) spelled out above, we expect the barrier $B$ between the two potential wells (cf.~Eq.~\eqref{eq:Barrier}) to follow~\cite{buchenau_prb_1992,episode_1_2020}
\begin{equation}\label{eq:barrier_scaling}
B \sim b_2^2 \sim \omega^4\,.
\end{equation}

Incorporating next the universal quartic law ${\cal D}(\omega)\!\sim\!\omega^4$ together with Eqs.~\eqref{eq:b3_b2_relation}-\eqref{eq:barrier_scaling}, we arrive at a prediction for the distribution of energy barriers $B$ on the potential energy landscape, in the $B\!\to\!0$ limit, which reads
\begin{equation}
    p(B) \sim B^{1/4}\,.
\end{equation}
This prediction assumes that the frequency distribution of QLEs with $b_3^2\!>\!(8/3)b_2b_4$ follows the universal quartic law. Future research should validate or refute this prediction, which we expect to hold for any structural glass quenched from a melt.

\subsection{QLEs as carriers of plastic flow}
\label{sec:stzs}

QLEs $\piv$ obtained from the nonlinear framework discussed in the previous Subsection are natural candidates to serve as the `Shear-Transformation-Zones' (STZs) envisioned by Falk and Langer in the late 1990s~\cite{falk_langer_stz}. In addition to their robustness against hybridization with phononic excitations, the stiffness $b_2\!=\!\frac{\partial^2U}{\partial\xv\partial\xv}\!:\!\piv\piv$ associated with QLEs $\piv$ follows a compact and physically transparent equation of motion with respect to the imposed deformations in the athermal, quasistatic limit~\cite{plastic_modes_prerc,micromechanics2016}; to leading order, it reads
\begin{equation}\label{eq:plastic_modes_eom}
    \frac{db_2}{d\bm{\epsilon}} \simeq -\frac{b_3\,\calBold{F}_{\piv}}{b_2}\,,
\end{equation}
where $\calBold{F}_{\piv}$ and $b_3$ were defined above in Eqs.~(\ref{eq:cal_F_definition}) and (\ref{eq:tau_definition}), respectively.

\begin{figure}[ht!]
  \includegraphics[width = 0.49\textwidth]{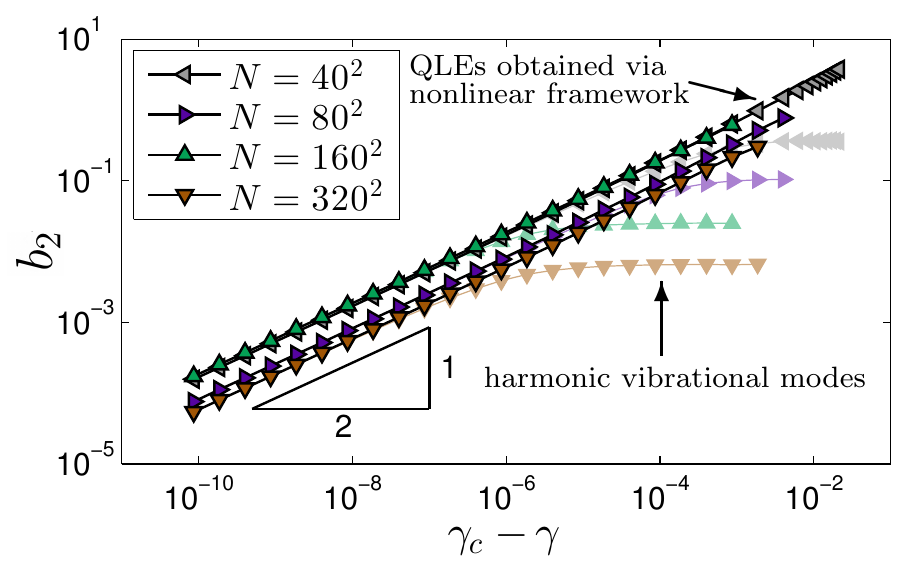}
  \caption{\footnotesize The strain-induced softening of the stiffness $b_2$ of QLEs near plastic instabilities (occurring at $\gamma_c$) observed in 2D computer glasses under athermal, quasistatic simple shear~\cite{micromechanics2016}. The outlined symbols represent QLEs measured using the nonlinear framework, whereas the pale symbols represent the stiffness of the lowest harmonic vibrational mode in the glass.}
  \label{fig:qle_eom_with_strain}
\end{figure}

If one considers an imposed-deformation tensor $\bm{\epsilon}(\gamma)$ parameterized by a strain parameter $\gamma$, Eq.~(\ref{eq:plastic_modes_eom}) would imply that near a plastic instability at a critical strain $\gamma_c$, one expects $b_2\!\sim\!\sqrt{\gamma_c\!-\!\gamma}$~\cite{lemaitre2004}, as indeed verified for a variety of system sizes in Fig.~\ref{fig:qle_eom_with_strain} (outlined symbols). We note that the strain interval $\gamma_c\!-\!\gamma$ over which this scaling holds is much larger than what is seen for deformation-induced destabilizing harmonic vibrational modes (pale symbols in Fig.~\ref{fig:qle_eom_with_strain}) due to phonon-hybridizations~\cite{micromechanics2016}. In addition, in~\cite{pseudo_harmonic_prl} it was shown that QLEs that are activated in plastic instabilities can be detected using the nonlinear framework discussed above at strains of order 5\% away from plastic instabilities, establishing that STZs are a priori encoded in a glass's microstructure, at odds with the claims of~\cite{Gendelman_2015}. We finally note again that the prefactor ${\cal A}_{\rm g}$ of the nonphononic VDoS was shown in~\cite{david_fracture_2021} to control the fracture toughness of computer glasses subjected to athermal, quasistatic external loading, further reinforcing the role of QLEs as the carriers of plastic flow in structural glasses.

\subsection{The effect of QLEs on physical observables in the low-temperature limit}
\label{sec:LHC}

In Subsect.~\ref{sec:qle_properties}, the effect of QLEs --- in particular of their number --- on the nonlinear and dissipative properties of glasses has been briefly discussed. QLEs also have significant effects on the statistical and spatial properties of physical observables defined with respect to glassy inherent structures (local minima of $U({\xv})$) in the low-temperature limit. To see this, let us consider a general physical observable ${\cal O}(\xv)$ that depends on the coordinates $\xv$ (strictly speaking, the components of the vector ${\xv}$ represent here the deviations of the system's degrees of freedom from a local minimum of its potential energy $U$), under constant volume. The thermal average of ${\cal O}$ is given by $\langle{\cal O}\rangle_{_T}\!=\!{\C Z}(T)^{-1}\!\int\!{\cal O}({\xv})\exp\!\left(\!{-\frac{U({\xv})}{k_{\rm B} T}}\!\right) d{\xv}$, where ${\C Z}(T)\!=\!\int\!\exp\!\left(\!{-\frac{U({\xv})}{k_{\rm B} T}}\!\right) d{\xv}$ is the partition function and $k_{\rm B}$ is Boltzmann's constant.

Performing then an expansion of $\langle{\cal O}\rangle_{_T}$ to leading order in $T$ and taking its derivative with respect to $T$ (which is equivalent to taking the general $T$ dependence and evaluating the derivative at $T\=0$), one obtains~\cite{lte_pnas}
\begin{equation}
 \label{eq:expansion}
\frac{1}{\tfrac{1}{2}k_{\rm B}}\!\frac{d\langle {\cal O} \rangle_{_T}}{dT}\bigg\rvert_{_{T=0}}\!\!\!\!\!\simeq\! \frac{\partial^2\!{\cal O}}{\partial\xv\partial\xv}\!:\!\calBold{M}^{-1}\! - \frac{\partial {\cal O}}{\partial\xv}\cdot\calBold{M}^{-1}\!\cdot\calBold{U}'''\!:\!\calBold{M}^{-1} \ ,
\end{equation}
where ${\calBold M}\!\equiv\!\frac{\partial^2U}{\partial\xv\partial\xv}$ the dynamical (Hessian) matrix (that defines the eigenvalue equation $\calBold{M}\cdot\mathBold{\Psi}\=\omega^2\,\mathBold{\Psi}$, where $\mathBold{\Psi}$ is the eigenmode),
$\calBold{U}'''\!\equiv\!\frac{\partial^3 U}{\partial\xv\partial\xv\partial\xv}$ is a third-order anharmonicity tensor and all derivatives are evaluated at a minimum of $U(\xv)$.

The structure of Eq.~\eqref{eq:expansion} immediately provides some physical insight. First, it is observed for any observable ${\cal O}$ that features $\partial {\cal O}/\partial\xv\!\ne\!\zerovector$, which is typical for local observables (i.e.~observables defined at the interaction scale), anharmonicity contributes. Second, the anharmonic contribution is expected to dominate the harmonic one in many situations and to be controlled by QLEs, implying strong spatial localization/heterogeneity and anomalous statistics. To see this, first note that $\calBold{M}^{-1}\!\sim\!\omega^{-2}$, i.e.~$\calBold{M}^{-1}$ is dominated by soft excitations. The question then is whether low-frequency phonons and QLEs, both featuring $\omega\!\to\!0$, make markedly different contributions to the anharmonic term in Eq.~\eqref{eq:expansion}. The point is that typically for local observables, spatial gradients (e.g.~as appearing in $\calBold{U}'''$ and $\partial {\cal O}/\partial\xv$) are vanishingly small for low-frequency phonons (which are extended objects, featuring long wavelengths), but finite for QLEs. Consequently, the anharmonic term in Eq.~\eqref{eq:expansion} is expected to be dominated by QLEs and feature anomalously large values due to extremely soft QLEs, $\omega\!\to\!0$.

To demonstrate this in the context of a fundamental physical observable, let us consider ${\cal O}\=\varepsilon_\alpha$~\cite{lte_pnas,zohar_prerc}, where $\varepsilon_\alpha$ is the interaction energy of a pair of particles denoted by $\alpha$, such that the total potential energy is given as a sum over pairwise interactions, $U\=\sum_\alpha \varepsilon_\alpha$. In light of Eq.~\eqref{eq:expansion}, we then consider $\tfrac{1}{2}k_{\rm B} c_\alpha\!\equiv\!\frac{d\langle \varepsilon_\alpha \rangle_{_T}}{dT}\big\rvert_{_{T=0}}$, where $c_\alpha$ can be viewed as the classical (non-quantum) local heat capacity such that $\sum_\alpha c_\alpha$ is the global specific heat in the classical limit. The global specific heat, taking into account quantum effects, is discussed in Sect.~\ref{sec:exp} in the context of experimental evidence for QLEs. For ${\cal O}\=\varepsilon_\alpha$, $\partial {\cal O}/\partial\xv$ is nothing but the internal force $f_\alpha$, which is generically finite in glasses due to glassy frustration (note that mechanical equilibrium implies that the sum of $f_\alpha$ per particle vanishes, not the individual contributions~\cite{lte_pnas}). Consequently, we expect the local heat capacity $c_\alpha$ to feature anomalous statistics due to QLEs and their $\omega^4$ VDoS, and to feature strong spatial localization, i.e.~to attain anomalously large values in spatial locations where soft QLEs reside.

\begin{figure}[ht!]
  \includegraphics[width = 0.4\textwidth]{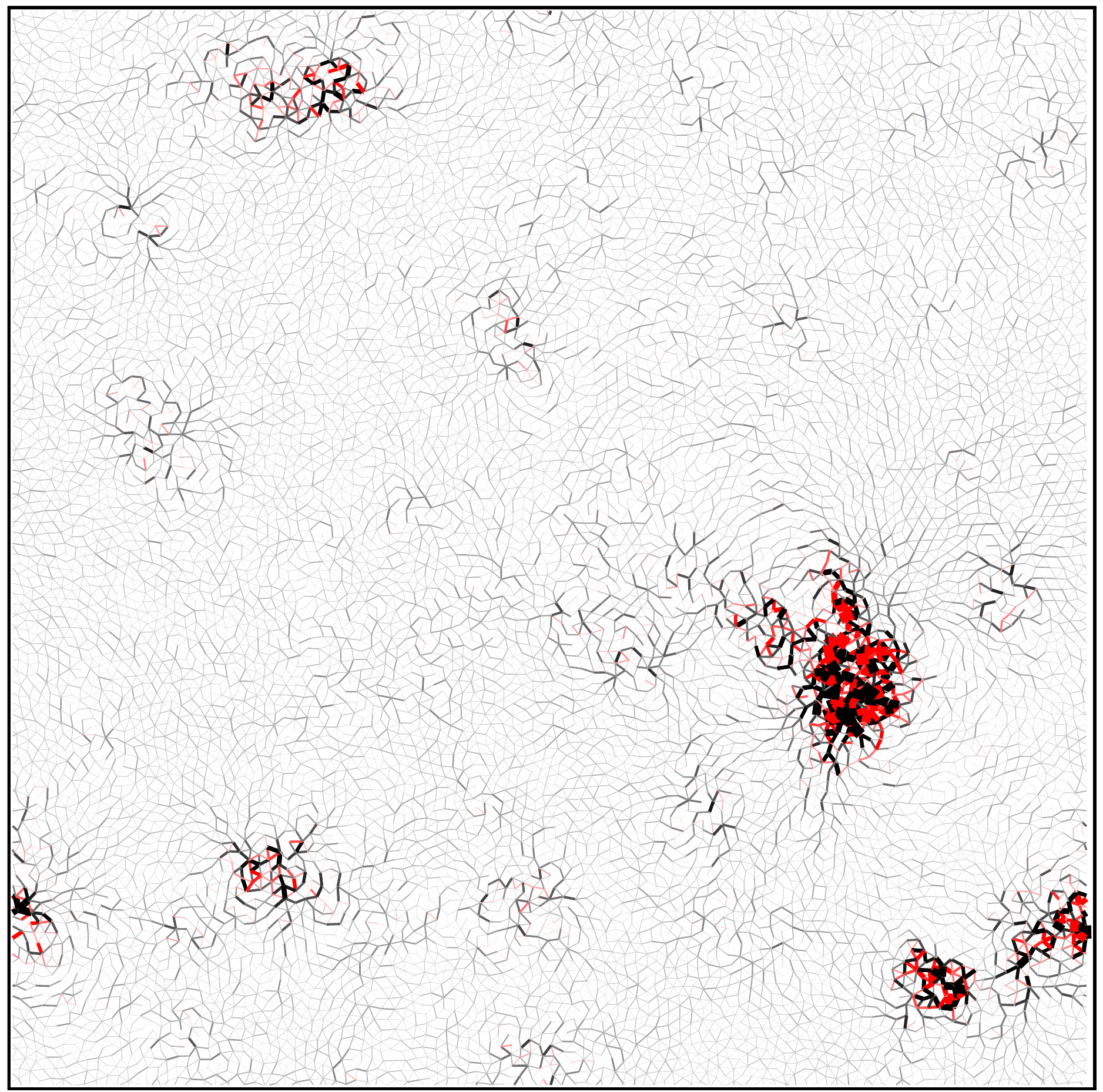}
  \caption{\footnotesize A map of the local heat capacity $c_\alpha$ in a 2D computer glass~\cite{lte_pnas}, see text for definitions and discussion, and~\cite{cge_paper} for model details. Each interacting pair of particles $\alpha$ is represented by a line whose thickness
  stands for the magnitude of $c_\alpha$, with red (black) representing negative (positive) values. Note the resemblance between this figure and Fig.~\ref{fig:Tp_dependence}a, both revealing the existence of QLEs.}
  \label{fig:LTE}
\end{figure}

The anomalous statistics of $c_\alpha$, i.e.~the fat-tailed (power-law) nature of the probability distribution function $p(c_\alpha)$, and its quantitative relation to the universal VDoS $\omega^4$ of QLEs have been demonstrated and discussed in~\cite{lte_pnas}. In Fig.~\ref{fig:LTE}, we present a spatial map of $c_\alpha$ for a 2D computer glass composed of 10,000 particles, where each interacting pair of particles $\alpha$ is represented by a line whose thickness stands for the magnitude of $c_\alpha$, with red (black) representing negative (positive) values. It is observed that the spatial distribution of $c_\alpha$ is highly heterogeneous, featuring strong localization in positions where soft QLEs reside; note in this context the visual resemblance between Figs.~\ref{fig:LTE} and~\ref{fig:Tp_dependence}a. Other physical observables, e.g.~the local thermal expansion coefficient, are expected to feature similar anomalous statistics and spatial heterogeneity due to QLEs.

The strong spatial heterogeneity of physical observables such as $c_\alpha$ implies heterogeneous dynamics, e.g.~structural relaxation under finite temperatures $T$ or irreversible rearrangements under the application of external forces. Indeed, it has been shown that locations of anomalously high $c_\alpha$ values are susceptible to local irreversible rearrangements when the glass is sheared externally~\cite{lte_pnas}, further strengthening the relation between QLEs and STZs, discussed in Subsect.~\ref{sec:stzs}. Furthermore, it has been shown that QLEs are in fact anisotropic objects that feature orientation-dependent coupling to driving forces~\cite{zohar_prerc}, a property that is important for their effect on plastic deformation. Finally, we note that strictly athermal quantities (i.e.~physical observables that are defined in inherent structures without involving any thermal averaging) are also strongly affected by QLEs. For example, it has been recently shown that fluctuations in the athermal shear modulus --- that are relevant for sound attenuation at low temperatures~\cite{Schirmacher_2013_boson_peak} --- feature anomalous statistics due to QLEs and their universal $\omega^4$ VDoS~\cite{scattering_jcp}.

\subsection{QLEs and double-well potentials}
\label{sec:qle_and_tls}

Within the Soft Potential Model (SPM) and the reconstruction picture, which were concisely reviewed in Sect.~\ref{sec:history}, soft QLEs and tunneling two-level systems (TLSs) are intimately related. Indeed, in~\cite{soft_potential_model_prb_1993} it was argued that experimental observations support a common basis for the universal properties of glasses, both in the extremely low temperatures regime --- where quantum tunneling dominates --- and slightly above it. Further experimental support for this argument was provided in~\cite{Grace_prb_1989,Brand_1991}, which concluded that tunneling TLSs and QLEs that give rise to the increase in the reduced heat capacity $C/T^3$ at $T\!\approx\!10\mbox{K}$ (cf.~Fig.~\ref{fig:ramos_fig}a) --- share a common origin.

Despite these claims and experimental evidence, some authors~\cite{zamponi_qle_tls_prm2021,itamar_where_are_tls_2021} recently argued that the connection between these two types of micromechanical objects may not be that clear. While we do not discuss this interesting issue in a comprehensive manner here, we do report on new (i.e.~not previously published) results that shed light on it. To set the stage for presenting these results, we first note that tunneling TLSs are a subset of double-well potentials whose quantum energy splitting is relevant for a given temperature. The energy splitting is determined by the asymmetry of the double-well potential and the associated tunneling amplitude, which in turn depends on the properties of the barrier (height and width) that separates the two potential wells. In what follows, we set aside the question of which double-well potentials significantly contribute to tunneling, and focus on the relations between QLEs and double-well potentials.

\begin{figure}[ht!]
  \includegraphics[width = 0.49\textwidth]{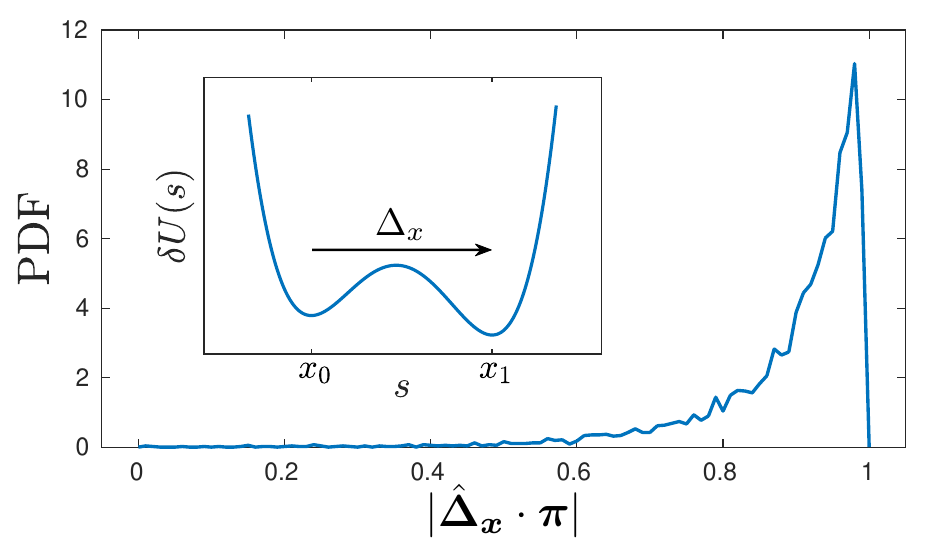}
  \caption{\footnotesize Probability distribution function (PDF) of the overlap between normalized displacement $\hat{\bm{\Delta}}_{\xv}\!\equiv\!(\xv_1\!-\!\xv_0)/|\xv_1\!-\!\xv_0|$ separating pairs $\xv_0$ and $\xv_1$ of adjacent potential energy landscape minima, and the QLE $\piv$ used to detect those pairs. The inset schematically illustrates the notations employed, see text for more details. The results presented in the figure have not been published elsewhere.}
  \label{fig:qle_tls_fig}
\end{figure}

Consider then the one-dimensional sketch of a double-well potential presented in the inset of Fig.~\ref{fig:qle_tls_fig}, whose minima occur at the positions $x_0$ and $x_1$, and denote by $\Delta_x$ the displacement between them, i.e.~$\Delta_x\!\equiv\!x_1-x_0$. The corresponding quantity in a glass is the displacement vector $\bm{\Delta}_\xv\!\equiv\!\xv_1\!-\!\xv_0$, where $\xv_0$ and $\xv_1$ are the particle position vectors in two adjacent mechanically-stable states. Consequently, a double-well potential can be characterized its normalized displacement vector $\hat{\bm{\Delta}}_\xv\!\equiv\!\bm{\Delta}_\xv/|\bm{\Delta}_\xv|$, which represents a direction in the multi-dimensional configurational space of a glass. The normalized displacement vector $\piv$ of a QLE corresponding to one of the potential wells, say the one at $\xv_0$, is also a direction in the multi-dimensional configurational space of a glass. Consequently, one measure of similarity between QLEs and double-well potentials is the degree of overlap between these two directions, i.e.~$|\hat{\bm{\Delta}}_\xv\!\cdot\!\piv|$.

To quantify the latter, we employ a close variant of the polydisperse, soft-spheres model put forward in~\cite{LB_swap_prx}, which can be subjected to extreme supercooling via the Swap-Monte-Carlo method. Further details about the model can be found in~\cite{boring_paper}. We studied 40,000 glasses of $N\!=\!\mbox{16,000}$ particles that were instantaneously quenched to $T\!=\!0$ from supercooled states equilibrated at $T_{\rm p}\!=\!0.4$; as a reference, the onset (crossover) temperature of this system was estimated at $\Tx\!\approx\!0.66$~\cite{karina_Tx_jcp_2020}.

In each glassy sample we pick up one of the softest QLEs $\piv$ using the nonlinear framework described above (initial conditions for detecting QLEs were chosen as described in Appendix~B of~\cite{sticky_spheres1_karina_pre2021}). Once we have a soft QLE $\piv$ at hand, starting from the initial state $\xv_0$ we displace particles a distance $s$ along $\piv$, and follow this displacement by an energy minimization. If a new mechanically-stable state $\xv_1$ is detected, the overlap $|\hat{\bm{\Delta}}_\xv\!\cdot\!\piv|$ is computed. In the search for the second stable state $\xv_1$ in each glass, we varied $s$ with small increments between $(0,a_0]$, where $a_0$ is a typical interparticle distance.

In Fig.~\ref{fig:qle_tls_fig}, we show the probability distribution function (PDF) of $|\hat{\bm{\Delta}}_\xv\cdot\piv|$. It is observed to be sharply peaked near unity, being predominantly supported over the relatively small range $|\hat{\bm{\Delta}}_\xv\!\cdot\!\piv|\!>\!0.8$. These results indicate that nearby minima in the potential energy landscape of a glass can be found by following QLEs, and that a high degree of overlap between the normalized displacements of QLEs and of double-well potentials (quantified by $\hat{\bm{\Delta}}_\xv$) exists. The latter suggest that QLEs and double-well potentials are closely related, at least as far as $\hat{\bm{\Delta}}_\xv$ and $\piv$ are concerned. Finally, we note that ($i$) the mean displacement amplitude between $\xv_0$ and $\xv_1$ is found to satisfy $\langle|\bm{\Delta}_\xv|\rangle/a_0\!\approx\!0.5$, and ($ii$) the fraction of QLEs that yielded a second mechanically-stable state $\xv_1$ by following their respective configurational directions (as described above) is $\approx\!14\%$; preliminary data (not shown) indicate that these fractions can substantially increase in systems with strong attractive forces, whose potential-energy-landscapes are highly fragmented~\cite{jeppe_project_jcp}.

\subsection{Additional efforts}
\label{sec:additional_efforts}

While we are not able to provide an exhaustive review of the literature relevant to understanding the emergence of low-energy QLEs in structural glasses in its entirety, we do concisely discuss here some additional, relevant recent efforts.

Several workers~\cite{MW_theta_and_omega,Harukuni_pre_2019,atsushi_dipole_instability_soft_matter2020,novel_instability_atsushi_2021} put forward theoretical frameworks that propose links between the statistical properties of various observables in a glass and QLEs. For example, in~\cite{MW_theta_and_omega} an attempt to relate the distribution $p(x)\!\sim\!x^{\theta}$ of local strain instability thresholds $x$ to the nonphononic VDoS of QLEs was presented, based on a picture of interacting soft anharmonic oscillators that bears similarities to the KHGPS model discussed in Sect.~\ref{sec:mean_field}. The proposed theory, accompanied by supporting numerical simulations, predicts that ${\cal D}(\omega)\!\sim\!\omega^{3 + 4\theta}$ for $\theta\!<\!1/4$, and ${\cal D}(\omega)\!\sim\!\omega^4$ for $\theta\!>\!1/4$. Similar relations between the $\omega^4$ law of QLEs and ($i$) the statistics of `finite-$\dbar$ fluctuations' attributed to vibrational-frequencies~\cite{Harukuni_pre_2019} ($ii$) the distribution of pairwise bond-stiffnesses~\cite{atsushi_dipole_instability_soft_matter2020,novel_instability_atsushi_2021} were proposed. However, these works provided no clear physical motivation for the form of the assumed input distributions. Furthermore, these predictions potentially disagree with a recent numerical demonstration~\cite{lerner2019finite} that observations of ${\cal D}(\omega)\!\sim\!\omega^\beta$ with $\beta\!<\!4$ suffer from finite-size effects, even for computer-glasses instantaneously quenched from high temperature liquid states.

A different route was taken in~\cite{eric_random_quench_dipoles_arXiv2020}, where a generic field-theoretic model of a quenched glass was put forward, featuring a nonphononic VDoS of excess modes that grows from zero frequency as $\sim\!\omega^3$. The model assumes an initial random stress field that is then driven by overdamped dynamics to a mechanical-equilibrium inherent structure; this quench procedure gives rise to various emergent properties such as long-range spatial correlations in the resulting stress field, as discussed previously~\cite{lemaitre_stres_correlations,eric_field_theory_prl_2018}.

Other workers put forward simplified, \emph{atomistic} models attempting to explain the origin of QLEs and their statistical properties. In~\cite{lisa_random_matrix_2019}, an atomistic model of an effectively one-dimensional, weighted spring-network with additional bonds --- that pushes the system away from isostaticity --- was put forward. It was established numerically that the model features a $\sim\!\omega^4$ scaling of the VDoS in an intermediate frequency regime. In~\cite{avraham_minimal_complexes_2021}, it was shown that mechanically frustrated and positionally disordered local structures --- termed `minimal complexes' --- constitute a minimal model of QLEs. It was further demonstrated that ensembles of marginally stable minimal complexes
feature a $\sim\!\omega^4$ VDoS.

Finally, we highlight a few recent studies of QLE-properties in generic computer-glass models. In~\cite{ning_xu_prl_2017}, a contribution $\sim\!\omega^3$ to the nonphononic VDoS was predicted (but not observed) based on a combination of numerical simulations of soft-sphere packings and an analysis of compression- and shear-induced plastic instabilities. In~\cite{paoluzzi_prl_2019,paoluzzi_prr_2020}, it was claimed that extended ``phonon-like" modes always exist below the lowest phonon frequency for high-$T_{\rm p}$ glasses, and that lowering $T_{\rm p}$ leads to the attenuation of the Debye spectrum in favor of the non-Debye one --- claims that stand in stark contradiction with the results reviewed here.

In~\cite{finite_T_VDoS_IP} it was shown that the $\omega^4$ nonphononic VDoS holds at finite temperatures deep in the glass phase.
In~\cite{smarajit_2021_modes_arXiv}, it was argued based on numerical simulations that the low-frequency nonphononic VDoS is crucially sensitive to the stress-ensemble considered. In particular, it was shown that the sample-to-sample minimal vibrational frequency of 2D glasses from which residual macroscopic stress was removed grows as $\omega_{\mbox{\tiny min}}^5$ instead of the expected $\omega_{\mbox{\tiny min}}^4$ for as-cast computer glasses~\cite{modes_prl_2016}. Even more recently, a numerical study of low-frequency vibrations in 2D computer glasses was presented~\cite{grzegorz_2D_modes_arXiv}; in that work, it was claimed that ${\cal D}(\omega)\!\sim\!\omega^3$ at frequencies below the lowest phonon frequency, however finite-size and glass-formation-protocol effects were not tested for (see also Fig.~\ref{fig:omega_min_2D} above). In the same work, an analysis of the intermediate-frequency vibrations revealed an excess $\sim\!\omega^2$ nonphononic VDoS, in agreement with previous predictions~\cite{eric_boson_peak_emt,silvio} and some simulations~\cite{non_debye_prl_2016,Atsushi_high_d_pre_2020}.

\section{Direct experimental evidence}
\label{sec:exp}

The search for QLEs, and the pressing need to elucidate their statistical and mechanical properties, was strongly motivated by various experimental observations in glasses, as explained in detail in the Introduction (cf.~Sect.~\ref{sec:intro}). Moreover, it is clear that QLEs significantly affect various glass properties and glassy phenomena. At the same time, up until now we did not discuss {\em direct and quantitative} experimental evidence for the existence of QLEs and their properties, but rather mostly focused on the study of computer glasses.

A natural physical quantity that may provide a route for obtaining direct and quantitative experimental evidence for the VDoS of QLEs is the specific heat $C(T)$, in particular its $T$ dependence in the low temperature limit. Other physical quantities, such as the sound absorption coefficient (cf.~the recent review in~\cite{buchenau2021sound}), can also be considered. The reason for the direct relevance of the specific heat is that it is given as $C(T)\=k_{\rm B}\!\int\!{\cal D}(\omega)\,x^2 e^{-x}(1-e^{-x})^{-2} d\omega$,
where $k_{\rm B}\,x^2e^{-x}(1-e^{-x})^{-2}$ is the single harmonic (normal) mode contribution, with $x\!\equiv\!\frac{\hbar\omega}{k_{\rm B}T}$. Consequently, a contribution ${\cal D}(\omega)\!\sim\!\omega^\delta$ to the VDoS translates into a contribution $C(T)\!\sim\!T^{\delta+1}$ to the specific heat~\cite{kittel2005introduction}.
\begin{figure}[ht!]
  \includegraphics[width = 0.49\textwidth]{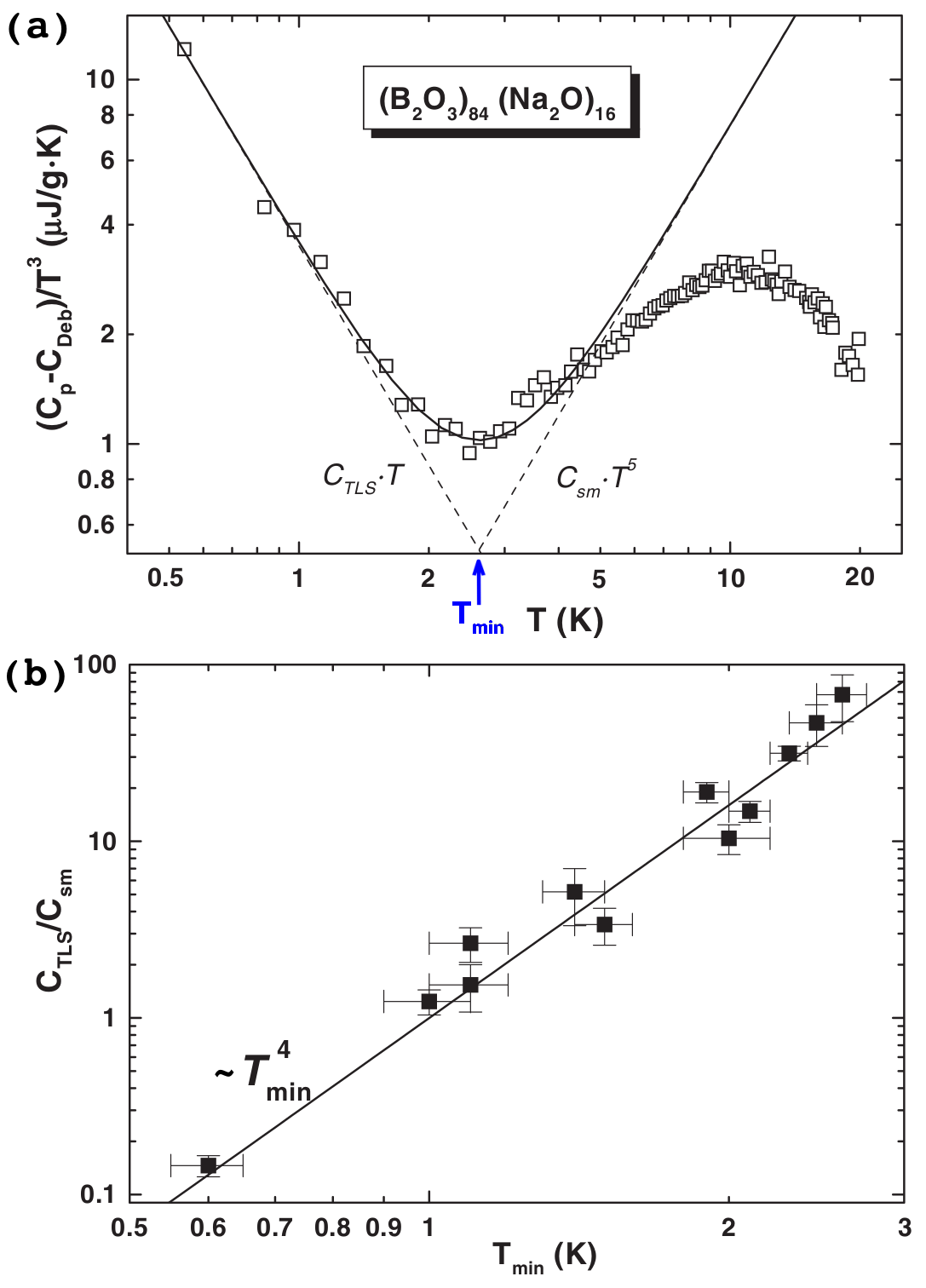}
  \caption{\footnotesize Analysis of the measured low-temperature specific heat of glasses in relation to Eq.~\eqref{eq:Cp}. Note that the experimental data are for the constant pressure specific heat $C_{\rm p}(T)$, and following previous authors (e.g.~\cite{ramos_2004,Schober_PRB_2014,Deformed_BMG_2014}), we assume it follows the same temperature dependence of the constant volume specific heat of Eq.~\eqref{eq:Cp}. (a) The reduced specific heat $(C_{\rm p}(T)\!-\!C_{\rm Deb}(T))/T^3$ of the (B$_{\rm 2}$O$_{\rm 3}$)$_{\rm 84}$(Na$_{\rm 2}$O)$_{\rm 16}$ glass is plotted against $T$ on a double-logarithmic scale~\cite{ramos_2004}. The solid and dashed lines correspond to a fit to Eq.~\eqref{eq:Cp}, see text and~\cite{ramos_2004} for details. The intersection of the dashed lines is denoted by $T_{\rm min}$. (b) The extracted ratio $C_{\rm TLS}/C_{\rm sm}$ for 12 different glasses (obtained from the fitting procedure presented in Fig.~\ref{fig:ramos_fig}a, see Eq.~\eqref{eq:Cp}, text and~\cite{ramos_2004} for details) is plotted against the {\em experimentally measured} $T_{\rm min}$. The solid line corresponds to $C_{\rm TLS}/C_{\rm sm}\!\sim\!T_{\rm min}^4$. Both panels are adapted with permission from M.A.~Ramos, Philos.~Mag.~\textbf{84}, 1313 (2004). Copyright 2004 Taylor \& Francis.}
  \label{fig:ramos_fig}
\end{figure}

The latter relations imply that the experimentally observed linear $T$ dependence of $C(T)$ in the limit of extremely low temperatures (below $1$K, see Introduction,~Sect.~\ref{sec:intro}) is associated with a constant VDoS ${\cal D}(\omega)$. Indeed, this approximately linear $T$ dependence of $C(T)$, which is attributed to tunneling TLSs, is associated with a constant TLSs VDoS (i.e.~$\delta\=0$)~\cite{anderson1972anomalous,soft_potential_model_1991}. Consequently, the tunneling TLSs contribution to the specific heat can be denoted as $C_{\rm TLS}T$. Debye's VDoS of low-frequency phonons takes the form ${\cal D}(\omega)\={\cal A}_{\rm D}\,\omega^2$ (in 3D, i.e.~here $\delta\=2$). Consequently, the low-frequency phonons contribution to the specific heat can be denoted as $C_{\rm Deb}(T)\=C_{\rm D}T^3$, which is the famous Debye's prediction~\cite{kittel2005introduction}. Taken together, the combined contributions of TLSs and phonons to $C(T)$ take the form $C_{\rm TLS}T+C_{\rm D}T^3$.

If one assumes there exist no other excitations in the low-frequency limit, $\omega\!\to\!0$, then $C(T)\!\simeq\!C_{\rm TLS}T+C_{\rm D}T^3$ is expected to accurately describe low temperature calorimetric measurements. In fact, since $C_{\rm D}\!\propto\!{\cal A}_{\rm D}$ and since ${\cal A}_{\rm D}$ depends on the linear elastic coefficients of a glass (or the elastic wave-speeds), $C_{\rm D}$ can be independently obtained from elastic measurements. Consequently, $C(T)\!\simeq\!C_{\rm TLS}T+C_{\rm D}T^3$ should both accurately describe calorimetric measurements of $C(T)$ and the resulting coefficient $C_{\rm D}$ should be consistent with the outcome of independent elastic measurements. However, it has been shown that fitting the measured specific heat to $C(T)\!\simeq\!C_{\rm TLS}T+C_{\rm D}T^3$ leads to systematic differences between the calorimetric and elastic estimates of $C_{\rm D}$ (where the former exceeds the latter)~\cite{ramos_2004}, clearly indicating that $C(T)\!\simeq\!C_{\rm TLS}T+C_{\rm D}T^3$ is incomplete, i.e.~that additional low-frequency excitations exist.

The VDoS of QLEs, ${\cal D}(\omega)\={\cal A}_{\rm g}\omega^4$, corresponds to $\delta\=4$ and implies a contribution to the specific heat of the form $C_{\rm sm}T^5$ (here `sm' corresponds to ``soft modes'', and we use $C_{\rm sm}$ instead of the more natural $C_{\rm QLE}$ in order to adhere to the notation of~\cite{ramos_2004}). Consequently, the low-temperature dependence of the specific heat of glasses is expected to take the form
\begin{equation}
\label{eq:Cp}
    C(T) \simeq C_{\rm TLS}T+C_{\rm D}T^3+C_{\rm sm}T^5 \ .
\end{equation}
In Fig.~\ref{fig:ramos_fig}a, we reproduce low-temperature measurements of $C(T)$ of the (B$_{\rm 2}$O$_{\rm 3}$)$_{\rm 84}$(Na$_{\rm 2}$O)$_{\rm 16}$ glass~\cite{ramos_2004}, where a fit to Eq.~\eqref{eq:Cp} is superimposed (solid line). In the figure, which is adapted from~\cite{ramos_2004}, Deybe's contribution $C_{\rm Deb}(T)$ is subtracted from $C(T)$ and its scaling with $T$ (i.e.~$C_{\rm Deb}(T)\!\propto\!T^3$) is used to normalize the outcome. Consequently, in this reduced representation of the specific heat, the tunneling TLSs contribution corresponds to a $T^{-2}$ dependence (left dashed line) and the QLEs contribution to a $T^{2}$ dependence (right dashed line), see~\cite{ramos_2004} for the details of the fitting procedure and the selection of the temperature range to be fitted.

The fitting procedure shown in Fig.~\ref{fig:ramos_fig}a closely follows the original procedure of Buchenau et al.~\cite{soft_potential_model_1991}, cf.~Figs.~2-3 therein. Repeating it for many glasses, cf.~Table 1 in~\cite{ramos_2004}, has shown that calorimetric and elastic estimates of $C_{\rm D}$ agree with each other within experimental error, eliminating the above-mentioned systematic deviations between the two independent estimates. Moreover, Eq.~\eqref{eq:Cp} suggests a self-consistency constraint on the analysis of the experimental data; it predicts that the minimum of the reduced specific heat $(C(T)\!-\!C_{\rm Deb}(T))/T^3$, occurring at $T_{\rm min}$ (cf.~the intersection of the two dashed lines in Fig.~\ref{fig:ramos_fig}a), satisfies $C_{\rm TLS}/C_{\rm sm}\!\sim\!T_{\rm min}^4$. In Fig.~\ref{fig:ramos_fig}b, we show the ratio $C_{\rm TLS}/C_{\rm sm}$ for 12 different glasses (obtained from the fitting procedure presented in Fig.~\ref{fig:ramos_fig}a) against the {\em experimentally measured} $T_{\rm min}$, revealing favorable agreement with the $C_{\rm TLS}/C_{\rm sm}\!\sim\!T_{\rm min}^4$ prediction~\cite{ramos_2004}. All in all, the analysis of the low-temperature specific heat of glasses appears to offer a rather direct experimental evidence for the existence of QLEs and their universal nonphononic VDoS $\sim\!\omega^4$.

\section{Outlook}
\label{sec:outlook}

We conclude this Perspective Article by briefly delineating several open questions and future research directions in relation to QLEs in structural glasses.

The existence of low-frequency phonons, i.e.~of soft extended excitations, is understood in general theoretical terms to emerge from the breakdown of global continuous symmetries, following Goldstone's theorem. As of now, there is no comparable theoretical understanding of the generic existence of QLEs, i.e.~of soft (quasi-) localized nonphononic excitations. Developing such a fundamental understanding is a challenge for future work.

Several theoretical frameworks, such as the GPS and KHGPS models discussed in Sects.~\ref{sec:reconstruction} and~\ref{sec:mean_field} respectively, a priori assume the existence of localized excitations and aim at offering some understanding of their VDoS. Such frameworks appear to be quite successful in rationalizing the emergence of a gapless VDoS that increases from zero frequency as $\omega^4$. These models also highlight some generic physical ingredients involved, such as long-range elastic interactions, anharmonicity and frustration-induced internal stresses. As these models focus on studying the minima of some effective Hamiltonians, it is tempting to interpret the latter as representing the liquid state from which a glass is formed and the minimization process to mimic the self-organizational processes taking place during vitrification in realistic finite dimensions. Yet, it is currently unclear how to substantiate such an interpretation and how to quantitatively support it. Likewise, it is currently unclear how to relate the parameters appearing in such models to physical properties of the ancestral liquid state and/or the nonequilibrium glassy state. Addressing these challenges seems important for developing a minimal model of QLEs.

QLEs populate the low-frequency tail of excess modes/excitations in glasses, i.e.~soft excitations that are not phononic in nature and that do not follow Debye's VDoS. It is now established, as discussed in this Perspective Article, that QLEs feature a characteristic frequency (or equivalently an energy) scale $\omega_{\rm g}$~\cite{cge_paper,pinching_pnas}. It would be interesting to clarify whether and how $\omega_{\rm g}$ might be related to other, previously identified frequency/energy scales in glasses. In particular, in the context of excess modes/excitations in glasses, a most well characterized quantity is the boson peak frequency $\omega_{\mbox{\tiny BP}}$. Consequently, a challenge for future work would be to elucidate the relation between $\omega_{\mbox{\tiny BP}}$ and the characteristic frequency $\omega_{\rm g}$ of QLEs (the possible existence of such relations has been raised in~\cite{pinching_pnas}).

Direct experimental evidence for the existence of QLEs and their universal properties has been discussed in Sect.~\ref{sec:exp}, focusing on low-temperature measurements of the specific heat of glasses. A combined theoretical and experimental challenge for future work is identifying additional, measurable low-temperature observables that can cleanly reveal QLEs. On the experimental side, it is also desirable to probe the physics of QLEs not just through their effect on low-temperature observables, but also through the identification of the glassy structure that underlie them, which in turn requires the development of experimental techniques featuring enhanced spatial, and possibly temporal, resolution.

Finally, future research should further elucidate the roles played by QLEs in various glass properties and dynamics. The latter include properties such as the fracture toughness~\cite{david_fracture_2021} and viscoelastic response functions, as well as the physics of plastic deformation. Another potentially interesting line of investigation would be the exploration of the roles played by QLEs in structural relaxation in equilibrium supercooled liquids~\cite{widmer2008irreversible,harrowell_2009,jeppe_project_jcp} and in out-of-equilibrium aging dynamics~\cite{schober1993_numerics,schober_correlate_modes_dynamics_1999}, with the hypothesis that relaxation/aging occurs at locations where QLEs reside and that the long-range elastic interactions between QLEs is of importance.\\

\acknowledgements
We warmly thank Karina Gonz\'alez-L\'opez, Geert Kapteijns, David Richard, Corrado Rainone, Gustavo D\"uring, Francesco Zamponi, Pierfrancesco Urbani, Avraham Moriel, Talya Vaknin and Robert Pater for their invaluable contributions to some the works reviewed here. We are also grateful to David Richard and Karina Gonz\'alez-L\'opez for their help with the graphics. We thank Omar Benzine, Zhiwen Pan and Lothar Wondraczek for calling our attention to the work of Ramos, discussed in Sect.~\ref{sec:exp}. We are grateful to Herbert Schober, Ulrich Buchenau, Geert Kapteijns, Karina Gonz\'alez-L\'opez and Avraham Moriel for reading an earlier version of the manuscript, and for providing useful and constructive comments. E.L.~acknowledges support from the NWO (Vidi grant no.~680-47-554/3259). E.B.~acknowledges support from the Ben May Center for Chemical Theory and Computation and the Harold Perlman Family.\\


\begin{thebibliography}{193}%
\makeatletter
\providecommand \@ifxundefined [1]{%
 \@ifx{#1\undefined}
}%
\providecommand \@ifnum [1]{%
 \ifnum #1\expandafter \@firstoftwo
 \else \expandafter \@secondoftwo
 \fi
}%
\providecommand \@ifx [1]{%
 \ifx #1\expandafter \@firstoftwo
 \else \expandafter \@secondoftwo
 \fi
}%
\providecommand \natexlab [1]{#1}%
\providecommand \enquote  [1]{``#1''}%
\providecommand \bibnamefont  [1]{#1}%
\providecommand \bibfnamefont [1]{#1}%
\providecommand \citenamefont [1]{#1}%
\providecommand \href@noop [0]{\@secondoftwo}%
\providecommand \href [0]{\begingroup \@sanitize@url \@href}%
\providecommand \@href[1]{\@@startlink{#1}\@@href}%
\providecommand \@@href[1]{\endgroup#1\@@endlink}%
\providecommand \@sanitize@url [0]{\catcode `\\12\catcode `\$12\catcode
  `\&12\catcode `\#12\catcode `\^12\catcode `\_12\catcode `\%12\relax}%
\providecommand \@@startlink[1]{}%
\providecommand \@@endlink[0]{}%
\providecommand \url  [0]{\begingroup\@sanitize@url \@url }%
\providecommand \@url [1]{\endgroup\@href {#1}{\urlprefix }}%
\providecommand \urlprefix  [0]{URL }%
\providecommand \Eprint [0]{\href }%
\providecommand \doibase [0]{https://doi.org/}%
\providecommand \selectlanguage [0]{\@gobble}%
\providecommand \bibinfo  [0]{\@secondoftwo}%
\providecommand \bibfield  [0]{\@secondoftwo}%
\providecommand \translation [1]{[#1]}%
\providecommand \BibitemOpen [0]{}%
\providecommand \bibitemStop [0]{}%
\providecommand \bibitemNoStop [0]{.\EOS\space}%
\providecommand \EOS [0]{\spacefactor3000\relax}%
\providecommand \BibitemShut  [1]{\csname bibitem#1\endcsname}%
\let\auto@bib@innerbib\@empty
\bibitem [{\citenamefont {Cavagna}(2009)}]{Cavagna200951}%
  \BibitemOpen
  \bibfield  {author} {\bibinfo {author} {\bibfnamefont {A.}~\bibnamefont
  {Cavagna}},\ }\bibfield  {title} {\bibinfo {title} {Supercooled liquids for
  pedestrians},\ }\href
  {https://doi.org/https://doi.org/10.1016/j.physrep.2009.03.003} {\bibfield
  {journal} {\bibinfo  {journal} {Phys. Rep.}\ }\textbf {\bibinfo {volume}
  {476}},\ \bibinfo {pages} {51 } (\bibinfo {year} {2009})}\BibitemShut
  {NoStop}%
\bibitem [{\citenamefont {Debenedetti}\ and\ \citenamefont
  {Stillinger}(2001)}]{Debenedetti2001}%
  \BibitemOpen
  \bibfield  {author} {\bibinfo {author} {\bibfnamefont {P.~G.}\ \bibnamefont
  {Debenedetti}}\ and\ \bibinfo {author} {\bibfnamefont {F.~H.}\ \bibnamefont
  {Stillinger}},\ }\bibfield  {title} {\bibinfo {title} {Supercooled liquids
  and the glass transition},\ }\href {https://doi.org/10.1038/35065704}
  {\bibfield  {journal} {\bibinfo  {journal} {Nature}\ }\textbf {\bibinfo
  {volume} {410}},\ \bibinfo {pages} {259} (\bibinfo {year}
  {2001})}\BibitemShut {NoStop}%
\bibitem [{\citenamefont {Wyart}\ and\ \citenamefont
  {Cates}(2017)}]{MW_cates_length_discussion_prl_2017}%
  \BibitemOpen
  \bibfield  {author} {\bibinfo {author} {\bibfnamefont {M.}~\bibnamefont
  {Wyart}}\ and\ \bibinfo {author} {\bibfnamefont {M.~E.}\ \bibnamefont
  {Cates}},\ }\bibfield  {title} {\bibinfo {title} {Does a growing static
  length scale control the glass transition?},\ }\href
  {https://doi.org/10.1103/PhysRevLett.119.195501} {\bibfield  {journal}
  {\bibinfo  {journal} {Phys. Rev. Lett.}\ }\textbf {\bibinfo {volume} {119}},\
  \bibinfo {pages} {195501} (\bibinfo {year} {2017})}\BibitemShut {NoStop}%
\bibitem [{\citenamefont {Berthier}\ \emph {et~al.}(2019)\citenamefont
  {Berthier}, \citenamefont {Biroli}, \citenamefont {Bouchaud},\ and\
  \citenamefont {Tarjus}}]{tarjus_no_length}%
  \BibitemOpen
  \bibfield  {author} {\bibinfo {author} {\bibfnamefont {L.}~\bibnamefont
  {Berthier}}, \bibinfo {author} {\bibfnamefont {G.}~\bibnamefont {Biroli}},
  \bibinfo {author} {\bibfnamefont {J.-P.}\ \bibnamefont {Bouchaud}},\ and\
  \bibinfo {author} {\bibfnamefont {G.}~\bibnamefont {Tarjus}},\ }\bibfield
  {title} {\bibinfo {title} {Can the glass transition be explained without a
  growing static length scale?},\ }\href {https://doi.org/10.1063/1.5086509}
  {\bibfield  {journal} {\bibinfo  {journal} {J. Chem. Phys.}\ }\textbf
  {\bibinfo {volume} {150}},\ \bibinfo {pages} {094501} (\bibinfo {year}
  {2019})}\BibitemShut {NoStop}%
\bibitem [{\citenamefont {Zeller}\ and\ \citenamefont
  {Pohl}(1971)}]{Zeller_and_Pohl_prb_1971}%
  \BibitemOpen
  \bibfield  {author} {\bibinfo {author} {\bibfnamefont {R.~C.}\ \bibnamefont
  {Zeller}}\ and\ \bibinfo {author} {\bibfnamefont {R.~O.}\ \bibnamefont
  {Pohl}},\ }\bibfield  {title} {\bibinfo {title} {Thermal conductivity and
  specific heat of noncrystalline solids},\ }\href
  {https://doi.org/10.1103/PhysRevB.4.2029} {\bibfield  {journal} {\bibinfo
  {journal} {Phys. Rev. B}\ }\textbf {\bibinfo {volume} {4}},\ \bibinfo {pages}
  {2029} (\bibinfo {year} {1971})}\BibitemShut {NoStop}%
\bibitem [{\citenamefont {Hunklinger}\ and\ \citenamefont
  {Arnold}(1976)}]{HUNKLINGER1976155}%
  \BibitemOpen
  \bibfield  {author} {\bibinfo {author} {\bibfnamefont {S.}~\bibnamefont
  {Hunklinger}}\ and\ \bibinfo {author} {\bibfnamefont {W.}~\bibnamefont
  {Arnold}},\ }\bibfield  {title} {\bibinfo {title} {Ultrasonic properties of
  glasses at low temperatures}\ }(\bibinfo  {publisher} {Academic Press},\
  \bibinfo {year} {1976})\ pp.\ \bibinfo {pages} {155--215}\BibitemShut
  {NoStop}%
\bibitem [{\citenamefont {Pohl}\ \emph {et~al.}(2002)\citenamefont {Pohl},
  \citenamefont {Liu},\ and\ \citenamefont {Thompson}}]{Pohl_review_2002}%
  \BibitemOpen
  \bibfield  {author} {\bibinfo {author} {\bibfnamefont {R.~O.}\ \bibnamefont
  {Pohl}}, \bibinfo {author} {\bibfnamefont {X.}~\bibnamefont {Liu}},\ and\
  \bibinfo {author} {\bibfnamefont {E.}~\bibnamefont {Thompson}},\ }\bibfield
  {title} {\bibinfo {title} {Low-temperature thermal conductivity and acoustic
  attenuation in amorphous solids},\ }\href
  {https://doi.org/10.1103/RevModPhys.74.991} {\bibfield  {journal} {\bibinfo
  {journal} {Rev. Mod. Phys.}\ }\textbf {\bibinfo {volume} {74}},\ \bibinfo
  {pages} {991} (\bibinfo {year} {2002})}\BibitemShut {NoStop}%
\bibitem [{\citenamefont {Kittel}(2005)}]{kittel2005introduction}%
  \BibitemOpen
  \bibfield  {author} {\bibinfo {author} {\bibfnamefont {C.}~\bibnamefont
  {Kittel}},\ }\href@noop {} {\emph {\bibinfo {title} {Introduction to solid
  state physics}}}\ (\bibinfo  {publisher} {Wiley},\ \bibinfo {year}
  {2005})\BibitemShut {NoStop}%
\bibitem [{\citenamefont {Phillips}(1972)}]{phillips1972tunneling}%
  \BibitemOpen
  \bibfield  {author} {\bibinfo {author} {\bibfnamefont {W.}~\bibnamefont
  {Phillips}},\ }\bibfield  {title} {\bibinfo {title} {Tunneling states in
  amorphous solids},\ }\href {https://doi.org/10.1007/BF00660072} {\bibfield
  {journal} {\bibinfo  {journal} {J. Low Temp. Phys.}\ }\textbf {\bibinfo
  {volume} {7}},\ \bibinfo {pages} {351} (\bibinfo {year} {1972})}\BibitemShut
  {NoStop}%
\bibitem [{\citenamefont {Anderson}\ \emph {et~al.}(1972)\citenamefont
  {Anderson}, \citenamefont {Halperin},\ and\ \citenamefont
  {Varma}}]{anderson1972anomalous}%
  \BibitemOpen
  \bibfield  {author} {\bibinfo {author} {\bibfnamefont {P.~W.}\ \bibnamefont
  {Anderson}}, \bibinfo {author} {\bibfnamefont {B.~I.}\ \bibnamefont
  {Halperin}},\ and\ \bibinfo {author} {\bibfnamefont {C.~M.}\ \bibnamefont
  {Varma}},\ }\bibfield  {title} {\bibinfo {title} {Anomalous low-temperature
  thermal properties of glasses and spin glasses},\ }\href
  {https://doi.org/10.1080/14786437208229210} {\bibfield  {journal} {\bibinfo
  {journal} {Philos. Mag.}\ }\textbf {\bibinfo {volume} {25}},\ \bibinfo
  {pages} {1} (\bibinfo {year} {1972})}\BibitemShut {NoStop}%
\bibitem [{\citenamefont {Shintani}\ and\ \citenamefont
  {Tanaka}(2008)}]{Shintani2008}%
  \BibitemOpen
  \bibfield  {author} {\bibinfo {author} {\bibfnamefont {H.}~\bibnamefont
  {Shintani}}\ and\ \bibinfo {author} {\bibfnamefont {H.}~\bibnamefont
  {Tanaka}},\ }\bibfield  {title} {\bibinfo {title} {Universal link between the
  boson peak and transverse phonons in glass},\ }\href
  {https://doi.org/10.1038/nmat2293} {\bibfield  {journal} {\bibinfo  {journal}
  {Nat. Mater.}\ }\textbf {\bibinfo {volume} {7}},\ \bibinfo {pages} {870}
  (\bibinfo {year} {2008})}\BibitemShut {NoStop}%
\bibitem [{\citenamefont {Ruocco}(2008)}]{Ruocco2008}%
  \BibitemOpen
  \bibfield  {author} {\bibinfo {author} {\bibfnamefont {G.}~\bibnamefont
  {Ruocco}},\ }\bibfield  {title} {\bibinfo {title} {When disorder helps},\
  }\href {https://doi.org/10.1038/nmat2311} {\bibfield  {journal} {\bibinfo
  {journal} {Nat. Mater.}\ }\textbf {\bibinfo {volume} {7}},\ \bibinfo {pages}
  {842} (\bibinfo {year} {2008})}\BibitemShut {NoStop}%
\bibitem [{\citenamefont {Kirillov}(1999)}]{KIRILLOV1999279}%
  \BibitemOpen
  \bibfield  {author} {\bibinfo {author} {\bibfnamefont {S.~A.}\ \bibnamefont
  {Kirillov}},\ }\bibfield  {title} {\bibinfo {title} {Spatial disorder and
  low-frequency raman patterns of amorphous solid, with special reference to
  quasi-elastic scattering and its relation to boson peak},\ }\href
  {https://doi.org/https://doi.org/10.1016/S0022-2860(98)00879-5} {\bibfield
  {journal} {\bibinfo  {journal} {J. Mol. Struct.}\ }\textbf {\bibinfo {volume}
  {479}},\ \bibinfo {pages} {279} (\bibinfo {year} {1999})}\BibitemShut
  {NoStop}%
\bibitem [{\citenamefont {Marruzzo}\ \emph {et~al.}(2013)\citenamefont
  {Marruzzo}, \citenamefont {Schirmacher}, \citenamefont {Fratalocchi},\ and\
  \citenamefont {Ruocco}}]{Schirmacher_2013_boson_peak}%
  \BibitemOpen
  \bibfield  {author} {\bibinfo {author} {\bibfnamefont {A.}~\bibnamefont
  {Marruzzo}}, \bibinfo {author} {\bibfnamefont {W.}~\bibnamefont
  {Schirmacher}}, \bibinfo {author} {\bibfnamefont {A.}~\bibnamefont
  {Fratalocchi}},\ and\ \bibinfo {author} {\bibfnamefont {G.}~\bibnamefont
  {Ruocco}},\ }\bibfield  {title} {\bibinfo {title} {Heterogeneous shear
  elasticity of glasses: the origin of the boson peak},\ }\href
  {https://doi.org/10.1038/srep01407} {\bibfield  {journal} {\bibinfo
  {journal} {Sci. Rep.}\ }\textbf {\bibinfo {volume} {3}},\ \bibinfo {pages}
  {1407 EP } (\bibinfo {year} {2013})}\BibitemShut {NoStop}%
\bibitem [{\citenamefont {Yang}\ \emph {et~al.}(2019)\citenamefont {Yang},
  \citenamefont {Wang}, \citenamefont {Ma}, \citenamefont {Zaccone},
  \citenamefont {Dai},\ and\ \citenamefont {Jiang}}]{Zaccone_prl_2019}%
  \BibitemOpen
  \bibfield  {author} {\bibinfo {author} {\bibfnamefont {J.}~\bibnamefont
  {Yang}}, \bibinfo {author} {\bibfnamefont {Y.-J.}\ \bibnamefont {Wang}},
  \bibinfo {author} {\bibfnamefont {E.}~\bibnamefont {Ma}}, \bibinfo {author}
  {\bibfnamefont {A.}~\bibnamefont {Zaccone}}, \bibinfo {author} {\bibfnamefont
  {L.~H.}\ \bibnamefont {Dai}},\ and\ \bibinfo {author} {\bibfnamefont {M.~Q.}\
  \bibnamefont {Jiang}},\ }\bibfield  {title} {\bibinfo {title} {Structural
  parameter of orientational order to predict the boson vibrational anomaly in
  glasses},\ }\href {https://doi.org/10.1103/PhysRevLett.122.015501} {\bibfield
   {journal} {\bibinfo  {journal} {Phys. Rev. Lett.}\ }\textbf {\bibinfo
  {volume} {122}},\ \bibinfo {pages} {015501} (\bibinfo {year}
  {2019})}\BibitemShut {NoStop}%
\bibitem [{\citenamefont {Malinovsky}\ and\ \citenamefont
  {Sokolov}(1986)}]{MALINOVSKY1986757}%
  \BibitemOpen
  \bibfield  {author} {\bibinfo {author} {\bibfnamefont {V.}~\bibnamefont
  {Malinovsky}}\ and\ \bibinfo {author} {\bibfnamefont {A.}~\bibnamefont
  {Sokolov}},\ }\bibfield  {title} {\bibinfo {title} {The nature of boson peak
  in raman scattering in glasses},\ }\href
  {https://doi.org/https://doi.org/10.1016/0038-1098(86)90854-9} {\bibfield
  {journal} {\bibinfo  {journal} {Solid State Commun.}\ }\textbf {\bibinfo
  {volume} {57}},\ \bibinfo {pages} {757} (\bibinfo {year} {1986})}\BibitemShut
  {NoStop}%
\bibitem [{\citenamefont {Schirmacher}\ \emph {et~al.}(1998)\citenamefont
  {Schirmacher}, \citenamefont {Diezemann},\ and\ \citenamefont
  {Ganter}}]{schirmacher_1998}%
  \BibitemOpen
  \bibfield  {author} {\bibinfo {author} {\bibfnamefont {W.}~\bibnamefont
  {Schirmacher}}, \bibinfo {author} {\bibfnamefont {G.}~\bibnamefont
  {Diezemann}},\ and\ \bibinfo {author} {\bibfnamefont {C.}~\bibnamefont
  {Ganter}},\ }\bibfield  {title} {\bibinfo {title} {Harmonic vibrational
  excitations in disordered solids and the ``boson peak''},\ }\href
  {https://doi.org/10.1103/PhysRevLett.81.136} {\bibfield  {journal} {\bibinfo
  {journal} {Phys. Rev. Lett.}\ }\textbf {\bibinfo {volume} {81}},\ \bibinfo
  {pages} {136} (\bibinfo {year} {1998})}\BibitemShut {NoStop}%
\bibitem [{\citenamefont {Taraskin}\ \emph {et~al.}(2001)\citenamefont
  {Taraskin}, \citenamefont {Loh}, \citenamefont {Natarajan},\ and\
  \citenamefont {Elliott}}]{taraskin2001origin}%
  \BibitemOpen
  \bibfield  {author} {\bibinfo {author} {\bibfnamefont {S.~N.}\ \bibnamefont
  {Taraskin}}, \bibinfo {author} {\bibfnamefont {Y.~L.}\ \bibnamefont {Loh}},
  \bibinfo {author} {\bibfnamefont {G.}~\bibnamefont {Natarajan}},\ and\
  \bibinfo {author} {\bibfnamefont {S.~R.}\ \bibnamefont {Elliott}},\
  }\bibfield  {title} {\bibinfo {title} {Origin of the boson peak in systems
  with lattice disorder},\ }\href {https://doi.org/10.1103/PhysRevLett.86.1255}
  {\bibfield  {journal} {\bibinfo  {journal} {Phys. Rev. Lett.}\ }\textbf
  {\bibinfo {volume} {86}},\ \bibinfo {pages} {1255} (\bibinfo {year}
  {2001})}\BibitemShut {NoStop}%
\bibitem [{\citenamefont {Lubchenko}\ and\ \citenamefont
  {Wolynes}(2003)}]{Lubchenko1515}%
  \BibitemOpen
  \bibfield  {author} {\bibinfo {author} {\bibfnamefont {V.}~\bibnamefont
  {Lubchenko}}\ and\ \bibinfo {author} {\bibfnamefont {P.~G.}\ \bibnamefont
  {Wolynes}},\ }\bibfield  {title} {\bibinfo {title} {The origin of the boson
  peak and thermal conductivity plateau in low-temperature glasses},\ }\href
  {https://doi.org/10.1073/pnas.252786999} {\bibfield  {journal} {\bibinfo
  {journal} {Proc. Natl. Acad. Sci. U.S.A.}\ }\textbf {\bibinfo {volume}
  {100}},\ \bibinfo {pages} {1515} (\bibinfo {year} {2003})}\BibitemShut
  {NoStop}%
\bibitem [{\citenamefont {Chumakov}\ \emph
  {et~al.}(2011{\natexlab{a}})\citenamefont {Chumakov}, \citenamefont {Monaco},
  \citenamefont {Monaco}, \citenamefont {Crichton}, \citenamefont {Bosak},
  \citenamefont {R\"uffer}, \citenamefont {Meyer}, \citenamefont {Kargl},
  \citenamefont {Comez}, \citenamefont {Fioretto}, \citenamefont {Giefers},
  \citenamefont {Roitsch}, \citenamefont {Wortmann}, \citenamefont {Manghnani},
  \citenamefont {Hushur}, \citenamefont {Williams}, \citenamefont {Balogh},
  \citenamefont {Parli\ifmmode~\acute{n}\else \'{n}\fi{}ski}, \citenamefont
  {Jochym},\ and\ \citenamefont {Piekarz}}]{Chumakov_2011_bosonPeak}%
  \BibitemOpen
  \bibfield  {author} {\bibinfo {author} {\bibfnamefont {A.~I.}\ \bibnamefont
  {Chumakov}}, \bibinfo {author} {\bibfnamefont {G.}~\bibnamefont {Monaco}},
  \bibinfo {author} {\bibfnamefont {A.}~\bibnamefont {Monaco}}, \bibinfo
  {author} {\bibfnamefont {W.~A.}\ \bibnamefont {Crichton}}, \bibinfo {author}
  {\bibfnamefont {A.}~\bibnamefont {Bosak}}, \bibinfo {author} {\bibfnamefont
  {R.}~\bibnamefont {R\"uffer}}, \bibinfo {author} {\bibfnamefont
  {A.}~\bibnamefont {Meyer}}, \bibinfo {author} {\bibfnamefont
  {F.}~\bibnamefont {Kargl}}, \bibinfo {author} {\bibfnamefont
  {L.}~\bibnamefont {Comez}}, \bibinfo {author} {\bibfnamefont
  {D.}~\bibnamefont {Fioretto}}, \bibinfo {author} {\bibfnamefont
  {H.}~\bibnamefont {Giefers}}, \bibinfo {author} {\bibfnamefont
  {S.}~\bibnamefont {Roitsch}}, \bibinfo {author} {\bibfnamefont
  {G.}~\bibnamefont {Wortmann}}, \bibinfo {author} {\bibfnamefont {M.~H.}\
  \bibnamefont {Manghnani}}, \bibinfo {author} {\bibfnamefont {A.}~\bibnamefont
  {Hushur}}, \bibinfo {author} {\bibfnamefont {Q.}~\bibnamefont {Williams}},
  \bibinfo {author} {\bibfnamefont {J.}~\bibnamefont {Balogh}}, \bibinfo
  {author} {\bibfnamefont {K.}~\bibnamefont {Parli\ifmmode~\acute{n}\else
  \'{n}\fi{}ski}}, \bibinfo {author} {\bibfnamefont {P.}~\bibnamefont
  {Jochym}},\ and\ \bibinfo {author} {\bibfnamefont {P.}~\bibnamefont
  {Piekarz}},\ }\bibfield  {title} {\bibinfo {title} {Equivalence of the boson
  peak in glasses to the transverse acoustic van hove singularity in
  crystals},\ }\href {https://doi.org/10.1103/PhysRevLett.106.225501}
  {\bibfield  {journal} {\bibinfo  {journal} {Phys. Rev. Lett.}\ }\textbf
  {\bibinfo {volume} {106}},\ \bibinfo {pages} {225501} (\bibinfo {year}
  {2011}{\natexlab{a}})}\BibitemShut {NoStop}%
\bibitem [{\citenamefont {DeGiuli}\ \emph
  {et~al.}(2014{\natexlab{a}})\citenamefont {DeGiuli}, \citenamefont
  {Laversanne-Finot}, \citenamefont {During}, \citenamefont {Lerner},\ and\
  \citenamefont {Wyart}}]{eric_boson_peak_emt}%
  \BibitemOpen
  \bibfield  {author} {\bibinfo {author} {\bibfnamefont {E.}~\bibnamefont
  {DeGiuli}}, \bibinfo {author} {\bibfnamefont {A.}~\bibnamefont
  {Laversanne-Finot}}, \bibinfo {author} {\bibfnamefont {G.}~\bibnamefont
  {During}}, \bibinfo {author} {\bibfnamefont {E.}~\bibnamefont {Lerner}},\
  and\ \bibinfo {author} {\bibfnamefont {M.}~\bibnamefont {Wyart}},\ }\bibfield
   {title} {\bibinfo {title} {Effects of coordination and pressure on sound
  attenuation{,} boson peak and elasticity in amorphous solids},\ }\href
  {https://doi.org/10.1039/C4SM00561A} {\bibfield  {journal} {\bibinfo
  {journal} {Soft Matter}\ }\textbf {\bibinfo {volume} {10}},\ \bibinfo {pages}
  {5628} (\bibinfo {year} {2014}{\natexlab{a}})}\BibitemShut {NoStop}%
\bibitem [{\citenamefont {Gurevich}\ \emph {et~al.}(2003)\citenamefont
  {Gurevich}, \citenamefont {Parshin},\ and\ \citenamefont
  {Schober}}]{Gurevich2003}%
  \BibitemOpen
  \bibfield  {author} {\bibinfo {author} {\bibfnamefont {V.~L.}\ \bibnamefont
  {Gurevich}}, \bibinfo {author} {\bibfnamefont {D.~A.}\ \bibnamefont
  {Parshin}},\ and\ \bibinfo {author} {\bibfnamefont {H.~R.}\ \bibnamefont
  {Schober}},\ }\bibfield  {title} {\bibinfo {title} {Anharmonicity,
  vibrational instability, and the boson peak in glasses},\ }\href
  {https://doi.org/10.1103/PhysRevB.67.094203} {\bibfield  {journal} {\bibinfo
  {journal} {Phys. Rev. B}\ }\textbf {\bibinfo {volume} {67}},\ \bibinfo
  {pages} {094203} (\bibinfo {year} {2003})}\BibitemShut {NoStop}%
\bibitem [{\citenamefont {Gurevich}\ \emph {et~al.}(2005)\citenamefont
  {Gurevich}, \citenamefont {Parshin},\ and\ \citenamefont
  {Schober}}]{Gurevich2005}%
  \BibitemOpen
  \bibfield  {author} {\bibinfo {author} {\bibfnamefont {V.~L.}\ \bibnamefont
  {Gurevich}}, \bibinfo {author} {\bibfnamefont {D.~A.}\ \bibnamefont
  {Parshin}},\ and\ \bibinfo {author} {\bibfnamefont {H.~R.}\ \bibnamefont
  {Schober}},\ }\bibfield  {title} {\bibinfo {title} {Pressure dependence of
  the boson peak in glasses},\ }\href
  {https://doi.org/10.1103/PhysRevB.71.014209} {\bibfield  {journal} {\bibinfo
  {journal} {Phys. Rev. B}\ }\textbf {\bibinfo {volume} {71}},\ \bibinfo
  {pages} {014209} (\bibinfo {year} {2005})}\BibitemShut {NoStop}%
\bibitem [{\citenamefont {Parshin}\ \emph {et~al.}(2007)\citenamefont
  {Parshin}, \citenamefont {Schober},\ and\ \citenamefont
  {Gurevich}}]{Gurevich2007}%
  \BibitemOpen
  \bibfield  {author} {\bibinfo {author} {\bibfnamefont {D.~A.}\ \bibnamefont
  {Parshin}}, \bibinfo {author} {\bibfnamefont {H.~R.}\ \bibnamefont
  {Schober}},\ and\ \bibinfo {author} {\bibfnamefont {V.~L.}\ \bibnamefont
  {Gurevich}},\ }\bibfield  {title} {\bibinfo {title} {Vibrational instability,
  two-level systems, and the boson peak in glasses},\ }\href
  {https://doi.org/10.1103/PhysRevB.76.064206} {\bibfield  {journal} {\bibinfo
  {journal} {Phys. Rev. B}\ }\textbf {\bibinfo {volume} {76}},\ \bibinfo
  {pages} {064206} (\bibinfo {year} {2007})}\BibitemShut {NoStop}%
\bibitem [{\citenamefont {Tanguy}\ \emph {et~al.}(2002)\citenamefont {Tanguy},
  \citenamefont {Wittmer}, \citenamefont {Leonforte},\ and\ \citenamefont
  {Barrat}}]{barrat_prb_2002}%
  \BibitemOpen
  \bibfield  {author} {\bibinfo {author} {\bibfnamefont {A.}~\bibnamefont
  {Tanguy}}, \bibinfo {author} {\bibfnamefont {J.~P.}\ \bibnamefont {Wittmer}},
  \bibinfo {author} {\bibfnamefont {F.}~\bibnamefont {Leonforte}},\ and\
  \bibinfo {author} {\bibfnamefont {J.-L.}\ \bibnamefont {Barrat}},\ }\bibfield
   {title} {\bibinfo {title} {Continuum limit of amorphous elastic bodies: A
  finite-size study of low-frequency harmonic vibrations},\ }\href
  {https://doi.org/10.1103/PhysRevB.66.174205} {\bibfield  {journal} {\bibinfo
  {journal} {Phys. Rev. B}\ }\textbf {\bibinfo {volume} {66}},\ \bibinfo
  {pages} {174205} (\bibinfo {year} {2002})}\BibitemShut {NoStop}%
\bibitem [{\citenamefont {Lema\^{\i}tre}\ and\ \citenamefont
  {Maloney}(2006)}]{lemaitre_sum_rules_2006}%
  \BibitemOpen
  \bibfield  {author} {\bibinfo {author} {\bibfnamefont {A.}~\bibnamefont
  {Lema\^{\i}tre}}\ and\ \bibinfo {author} {\bibfnamefont {C.}~\bibnamefont
  {Maloney}},\ }\bibfield  {title} {\bibinfo {title} {Sum rules for the
  quasi-static and visco-elastic response of disordered solids at zero
  temperature},\ }\href {https://doi.org/10.1007/s10955-005-9015-5} {\bibfield
  {journal} {\bibinfo  {journal} {J. Stat. Phys.}\ }\textbf {\bibinfo {volume}
  {123}},\ \bibinfo {pages} {415} (\bibinfo {year} {2006})}\BibitemShut
  {NoStop}%
\bibitem [{\citenamefont {Mosayebi}\ \emph {et~al.}(2010)\citenamefont
  {Mosayebi}, \citenamefont {Del~Gado}, \citenamefont {Ilg},\ and\
  \citenamefont {\"Ottinger}}]{delgado_prl_2010}%
  \BibitemOpen
  \bibfield  {author} {\bibinfo {author} {\bibfnamefont {M.}~\bibnamefont
  {Mosayebi}}, \bibinfo {author} {\bibfnamefont {E.}~\bibnamefont {Del~Gado}},
  \bibinfo {author} {\bibfnamefont {P.}~\bibnamefont {Ilg}},\ and\ \bibinfo
  {author} {\bibfnamefont {H.~C.}\ \bibnamefont {\"Ottinger}},\ }\bibfield
  {title} {\bibinfo {title} {Probing a critical length scale at the glass
  transition},\ }\href {https://doi.org/10.1103/PhysRevLett.104.205704}
  {\bibfield  {journal} {\bibinfo  {journal} {Phys. Rev. Lett.}\ }\textbf
  {\bibinfo {volume} {104}},\ \bibinfo {pages} {205704} (\bibinfo {year}
  {2010})}\BibitemShut {NoStop}%
\bibitem [{\citenamefont {Argon}(1979)}]{argon_st}%
  \BibitemOpen
  \bibfield  {author} {\bibinfo {author} {\bibfnamefont {A.}~\bibnamefont
  {Argon}},\ }\bibfield  {title} {\bibinfo {title} {Plastic deformation in
  metallic glasses},\ }\href
  {https://doi.org/http://dx.doi.org/10.1016/0001-6160(79)90055-5} {\bibfield
  {journal} {\bibinfo  {journal} {Acta Metall.}\ }\textbf {\bibinfo
  {volume} {27}},\ \bibinfo {pages} {47 } (\bibinfo {year} {1979})}\BibitemShut
  {NoStop}%
\bibitem [{\citenamefont {Argon}\ and\ \citenamefont
  {Kuo}(1979)}]{argon_bubble_raft}%
  \BibitemOpen
  \bibfield  {author} {\bibinfo {author} {\bibfnamefont {A.}~\bibnamefont
  {Argon}}\ and\ \bibinfo {author} {\bibfnamefont {H.}~\bibnamefont {Kuo}},\
  }\bibfield  {title} {\bibinfo {title} {Plastic flow in a disordered bubble
  raft (an analog of a metallic glass)},\ }\href
  {https://doi.org/http://dx.doi.org/10.1016/0025-5416(79)90174-5} {\bibfield
  {journal} {\bibinfo  {journal} {Mater. Sci. Eng.}\ }\textbf {\bibinfo
  {volume} {39}},\ \bibinfo {pages} {101 } (\bibinfo {year}
  {1979})}\BibitemShut {NoStop}%
\bibitem [{\citenamefont {Falk}\ and\ \citenamefont
  {Langer}(1998)}]{falk_langer_stz}%
  \BibitemOpen
  \bibfield  {author} {\bibinfo {author} {\bibfnamefont {M.~L.}\ \bibnamefont
  {Falk}}\ and\ \bibinfo {author} {\bibfnamefont {J.~S.}\ \bibnamefont
  {Langer}},\ }\bibfield  {title} {\bibinfo {title} {Dynamics of viscoplastic
  deformation in amorphous solids},\ }\href
  {https://doi.org/10.1103/PhysRevE.57.7192} {\bibfield  {journal} {\bibinfo
  {journal} {Phys. Rev. E}\ }\textbf {\bibinfo {volume} {57}},\ \bibinfo
  {pages} {7192} (\bibinfo {year} {1998})}\BibitemShut {NoStop}%
\bibitem [{\citenamefont {Spaepen}(1977)}]{spaepen_1977}%
  \BibitemOpen
  \bibfield  {author} {\bibinfo {author} {\bibfnamefont {F.}~\bibnamefont
  {Spaepen}},\ }\bibfield  {title} {\bibinfo {title} {A microscopic mechanism
  for steady state inhomogeneous flow in metallic glasses},\ }\href
  {https://doi.org/https://doi.org/10.1016/0001-6160(77)90232-2} {\bibfield
  {journal} {\bibinfo  {journal} {Acta Metall.}\ }\textbf {\bibinfo {volume}
  {25}},\ \bibinfo {pages} {407} (\bibinfo {year} {1977})}\BibitemShut
  {NoStop}%
\bibitem [{\citenamefont {Deng}\ \emph {et~al.}(1989)\citenamefont {Deng},
  \citenamefont {Argon},\ and\ \citenamefont {Yip}}]{argon_simulations}%
  \BibitemOpen
  \bibfield  {author} {\bibinfo {author} {\bibfnamefont {D.}~\bibnamefont
  {Deng}}, \bibinfo {author} {\bibfnamefont {A.~S.}\ \bibnamefont {Argon}},\
  and\ \bibinfo {author} {\bibfnamefont {S.}~\bibnamefont {Yip}},\ }\bibfield
  {title} {\bibinfo {title} {Simulation of plastic deformation in a
  two-dimensional atomic glass by molecular dynamics iv},\ }\href
  {https://doi.org/10.1098/rsta.1989.0092} {\bibfield  {journal} {\bibinfo
  {journal} {Philos. Trans. Royal Soc. A}\ }\textbf {\bibinfo {volume} {329}},\
  \bibinfo {pages} {613} (\bibinfo {year} {1989})}\BibitemShut {NoStop}%
\bibitem [{\citenamefont {Malandro}\ and\ \citenamefont
  {Lacks}(1999)}]{Malandro_Lacks}%
  \BibitemOpen
  \bibfield  {author} {\bibinfo {author} {\bibfnamefont {D.~L.}\ \bibnamefont
  {Malandro}}\ and\ \bibinfo {author} {\bibfnamefont {D.~J.}\ \bibnamefont
  {Lacks}},\ }\bibfield  {title} {\bibinfo {title} {Relationships of
  shear-induced changes in the potential energy landscape to the mechanical
  properties of ductile glasses},\ }\href
  {https://doi.org/http://dx.doi.org/10.1063/1.478340} {\bibfield  {journal}
  {\bibinfo  {journal} {J. Chem. Phys.}\ }\textbf {\bibinfo {volume} {110}},\
  \bibinfo {pages} {4593} (\bibinfo {year} {1999})}\BibitemShut {NoStop}%
\bibitem [{\citenamefont {Maloney}\ and\ \citenamefont
  {Lema\^{\i}tre}(2004)}]{lemaitre2004}%
  \BibitemOpen
  \bibfield  {author} {\bibinfo {author} {\bibfnamefont {C.}~\bibnamefont
  {Maloney}}\ and\ \bibinfo {author} {\bibfnamefont {A.}~\bibnamefont
  {Lema\^{\i}tre}},\ }\bibfield  {title} {\bibinfo {title} {Universal breakdown
  of elasticity at the onset of material failure},\ }\href
  {https://doi.org/10.1103/PhysRevLett.93.195501} {\bibfield  {journal}
  {\bibinfo  {journal} {Phys. Rev. Lett.}\ }\textbf {\bibinfo {volume} {93}},\
  \bibinfo {pages} {195501} (\bibinfo {year} {2004})}\BibitemShut {NoStop}%
\bibitem [{\citenamefont {Demkowicz}\ and\ \citenamefont
  {Argon}(2005)}]{Argon_prb_2005}%
  \BibitemOpen
  \bibfield  {author} {\bibinfo {author} {\bibfnamefont {M.~J.}\ \bibnamefont
  {Demkowicz}}\ and\ \bibinfo {author} {\bibfnamefont {A.~S.}\ \bibnamefont
  {Argon}},\ }\bibfield  {title} {\bibinfo {title} {Liquidlike atomic
  environments act as plasticity carriers in amorphous silicon},\ }\href
  {https://doi.org/10.1103/PhysRevB.72.245205} {\bibfield  {journal} {\bibinfo
  {journal} {Phys. Rev. B}\ }\textbf {\bibinfo {volume} {72}},\ \bibinfo
  {pages} {245205} (\bibinfo {year} {2005})}\BibitemShut {NoStop}%
\bibitem [{\citenamefont {Maloney}\ and\ \citenamefont
  {Lema\^{\i}tre}(2006)}]{lemaitre2006_avalanches}%
  \BibitemOpen
  \bibfield  {author} {\bibinfo {author} {\bibfnamefont {C.~E.}\ \bibnamefont
  {Maloney}}\ and\ \bibinfo {author} {\bibfnamefont {A.}~\bibnamefont
  {Lema\^{\i}tre}},\ }\bibfield  {title} {\bibinfo {title} {Amorphous systems
  in athermal, quasistatic shear},\ }\href
  {https://doi.org/10.1103/PhysRevE.74.016118} {\bibfield  {journal} {\bibinfo
  {journal} {Phys. Rev. E}\ }\textbf {\bibinfo {volume} {74}},\ \bibinfo
  {pages} {016118} (\bibinfo {year} {2006})}\BibitemShut {NoStop}%
\bibitem [{\citenamefont {Chikkadi}\ \emph {et~al.}(2011)\citenamefont
  {Chikkadi}, \citenamefont {Wegdam}, \citenamefont {Bonn}, \citenamefont
  {Nienhuis},\ and\ \citenamefont {Schall}}]{schall_stz_colloids}%
  \BibitemOpen
  \bibfield  {author} {\bibinfo {author} {\bibfnamefont {V.}~\bibnamefont
  {Chikkadi}}, \bibinfo {author} {\bibfnamefont {G.}~\bibnamefont {Wegdam}},
  \bibinfo {author} {\bibfnamefont {D.}~\bibnamefont {Bonn}}, \bibinfo {author}
  {\bibfnamefont {B.}~\bibnamefont {Nienhuis}},\ and\ \bibinfo {author}
  {\bibfnamefont {P.}~\bibnamefont {Schall}},\ }\bibfield  {title} {\bibinfo
  {title} {Long-range strain correlations in sheared colloidal glasses},\
  }\href {https://doi.org/10.1103/PhysRevLett.107.198303} {\bibfield  {journal}
  {\bibinfo  {journal} {Phys. Rev. Lett.}\ }\textbf {\bibinfo {volume} {107}},\
  \bibinfo {pages} {198303} (\bibinfo {year} {2011})}\BibitemShut {NoStop}%
\bibitem [{\citenamefont {Manning}\ and\ \citenamefont
  {Liu}(2011)}]{manning2011}%
  \BibitemOpen
  \bibfield  {author} {\bibinfo {author} {\bibfnamefont {M.~L.}\ \bibnamefont
  {Manning}}\ and\ \bibinfo {author} {\bibfnamefont {A.~J.}\ \bibnamefont
  {Liu}},\ }\bibfield  {title} {\bibinfo {title} {Vibrational modes identify
  soft spots in a sheared disordered packing},\ }\href
  {https://doi.org/10.1103/PhysRevLett.107.108302} {\bibfield  {journal}
  {\bibinfo  {journal} {Phys. Rev. Lett.}\ }\textbf {\bibinfo {volume} {107}},\
  \bibinfo {pages} {108302} (\bibinfo {year} {2011})}\BibitemShut {NoStop}%
\bibitem [{\citenamefont {Dasgupta}\ \emph {et~al.}(2013)\citenamefont
  {Dasgupta}, \citenamefont {Gendelman}, \citenamefont {Mishra}, \citenamefont
  {Procaccia},\ and\ \citenamefont {Shor}}]{carmel_pre_2013}%
  \BibitemOpen
  \bibfield  {author} {\bibinfo {author} {\bibfnamefont {R.}~\bibnamefont
  {Dasgupta}}, \bibinfo {author} {\bibfnamefont {O.}~\bibnamefont {Gendelman}},
  \bibinfo {author} {\bibfnamefont {P.}~\bibnamefont {Mishra}}, \bibinfo
  {author} {\bibfnamefont {I.}~\bibnamefont {Procaccia}},\ and\ \bibinfo
  {author} {\bibfnamefont {C.~A. B.~Z.}\ \bibnamefont {Shor}},\ }\bibfield
  {title} {\bibinfo {title} {Shear localization in three-dimensional amorphous
  solids},\ }\href {https://doi.org/10.1103/PhysRevE.88.032401} {\bibfield
  {journal} {\bibinfo  {journal} {Phys. Rev. E}\ }\textbf {\bibinfo {volume}
  {88}},\ \bibinfo {pages} {032401} (\bibinfo {year} {2013})}\BibitemShut
  {NoStop}%
\bibitem [{\citenamefont {Ding}\ \emph {et~al.}(2014)\citenamefont {Ding},
  \citenamefont {Patinet}, \citenamefont {Falk}, \citenamefont {Cheng},\ and\
  \citenamefont {Ma}}]{falk_qlm_2014}%
  \BibitemOpen
  \bibfield  {author} {\bibinfo {author} {\bibfnamefont {J.}~\bibnamefont
  {Ding}}, \bibinfo {author} {\bibfnamefont {S.}~\bibnamefont {Patinet}},
  \bibinfo {author} {\bibfnamefont {M.~L.}\ \bibnamefont {Falk}}, \bibinfo
  {author} {\bibfnamefont {Y.}~\bibnamefont {Cheng}},\ and\ \bibinfo {author}
  {\bibfnamefont {E.}~\bibnamefont {Ma}},\ }\bibfield  {title} {\bibinfo
  {title} {Soft spots and their structural signature in a metallic glass},\
  }\href {https://doi.org/10.1073/pnas.1412095111} {\bibfield  {journal}
  {\bibinfo  {journal} {Proc. Natl. Acad. Sci. U.S.A.}\ }\textbf {\bibinfo
  {volume} {111}},\ \bibinfo {pages} {14052} (\bibinfo {year}
  {2014})}\BibitemShut {NoStop}%
\bibitem [{\citenamefont {Patinet}\ \emph {et~al.}(2016)\citenamefont
  {Patinet}, \citenamefont {Vandembroucq},\ and\ \citenamefont
  {Falk}}]{falk_prl_2016}%
  \BibitemOpen
  \bibfield  {author} {\bibinfo {author} {\bibfnamefont {S.}~\bibnamefont
  {Patinet}}, \bibinfo {author} {\bibfnamefont {D.}~\bibnamefont
  {Vandembroucq}},\ and\ \bibinfo {author} {\bibfnamefont {M.~L.}\ \bibnamefont
  {Falk}},\ }\bibfield  {title} {\bibinfo {title} {Connecting local yield
  stresses with plastic activity in amorphous solids},\ }\href
  {https://doi.org/10.1103/PhysRevLett.117.045501} {\bibfield  {journal}
  {\bibinfo  {journal} {Phys. Rev. Lett.}\ }\textbf {\bibinfo {volume} {117}},\
  \bibinfo {pages} {045501} (\bibinfo {year} {2016})}\BibitemShut {NoStop}%
\bibitem [{\citenamefont {Richard}\ \emph
  {et~al.}(2020{\natexlab{a}})\citenamefont {Richard}, \citenamefont {Ozawa},
  \citenamefont {Patinet}, \citenamefont {Stanifer}, \citenamefont {Shang},
  \citenamefont {Ridout}, \citenamefont {Xu}, \citenamefont {Zhang},
  \citenamefont {Morse}, \citenamefont {Barrat}, \citenamefont {Berthier},
  \citenamefont {Falk}, \citenamefont {Guan}, \citenamefont {Liu},
  \citenamefont {Martens}, \citenamefont {Sastry}, \citenamefont
  {Vandembroucq}, \citenamefont {Lerner},\ and\ \citenamefont
  {Manning}}]{david_collaboration_2020}%
  \BibitemOpen
  \bibfield  {author} {\bibinfo {author} {\bibfnamefont {D.}~\bibnamefont
  {Richard}}, \bibinfo {author} {\bibfnamefont {M.}~\bibnamefont {Ozawa}},
  \bibinfo {author} {\bibfnamefont {S.}~\bibnamefont {Patinet}}, \bibinfo
  {author} {\bibfnamefont {E.}~\bibnamefont {Stanifer}}, \bibinfo {author}
  {\bibfnamefont {B.}~\bibnamefont {Shang}}, \bibinfo {author} {\bibfnamefont
  {S.~A.}\ \bibnamefont {Ridout}}, \bibinfo {author} {\bibfnamefont
  {B.}~\bibnamefont {Xu}}, \bibinfo {author} {\bibfnamefont {G.}~\bibnamefont
  {Zhang}}, \bibinfo {author} {\bibfnamefont {P.~K.}\ \bibnamefont {Morse}},
  \bibinfo {author} {\bibfnamefont {J.-L.}\ \bibnamefont {Barrat}}, \bibinfo
  {author} {\bibfnamefont {L.}~\bibnamefont {Berthier}}, \bibinfo {author}
  {\bibfnamefont {M.~L.}\ \bibnamefont {Falk}}, \bibinfo {author}
  {\bibfnamefont {P.}~\bibnamefont {Guan}}, \bibinfo {author} {\bibfnamefont
  {A.~J.}\ \bibnamefont {Liu}}, \bibinfo {author} {\bibfnamefont
  {K.}~\bibnamefont {Martens}}, \bibinfo {author} {\bibfnamefont
  {S.}~\bibnamefont {Sastry}}, \bibinfo {author} {\bibfnamefont
  {D.}~\bibnamefont {Vandembroucq}}, \bibinfo {author} {\bibfnamefont
  {E.}~\bibnamefont {Lerner}},\ and\ \bibinfo {author} {\bibfnamefont {M.~L.}\
  \bibnamefont {Manning}},\ }\bibfield  {title} {\bibinfo {title} {Predicting
  plasticity in disordered solids from structural indicators},\ }\href
  {https://doi.org/10.1103/PhysRevMaterials.4.113609} {\bibfield  {journal}
  {\bibinfo  {journal} {Phys. Rev. Materials}\ }\textbf {\bibinfo {volume}
  {4}},\ \bibinfo {pages} {113609} (\bibinfo {year}
  {2020}{\natexlab{a}})}\BibitemShut {NoStop}%
\bibitem [{\citenamefont {Ozawa}\ \emph {et~al.}(2018)\citenamefont {Ozawa},
  \citenamefont {Berthier}, \citenamefont {Biroli}, \citenamefont {Rosso},\
  and\ \citenamefont {Tarjus}}]{Ozawa6656}%
  \BibitemOpen
  \bibfield  {author} {\bibinfo {author} {\bibfnamefont {M.}~\bibnamefont
  {Ozawa}}, \bibinfo {author} {\bibfnamefont {L.}~\bibnamefont {Berthier}},
  \bibinfo {author} {\bibfnamefont {G.}~\bibnamefont {Biroli}}, \bibinfo
  {author} {\bibfnamefont {A.}~\bibnamefont {Rosso}},\ and\ \bibinfo {author}
  {\bibfnamefont {G.}~\bibnamefont {Tarjus}},\ }\bibfield  {title} {\bibinfo
  {title} {Random critical point separates brittle and ductile yielding
  transitions in amorphous materials},\ }\href
  {https://doi.org/10.1073/pnas.1806156115} {\bibfield  {journal} {\bibinfo
  {journal} {Proc. Natl. Acad. Sci. U.S.A.}\ }\textbf {\bibinfo {volume}
  {115}},\ \bibinfo {pages} {6656} (\bibinfo {year} {2018})}\BibitemShut
  {NoStop}%
\bibitem [{\citenamefont {Popovi\ifmmode~\acute{c}\else \'{c}\fi{}}\ \emph
  {et~al.}(2018)\citenamefont {Popovi\ifmmode~\acute{c}\else \'{c}\fi{}},
  \citenamefont {de~Geus},\ and\ \citenamefont
  {Wyart}}]{mw_sudden_failure_pre_2018}%
  \BibitemOpen
  \bibfield  {author} {\bibinfo {author} {\bibfnamefont {M.}~\bibnamefont
  {Popovi\ifmmode~\acute{c}\else \'{c}\fi{}}}, \bibinfo {author} {\bibfnamefont
  {T.~W.~J.}\ \bibnamefont {de~Geus}},\ and\ \bibinfo {author} {\bibfnamefont
  {M.}~\bibnamefont {Wyart}},\ }\bibfield  {title} {\bibinfo {title}
  {Elastoplastic description of sudden failure in athermal amorphous materials
  during quasistatic loading},\ }\href
  {https://doi.org/10.1103/PhysRevE.98.040901} {\bibfield  {journal} {\bibinfo
  {journal} {Phys. Rev. E}\ }\textbf {\bibinfo {volume} {98}},\ \bibinfo
  {pages} {040901} (\bibinfo {year} {2018})}\BibitemShut {NoStop}%
\bibitem [{\citenamefont {Barlow}\ \emph {et~al.}(2020)\citenamefont {Barlow},
  \citenamefont {Cochran},\ and\ \citenamefont {Fielding}}]{Fielding_prl_2020}%
  \BibitemOpen
  \bibfield  {author} {\bibinfo {author} {\bibfnamefont {H.~J.}\ \bibnamefont
  {Barlow}}, \bibinfo {author} {\bibfnamefont {J.~O.}\ \bibnamefont
  {Cochran}},\ and\ \bibinfo {author} {\bibfnamefont {S.~M.}\ \bibnamefont
  {Fielding}},\ }\bibfield  {title} {\bibinfo {title} {Ductile and brittle
  yielding in thermal and athermal amorphous materials},\ }\href
  {https://doi.org/10.1103/PhysRevLett.125.168003} {\bibfield  {journal}
  {\bibinfo  {journal} {Phys. Rev. Lett.}\ }\textbf {\bibinfo {volume} {125}},\
  \bibinfo {pages} {168003} (\bibinfo {year} {2020})}\BibitemShut {NoStop}%
\bibitem [{\citenamefont {Richard}\ \emph
  {et~al.}(2021{\natexlab{a}})\citenamefont {Richard}, \citenamefont
  {Rainone},\ and\ \citenamefont {Lerner}}]{david_macroscopic_shear_band}%
  \BibitemOpen
  \bibfield  {author} {\bibinfo {author} {\bibfnamefont {D.}~\bibnamefont
  {Richard}}, \bibinfo {author} {\bibfnamefont {C.}~\bibnamefont {Rainone}},\
  and\ \bibinfo {author} {\bibfnamefont {E.}~\bibnamefont {Lerner}},\
  }\bibfield  {title} {\bibinfo {title} {Finite-size study of the athermal
  quasistatic yielding transition in structural glasses},\ }\href
  {https://doi.org/10.1063/5.0053303} {\bibfield  {journal} {\bibinfo
  {journal} {J. Chem. Phys.}\ }\textbf {\bibinfo {volume} {155}},\ \bibinfo
  {pages} {056101} (\bibinfo {year} {2021}{\natexlab{a}})}\BibitemShut
  {NoStop}%
\bibitem [{\citenamefont {Conner}\ \emph {et~al.}(2003)\citenamefont {Conner},
  \citenamefont {Johnson}, \citenamefont {Paton},\ and\ \citenamefont
  {Nix}}]{johnson_shear_bands_2003}%
  \BibitemOpen
  \bibfield  {author} {\bibinfo {author} {\bibfnamefont {R.~D.}\ \bibnamefont
  {Conner}}, \bibinfo {author} {\bibfnamefont {W.~L.}\ \bibnamefont {Johnson}},
  \bibinfo {author} {\bibfnamefont {N.~E.}\ \bibnamefont {Paton}},\ and\
  \bibinfo {author} {\bibfnamefont {W.~D.}\ \bibnamefont {Nix}},\ }\bibfield
  {title} {\bibinfo {title} {Shear bands and cracking of metallic glass plates
  in bending},\ }\href {https://doi.org/10.1063/1.1582555} {\bibfield
  {journal} {\bibinfo  {journal} {J. Appl. Phys.}\ }\textbf {\bibinfo {volume}
  {94}},\ \bibinfo {pages} {904} (\bibinfo {year} {2003})}\BibitemShut
  {NoStop}%
\bibitem [{\citenamefont {Shi}\ and\ \citenamefont
  {Falk}(2005)}]{falk_shi_prl_2005}%
  \BibitemOpen
  \bibfield  {author} {\bibinfo {author} {\bibfnamefont {Y.}~\bibnamefont
  {Shi}}\ and\ \bibinfo {author} {\bibfnamefont {M.~L.}\ \bibnamefont {Falk}},\
  }\bibfield  {title} {\bibinfo {title} {Strain localization and percolation of
  stable structure in amorphous solids},\ }\href
  {https://doi.org/10.1103/PhysRevLett.95.095502} {\bibfield  {journal}
  {\bibinfo  {journal} {Phys. Rev. Lett.}\ }\textbf {\bibinfo {volume} {95}},\
  \bibinfo {pages} {095502} (\bibinfo {year} {2005})}\BibitemShut {NoStop}%
\bibitem [{\citenamefont {Schuh}\ \emph {et~al.}(2007)\citenamefont {Schuh},
  \citenamefont {Hufnagel},\ and\ \citenamefont
  {Ramamurty}}]{Schuh_review_2007}%
  \BibitemOpen
  \bibfield  {author} {\bibinfo {author} {\bibfnamefont {C.~A.}\ \bibnamefont
  {Schuh}}, \bibinfo {author} {\bibfnamefont {T.~C.}\ \bibnamefont
  {Hufnagel}},\ and\ \bibinfo {author} {\bibfnamefont {U.}~\bibnamefont
  {Ramamurty}},\ }\bibfield  {title} {\bibinfo {title} {Mechanical behavior of
  amorphous alloys},\ }\href
  {https://doi.org/http://dx.doi.org/10.1016/j.actamat.2007.01.052} {\bibfield
  {journal} {\bibinfo  {journal} {Acta Mater.}\ }\textbf {\bibinfo {volume}
  {55}},\ \bibinfo {pages} {4067 } (\bibinfo {year} {2007})}\BibitemShut
  {NoStop}%
\bibitem [{\citenamefont {Falk}\ and\ \citenamefont {Langer}(2011)}]{Falk2011}%
  \BibitemOpen
  \bibfield  {author} {\bibinfo {author} {\bibfnamefont {M.~L.}\ \bibnamefont
  {Falk}}\ and\ \bibinfo {author} {\bibfnamefont {J.}~\bibnamefont {Langer}},\
  }\bibfield  {title} {\bibinfo {title} {Deformation and failure of amorphous,
  solidlike materials},\ }\href
  {https://doi.org/10.1146/annurev-conmatphys-062910-140452} {\bibfield
  {journal} {\bibinfo  {journal} {Annu. Rev. Condens. Matter Phys.}\ }\textbf
  {\bibinfo {volume} {2}},\ \bibinfo {pages} {353} (\bibinfo {year}
  {2011})}\BibitemShut {NoStop}%
\bibitem [{\citenamefont {Parisi}\ \emph {et~al.}(2017)\citenamefont {Parisi},
  \citenamefont {Procaccia}, \citenamefont {Rainone},\ and\ \citenamefont
  {Singh}}]{itamar_yielding_pnas_2017}%
  \BibitemOpen
  \bibfield  {author} {\bibinfo {author} {\bibfnamefont {G.}~\bibnamefont
  {Parisi}}, \bibinfo {author} {\bibfnamefont {I.}~\bibnamefont {Procaccia}},
  \bibinfo {author} {\bibfnamefont {C.}~\bibnamefont {Rainone}},\ and\ \bibinfo
  {author} {\bibfnamefont {M.}~\bibnamefont {Singh}},\ }\bibfield  {title}
  {\bibinfo {title} {Shear bands as manifestation of a criticality in yielding
  amorphous solids},\ }\href {https://doi.org/10.1073/pnas.1700075114}
  {\bibfield  {journal} {\bibinfo  {journal} {Proc. Natl. Acad. Sci. U.S.A.}\
  }\textbf {\bibinfo {volume} {114}},\ \bibinfo {pages} {5577} (\bibinfo {year}
  {2017})}\BibitemShut {NoStop}%
\bibitem [{\citenamefont {Lerner}\ and\ \citenamefont
  {Bouchbinder}(2018)}]{cge_paper}%
  \BibitemOpen
  \bibfield  {author} {\bibinfo {author} {\bibfnamefont {E.}~\bibnamefont
  {Lerner}}\ and\ \bibinfo {author} {\bibfnamefont {E.}~\bibnamefont
  {Bouchbinder}},\ }\bibfield  {title} {\bibinfo {title} {A characteristic
  energy scale in glasses},\ }\href {https://doi.org/10.1063/1.5024776}
  {\bibfield  {journal} {\bibinfo  {journal} {J. Chem. Phys.}\ }\textbf
  {\bibinfo {volume} {148}},\ \bibinfo {pages} {214502} (\bibinfo {year}
  {2018})}\BibitemShut {NoStop}%
\bibitem [{\citenamefont {Kapteijns}\ \emph {et~al.}(2018)\citenamefont
  {Kapteijns}, \citenamefont {Bouchbinder},\ and\ \citenamefont
  {Lerner}}]{modes_prl_2018}%
  \BibitemOpen
  \bibfield  {author} {\bibinfo {author} {\bibfnamefont {G.}~\bibnamefont
  {Kapteijns}}, \bibinfo {author} {\bibfnamefont {E.}~\bibnamefont
  {Bouchbinder}},\ and\ \bibinfo {author} {\bibfnamefont {E.}~\bibnamefont
  {Lerner}},\ }\bibfield  {title} {\bibinfo {title} {Universal nonphononic
  density of states in 2D, 3D, and 4D glasses},\ }\href
  {https://doi.org/10.1103/PhysRevLett.121.055501} {\bibfield  {journal}
  {\bibinfo  {journal} {Phys. Rev. Lett.}\ }\textbf {\bibinfo {volume} {121}},\
  \bibinfo {pages} {055501} (\bibinfo {year} {2018})}\BibitemShut {NoStop}%
\bibitem [{\citenamefont {Moriel}\ \emph {et~al.}(2020)\citenamefont {Moriel},
  \citenamefont {Lubomirsky}, \citenamefont {Lerner},\ and\ \citenamefont
  {Bouchbinder}}]{avraham_core_properties_pre_2020}%
  \BibitemOpen
  \bibfield  {author} {\bibinfo {author} {\bibfnamefont {A.}~\bibnamefont
  {Moriel}}, \bibinfo {author} {\bibfnamefont {Y.}~\bibnamefont {Lubomirsky}},
  \bibinfo {author} {\bibfnamefont {E.}~\bibnamefont {Lerner}},\ and\ \bibinfo
  {author} {\bibfnamefont {E.}~\bibnamefont {Bouchbinder}},\ }\bibfield
  {title} {\bibinfo {title} {Extracting the properties of quasilocalized modes
  in computer glasses: Long-range continuum fields, contour integrals, and
  boundary effects},\ }\href {https://doi.org/10.1103/PhysRevE.102.033008}
  {\bibfield  {journal} {\bibinfo  {journal} {Phys. Rev. E}\ }\textbf {\bibinfo
  {volume} {102}},\ \bibinfo {pages} {033008} (\bibinfo {year}
  {2020})}\BibitemShut {NoStop}%
\bibitem [{\citenamefont {Rainone}\ \emph
  {et~al.}(2020{\natexlab{a}})\citenamefont {Rainone}, \citenamefont
  {Bouchbinder},\ and\ \citenamefont {Lerner}}]{pinching_pnas}%
  \BibitemOpen
  \bibfield  {author} {\bibinfo {author} {\bibfnamefont {C.}~\bibnamefont
  {Rainone}}, \bibinfo {author} {\bibfnamefont {E.}~\bibnamefont
  {Bouchbinder}},\ and\ \bibinfo {author} {\bibfnamefont {E.}~\bibnamefont
  {Lerner}},\ }\bibfield  {title} {\bibinfo {title} {Pinching a glass reveals
  key properties of its soft spots},\ }\href
  {https://doi.org/10.1073/pnas.1919958117} {\bibfield  {journal} {\bibinfo
  {journal} {Proc. Natl. Acad. Sci. U.S.A.}\ }\textbf {\bibinfo {volume}
  {117}},\ \bibinfo {pages} {5228} (\bibinfo {year}
  {2020}{\natexlab{a}})}\BibitemShut {NoStop}%
\bibitem [{\citenamefont {Wang}\ \emph {et~al.}(2019)\citenamefont {Wang},
  \citenamefont {Ninarello}, \citenamefont {Guan}, \citenamefont {Berthier},
  \citenamefont {Szamel},\ and\ \citenamefont {Flenner}}]{LB_modes_2019}%
  \BibitemOpen
  \bibfield  {author} {\bibinfo {author} {\bibfnamefont {L.}~\bibnamefont
  {Wang}}, \bibinfo {author} {\bibfnamefont {A.}~\bibnamefont {Ninarello}},
  \bibinfo {author} {\bibfnamefont {P.}~\bibnamefont {Guan}}, \bibinfo {author}
  {\bibfnamefont {L.}~\bibnamefont {Berthier}}, \bibinfo {author}
  {\bibfnamefont {G.}~\bibnamefont {Szamel}},\ and\ \bibinfo {author}
  {\bibfnamefont {E.}~\bibnamefont {Flenner}},\ }\bibfield  {title} {\bibinfo
  {title} {Low-frequency vibrational modes of stable glasses},\ }\href
  {https://doi.org/10.1038/s41467-018-07978-1} {\bibfield  {journal} {\bibinfo
  {journal} {Nat. Commun.}\ }\textbf {\bibinfo {volume} {10}},\ \bibinfo
  {pages} {26} (\bibinfo {year} {2019})}\BibitemShut {NoStop}%
\bibitem [{\citenamefont {Lerner}\ \emph {et~al.}(2016)\citenamefont {Lerner},
  \citenamefont {D\"uring},\ and\ \citenamefont
  {Bouchbinder}}]{modes_prl_2016}%
  \BibitemOpen
  \bibfield  {author} {\bibinfo {author} {\bibfnamefont {E.}~\bibnamefont
  {Lerner}}, \bibinfo {author} {\bibfnamefont {G.}~\bibnamefont {D\"uring}},\
  and\ \bibinfo {author} {\bibfnamefont {E.}~\bibnamefont {Bouchbinder}},\
  }\bibfield  {title} {\bibinfo {title} {Statistics and properties of
  low-frequency vibrational modes in structural glasses},\ }\href
  {https://doi.org/10.1103/PhysRevLett.117.035501} {\bibfield  {journal}
  {\bibinfo  {journal} {Phys. Rev. Lett.}\ }\textbf {\bibinfo {volume} {117}},\
  \bibinfo {pages} {035501} (\bibinfo {year} {2016})}\BibitemShut {NoStop}%
\bibitem [{\citenamefont {Mizuno}\ \emph {et~al.}(2017)\citenamefont {Mizuno},
  \citenamefont {Shiba},\ and\ \citenamefont {Ikeda}}]{ikeda_pnas}%
  \BibitemOpen
  \bibfield  {author} {\bibinfo {author} {\bibfnamefont {H.}~\bibnamefont
  {Mizuno}}, \bibinfo {author} {\bibfnamefont {H.}~\bibnamefont {Shiba}},\ and\
  \bibinfo {author} {\bibfnamefont {A.}~\bibnamefont {Ikeda}},\ }\bibfield
  {title} {\bibinfo {title} {Continuum limit of the vibrational properties of
  amorphous solids},\ }\href {https://doi.org/10.1073/pnas.1709015114}
  {\bibfield  {journal} {\bibinfo  {journal} {Proc. Natl. Acad. Sci. U.S.A.}\
  }\textbf {\bibinfo {volume} {114}},\ \bibinfo {pages} {E9767} (\bibinfo
  {year} {2017})}\BibitemShut {NoStop}%
\bibitem [{\citenamefont {Richard}\ \emph
  {et~al.}(2020{\natexlab{b}})\citenamefont {Richard}, \citenamefont
  {Gonz\'alez-L\'opez}, \citenamefont {Kapteijns}, \citenamefont {Pater},
  \citenamefont {Vaknin}, \citenamefont {Bouchbinder},\ and\ \citenamefont
  {Lerner}}]{modes_prl_2020}%
  \BibitemOpen
  \bibfield  {author} {\bibinfo {author} {\bibfnamefont {D.}~\bibnamefont
  {Richard}}, \bibinfo {author} {\bibfnamefont {K.}~\bibnamefont
  {Gonz\'alez-L\'opez}}, \bibinfo {author} {\bibfnamefont {G.}~\bibnamefont
  {Kapteijns}}, \bibinfo {author} {\bibfnamefont {R.}~\bibnamefont {Pater}},
  \bibinfo {author} {\bibfnamefont {T.}~\bibnamefont {Vaknin}}, \bibinfo
  {author} {\bibfnamefont {E.}~\bibnamefont {Bouchbinder}},\ and\ \bibinfo
  {author} {\bibfnamefont {E.}~\bibnamefont {Lerner}},\ }\bibfield  {title}
  {\bibinfo {title} {Universality of the nonphononic vibrational spectrum
  across different classes of computer glasses},\ }\href
  {https://doi.org/10.1103/PhysRevLett.125.085502} {\bibfield  {journal}
  {\bibinfo  {journal} {Phys. Rev. Lett.}\ }\textbf {\bibinfo {volume} {125}},\
  \bibinfo {pages} {085502} (\bibinfo {year} {2020}{\natexlab{b}})}\BibitemShut
  {NoStop}%
\bibitem [{\citenamefont {Il'in}\ \emph {et~al.}(1987)\citenamefont {Il'in},
  \citenamefont {Karpov},\ and\ \citenamefont
  {Parshin}}]{soft_potential_model_1987}%
  \BibitemOpen
  \bibfield  {author} {\bibinfo {author} {\bibfnamefont {M.}~\bibnamefont
  {Il'in}}, \bibinfo {author} {\bibfnamefont {V.}~\bibnamefont {Karpov}},\ and\
  \bibinfo {author} {\bibfnamefont {D.}~\bibnamefont {Parshin}},\ }\bibfield
  {title} {\bibinfo {title} {Parameters of soft atomic potentials in glasses},\
  }\href {http://www.jetp.ac.ru/cgi-bin/e/index/e/65/1/p165?a=list} {\bibfield
  {journal} {\bibinfo  {journal} {Zh. Eksp. Teor. Fiz.}\ }\textbf {\bibinfo
  {volume} {92}},\ \bibinfo {pages} {291} (\bibinfo {year} {1987})}\BibitemShut
  {NoStop}%
\bibitem [{\citenamefont {Galperin}\ \emph {et~al.}(1989)\citenamefont
  {Galperin}, \citenamefont {Karpov},\ and\ \citenamefont
  {Kozub}}]{soft_potential_model_1989}%
  \BibitemOpen
  \bibfield  {author} {\bibinfo {author} {\bibfnamefont {Y.}~\bibnamefont
  {Galperin}}, \bibinfo {author} {\bibfnamefont {V.}~\bibnamefont {Karpov}},\
  and\ \bibinfo {author} {\bibfnamefont {V.}~\bibnamefont {Kozub}},\ }\bibfield
   {title} {\bibinfo {title} {Localized states in glasses},\ }\href
  {https://doi.org/10.1080/00018738900101162} {\bibfield  {journal} {\bibinfo
  {journal} {Adv. Phys.}\ }\textbf {\bibinfo {volume} {38}},\ \bibinfo {pages}
  {669} (\bibinfo {year} {1989})}\BibitemShut {NoStop}%
\bibitem [{\citenamefont {Rosenstock}(1962)}]{rosenstock1962}%
  \BibitemOpen
  \bibfield  {author} {\bibinfo {author} {\bibfnamefont {H.~B.}\ \bibnamefont
  {Rosenstock}},\ }\bibfield  {title} {\bibinfo {title} {Anomalous specific
  heat of disordered solids},\ }\href
  {https://doi.org/https://doi.org/10.1016/0022-3697(62)90525-5} {\bibfield
  {journal} {\bibinfo  {journal} {J. Phys. Chem. Solids}\ }\textbf {\bibinfo
  {volume} {23}},\ \bibinfo {pages} {659} (\bibinfo {year} {1962})}\BibitemShut
  {NoStop}%
\bibitem [{\citenamefont {Anderson}(1959)}]{Anderson_1959}%
  \BibitemOpen
  \bibfield  {author} {\bibinfo {author} {\bibfnamefont {O.}~\bibnamefont
  {Anderson}},\ }\bibfield  {title} {\bibinfo {title} {The debye temperature of
  vitreous silica},\ }\href
  {https://doi.org/https://doi.org/10.1016/0022-3697(59)90250-1} {\bibfield
  {journal} {\bibinfo  {journal} {J. Phys. Chem. Solids}\ }\textbf {\bibinfo
  {volume} {12}},\ \bibinfo {pages} {41} (\bibinfo {year} {1959})}\BibitemShut
  {NoStop}%
\bibitem [{\citenamefont {Flubacher}\ \emph {et~al.}(1959)\citenamefont
  {Flubacher}, \citenamefont {Leadbetter}, \citenamefont {Morrison},\ and\
  \citenamefont {Stoicheff}}]{Flubacher_1959}%
  \BibitemOpen
  \bibfield  {author} {\bibinfo {author} {\bibfnamefont {P.}~\bibnamefont
  {Flubacher}}, \bibinfo {author} {\bibfnamefont {A.}~\bibnamefont
  {Leadbetter}}, \bibinfo {author} {\bibfnamefont {J.}~\bibnamefont
  {Morrison}},\ and\ \bibinfo {author} {\bibfnamefont {B.}~\bibnamefont
  {Stoicheff}},\ }\bibfield  {title} {\bibinfo {title} {The low-temperature
  heat capacity and the raman and brillouin spectra of vitreous silica},\
  }\href {https://doi.org/https://doi.org/10.1016/0022-3697(59)90251-3}
  {\bibfield  {journal} {\bibinfo  {journal} {J. Phys. Chem. Solids}\ }\textbf
  {\bibinfo {volume} {12}},\ \bibinfo {pages} {53} (\bibinfo {year}
  {1959})}\BibitemShut {NoStop}%
\bibitem [{\citenamefont {Leadbetter}\ and\ \citenamefont
  {Litchinsky}(1970)}]{Leadbetter1970}%
  \BibitemOpen
  \bibfield  {author} {\bibinfo {author} {\bibfnamefont {A.~J.}\ \bibnamefont
  {Leadbetter}}\ and\ \bibinfo {author} {\bibfnamefont {D.}~\bibnamefont
  {Litchinsky}},\ }\bibfield  {title} {\bibinfo {title} {Vibrational properties
  of vitreous germania by inelastic cold neutron scattering},\ }\href
  {https://doi.org/10.1039/DF9705000062} {\bibfield  {journal} {\bibinfo
  {journal} {Discuss. Faraday Soc.}\ }\textbf {\bibinfo {volume} {50}},\
  \bibinfo {pages} {62} (\bibinfo {year} {1970})}\BibitemShut {NoStop}%
\bibitem [{\citenamefont {Laird}\ and\ \citenamefont
  {Schober}(1991)}]{Schober_Laird_numerics_PRL}%
  \BibitemOpen
  \bibfield  {author} {\bibinfo {author} {\bibfnamefont {B.~B.}\ \bibnamefont
  {Laird}}\ and\ \bibinfo {author} {\bibfnamefont {H.~R.}\ \bibnamefont
  {Schober}},\ }\bibfield  {title} {\bibinfo {title} {Localized low-frequency
  vibrational modes in a simple model glass},\ }\href
  {https://doi.org/10.1103/PhysRevLett.66.636} {\bibfield  {journal} {\bibinfo
  {journal} {Phys. Rev. Lett.}\ }\textbf {\bibinfo {volume} {66}},\ \bibinfo
  {pages} {636} (\bibinfo {year} {1991})}\BibitemShut {NoStop}%
\bibitem [{\citenamefont {Dederichs}\ \emph {et~al.}(1973)\citenamefont
  {Dederichs}, \citenamefont {Lehmann},\ and\ \citenamefont
  {Scholz}}]{Resonant_Modes_1973}%
  \BibitemOpen
  \bibfield  {author} {\bibinfo {author} {\bibfnamefont {P.~H.}\ \bibnamefont
  {Dederichs}}, \bibinfo {author} {\bibfnamefont {C.}~\bibnamefont {Lehmann}},\
  and\ \bibinfo {author} {\bibfnamefont {A.}~\bibnamefont {Scholz}},\
  }\bibfield  {title} {\bibinfo {title} {Resonance modes of interstitial atoms
  in fcc metals},\ }\href {https://doi.org/10.1103/PhysRevLett.31.1130}
  {\bibfield  {journal} {\bibinfo  {journal} {Phys. Rev. Lett.}\ }\textbf
  {\bibinfo {volume} {31}},\ \bibinfo {pages} {1130} (\bibinfo {year}
  {1973})}\BibitemShut {NoStop}%
\bibitem [{\citenamefont {Phillips}(1978)}]{Phillips1978}%
  \BibitemOpen
  \bibfield  {author} {\bibinfo {author} {\bibfnamefont {W.}~\bibnamefont
  {Phillips}},\ }\bibfield  {title} {\bibinfo {title} {Structure and the low
  temperature properties of amorphous solids},\ }\href
  {https://doi.org/https://doi.org/10.1016/0022-3093(78)90108-4} {\bibfield
  {journal} {\bibinfo  {journal} {J. Non-Cryst. Solids}\ }\textbf {\bibinfo
  {volume} {31}},\ \bibinfo {pages} {267} (\bibinfo {year} {1978})} \BibitemShut {NoStop}%
\bibitem [{\citenamefont {Maeda}\ and\ \citenamefont
  {Takeuchi}(1978)}]{Maeda1978}%
  \BibitemOpen
  \bibfield  {author} {\bibinfo {author} {\bibfnamefont {K.}~\bibnamefont
  {Maeda}}\ and\ \bibinfo {author} {\bibfnamefont {S.}~\bibnamefont
  {Takeuchi}},\ }\bibfield  {title} {\bibinfo {title} {Computer simulation of
  deformation in two-dimensional amorphous structures},\ }\href
  {https://doi.org/https://doi.org/10.1002/pssa.2210490233} {\bibfield
  {journal} {\bibinfo  {journal} {Phys. Status Solidi A}\ }\textbf {\bibinfo
  {volume} {49}},\ \bibinfo {pages} {685} (\bibinfo {year} {1978})}\BibitemShut
  {NoStop}%
\bibitem [{\citenamefont {Egami}\ \emph {et~al.}(1980)\citenamefont {Egami},
  \citenamefont {Maeda},\ and\ \citenamefont {Vitek}}]{Egami_Maeda_Vitek_1980}%
  \BibitemOpen
  \bibfield  {author} {\bibinfo {author} {\bibfnamefont {T.}~\bibnamefont
  {Egami}}, \bibinfo {author} {\bibfnamefont {K.}~\bibnamefont {Maeda}},\ and\
  \bibinfo {author} {\bibfnamefont {V.}~\bibnamefont {Vitek}},\ }\bibfield
  {title} {\bibinfo {title} {Structural defects in amorphous solids a computer
  simulation study},\ }\href {https://doi.org/10.1080/01418618008243894}
  {\bibfield  {journal} {\bibinfo  {journal} {Philos. Mag. A}\ }\textbf
  {\bibinfo {volume} {41}},\ \bibinfo {pages} {883} (\bibinfo {year}
  {1980})}\BibitemShut {NoStop}%
\bibitem [{\citenamefont {Srolovitz}\ \emph {et~al.}(1981)\citenamefont
  {Srolovitz}, \citenamefont {Maeda}, \citenamefont {Vitek},\ and\
  \citenamefont {Egami}}]{Srolovitz1981}%
  \BibitemOpen
  \bibfield  {author} {\bibinfo {author} {\bibfnamefont {D.}~\bibnamefont
  {Srolovitz}}, \bibinfo {author} {\bibfnamefont {K.}~\bibnamefont {Maeda}},
  \bibinfo {author} {\bibfnamefont {V.}~\bibnamefont {Vitek}},\ and\ \bibinfo
  {author} {\bibfnamefont {T.}~\bibnamefont {Egami}},\ }\bibfield  {title}
  {\bibinfo {title} {Structural defects in amorphous solids statistical
  analysis of a computer model},\ }\href
  {https://doi.org/10.1080/01418618108239553} {\bibfield  {journal} {\bibinfo
  {journal} {Philos. Mag. A}\ }\textbf {\bibinfo {volume} {44}},\ \bibinfo
  {pages} {847} (\bibinfo {year} {1981})}\BibitemShut {NoStop}%
\bibitem [{\citenamefont {Srolovitz}\ \emph {et~al.}(1983)\citenamefont
  {Srolovitz}, \citenamefont {Vitek},\ and\ \citenamefont
  {Egami}}]{Srolovitz1983}%
  \BibitemOpen
  \bibfield  {author} {\bibinfo {author} {\bibfnamefont {D.}~\bibnamefont
  {Srolovitz}}, \bibinfo {author} {\bibfnamefont {V.}~\bibnamefont {Vitek}},\
  and\ \bibinfo {author} {\bibfnamefont {T.}~\bibnamefont {Egami}},\ }\bibfield
   {title} {\bibinfo {title} {An atomistic study of deformation of amorphous
  metals},\ }\href
  {https://doi.org/https://doi.org/10.1016/0001-6160(83)90110-4} {\bibfield
  {journal} {\bibinfo  {journal} {Acta Metall.}\ }\textbf {\bibinfo {volume}
  {31}},\ \bibinfo {pages} {335} (\bibinfo {year} {1983})}\BibitemShut
  {NoStop}%
\bibitem [{\citenamefont {Karpov}\ \emph {et~al.}(1982)\citenamefont {Karpov},
  \citenamefont {Klinger},\ and\ \citenamefont {Ignatiev}}]{karpov1982}%
  \BibitemOpen
  \bibfield  {author} {\bibinfo {author} {\bibfnamefont {V.}~\bibnamefont
  {Karpov}}, \bibinfo {author} {\bibfnamefont {M.}~\bibnamefont {Klinger}},\
  and\ \bibinfo {author} {\bibfnamefont {P.}~\bibnamefont {Ignatiev}},\
  }\bibfield  {title} {\bibinfo {title} {Atomic tunneling states and
  low-temperature anomalies of thermal properties in amorphous materials},\
  }\href {https://doi.org/https://doi.org/10.1016/0038-1098(82)90866-3}
  {\bibfield  {journal} {\bibinfo  {journal} {Solid State Commun.}\ }\textbf
  {\bibinfo {volume} {44}},\ \bibinfo {pages} {333} (\bibinfo {year}
  {1982})}\BibitemShut {NoStop}%
\bibitem [{\citenamefont {Klinger}(1983)}]{klinger1983}%
  \BibitemOpen
  \bibfield  {author} {\bibinfo {author} {\bibfnamefont {M.}~\bibnamefont
  {Klinger}},\ }\bibfield  {title} {\bibinfo {title} {Atomic quantum diffusion,
  tunnelling states and some related phenomena in condensed systems},\ }\href
  {https://doi.org/https://doi.org/10.1016/0370-1573(83)90012-1} {\bibfield
  {journal} {\bibinfo  {journal} {Phys. Rep.}\ }\textbf {\bibinfo {volume}
  {94}},\ \bibinfo {pages} {183} (\bibinfo {year} {1983})}\BibitemShut
  {NoStop}%
\bibitem [{\citenamefont {Karpov}\ \emph {et~al.}(1983)\citenamefont {Karpov},
  \citenamefont {Klinger},\ and\ \citenamefont {Ignat’Ev}}]{karpov1983}%
  \BibitemOpen
  \bibfield  {author} {\bibinfo {author} {\bibfnamefont {V.}~\bibnamefont
  {Karpov}}, \bibinfo {author} {\bibfnamefont {I.}~\bibnamefont {Klinger}},\
  and\ \bibinfo {author} {\bibfnamefont {F.}~\bibnamefont {Ignat’Ev}},\
  }\bibfield  {title} {\bibinfo {title} {Theory of the low-temperature
  anomalies in the thermal properties of amorphous structures},\ }\href@noop {}
  {\bibfield  {journal} {\bibinfo  {journal} {Zh. eksp. teor. Fiz}\ }\textbf
  {\bibinfo {volume} {84}},\ \bibinfo {pages} {760} (\bibinfo {year}
  {1983})}\BibitemShut {NoStop}%
\bibitem [{\citenamefont {Buchenau}\ \emph {et~al.}(1991)\citenamefont
  {Buchenau}, \citenamefont {Galperin}, \citenamefont {Gurevich},\ and\
  \citenamefont {Schober}}]{soft_potential_model_1991}%
  \BibitemOpen
  \bibfield  {author} {\bibinfo {author} {\bibfnamefont {U.}~\bibnamefont
  {Buchenau}}, \bibinfo {author} {\bibfnamefont {Y.~M.}\ \bibnamefont
  {Galperin}}, \bibinfo {author} {\bibfnamefont {V.~L.}\ \bibnamefont
  {Gurevich}},\ and\ \bibinfo {author} {\bibfnamefont {H.~R.}\ \bibnamefont
  {Schober}},\ }\bibfield  {title} {\bibinfo {title} {Anharmonic potentials and
  vibrational localization in glasses},\ }\href
  {https://doi.org/10.1103/PhysRevB.43.5039} {\bibfield  {journal} {\bibinfo
  {journal} {Phys. Rev. B}\ }\textbf {\bibinfo {volume} {43}},\ \bibinfo
  {pages} {5039} (\bibinfo {year} {1991})}\BibitemShut {NoStop}%
\bibitem [{\citenamefont {Buchenau}\ \emph {et~al.}(1992)\citenamefont
  {Buchenau}, \citenamefont {Galperin}, \citenamefont {Gurevich}, \citenamefont
  {Parshin}, \citenamefont {Ramos},\ and\ \citenamefont
  {Schober}}]{buchenau_prb_1992}%
  \BibitemOpen
  \bibfield  {author} {\bibinfo {author} {\bibfnamefont {U.}~\bibnamefont
  {Buchenau}}, \bibinfo {author} {\bibfnamefont {Y.~M.}\ \bibnamefont
  {Galperin}}, \bibinfo {author} {\bibfnamefont {V.~L.}\ \bibnamefont
  {Gurevich}}, \bibinfo {author} {\bibfnamefont {D.~A.}\ \bibnamefont
  {Parshin}}, \bibinfo {author} {\bibfnamefont {M.~A.}\ \bibnamefont {Ramos}},\
  and\ \bibinfo {author} {\bibfnamefont {H.~R.}\ \bibnamefont {Schober}},\
  }\bibfield  {title} {\bibinfo {title} {Interaction of soft modes and sound
  waves in glasses},\ }\href {https://doi.org/10.1103/PhysRevB.46.2798}
  {\bibfield  {journal} {\bibinfo  {journal} {Phys. Rev. B}\ }\textbf {\bibinfo
  {volume} {46}},\ \bibinfo {pages} {2798} (\bibinfo {year}
  {1992})}\BibitemShut {NoStop}%
\bibitem [{\citenamefont {Buchenau}(1992)}]{buchenau_Phil_Mag_1992}%
  \BibitemOpen
  \bibfield  {author} {\bibinfo {author} {\bibfnamefont {U.}~\bibnamefont
  {Buchenau}},\ }\bibfield  {title} {\bibinfo {title} {Soft localized
  vibrations in glasses and undercooled liquids},\ }\href
  {https://doi.org/10.1080/13642819208217904} {\bibfield  {journal} {\bibinfo
  {journal} {Philos. Mag B}\ }\textbf {\bibinfo {volume} {65}},\ \bibinfo
  {pages} {303} (\bibinfo {year} {1992})}\BibitemShut {NoStop}%
\bibitem [{\citenamefont {Gurevich}\ \emph {et~al.}(1993)\citenamefont
  {Gurevich}, \citenamefont {Parshin}, \citenamefont {Pelous},\ and\
  \citenamefont {Schober}}]{soft_potential_model_prb_1993}%
  \BibitemOpen
  \bibfield  {author} {\bibinfo {author} {\bibfnamefont {V.~L.}\ \bibnamefont
  {Gurevich}}, \bibinfo {author} {\bibfnamefont {D.~A.}\ \bibnamefont
  {Parshin}}, \bibinfo {author} {\bibfnamefont {J.}~\bibnamefont {Pelous}},\
  and\ \bibinfo {author} {\bibfnamefont {H.~R.}\ \bibnamefont {Schober}},\
  }\bibfield  {title} {\bibinfo {title} {Theory of low-energy raman scattering
  in glasses},\ }\href {https://doi.org/10.1103/PhysRevB.48.16318} {\bibfield
  {journal} {\bibinfo  {journal} {Phys. Rev. B}\ }\textbf {\bibinfo {volume}
  {48}},\ \bibinfo {pages} {16318} (\bibinfo {year} {1993})}\BibitemShut
  {NoStop}%
\bibitem [{\citenamefont {Gurarie}\ and\ \citenamefont
  {Chalker}(2003)}]{Chalker2003}%
  \BibitemOpen
  \bibfield  {author} {\bibinfo {author} {\bibfnamefont {V.}~\bibnamefont
  {Gurarie}}\ and\ \bibinfo {author} {\bibfnamefont {J.~T.}\ \bibnamefont
  {Chalker}},\ }\bibfield  {title} {\bibinfo {title} {Bosonic excitations in
  random media},\ }\href {https://doi.org/10.1103/PhysRevB.68.134207}
  {\bibfield  {journal} {\bibinfo  {journal} {Phys. Rev. B}\ }\textbf {\bibinfo
  {volume} {68}},\ \bibinfo {pages} {134207} (\bibinfo {year}
  {2003})}\BibitemShut {NoStop}%
\bibitem [{\citenamefont {Grannan}\ \emph
  {et~al.}(1990{\natexlab{a}})\citenamefont {Grannan}, \citenamefont
  {Randeria},\ and\ \citenamefont {Sethna}}]{sethna_prb_1990_1}%
  \BibitemOpen
  \bibfield  {author} {\bibinfo {author} {\bibfnamefont {E.~R.}\ \bibnamefont
  {Grannan}}, \bibinfo {author} {\bibfnamefont {M.}~\bibnamefont {Randeria}},\
  and\ \bibinfo {author} {\bibfnamefont {J.~P.}\ \bibnamefont {Sethna}},\
  }\bibfield  {title} {\bibinfo {title} {Low-temperature properties of a model
  glass. I. Elastic dipole model},\ }\href
  {https://doi.org/10.1103/PhysRevB.41.7784} {\bibfield  {journal} {\bibinfo
  {journal} {Phys. Rev. B}\ }\textbf {\bibinfo {volume} {41}},\ \bibinfo
  {pages} {7784} (\bibinfo {year} {1990}{\natexlab{a}})}\BibitemShut {NoStop}%
\bibitem [{\citenamefont {Grannan}\ \emph
  {et~al.}(1990{\natexlab{b}})\citenamefont {Grannan}, \citenamefont
  {Randeria},\ and\ \citenamefont {Sethna}}]{sethna_prb_1990_2}%
  \BibitemOpen
  \bibfield  {author} {\bibinfo {author} {\bibfnamefont {E.~R.}\ \bibnamefont
  {Grannan}}, \bibinfo {author} {\bibfnamefont {M.}~\bibnamefont {Randeria}},\
  and\ \bibinfo {author} {\bibfnamefont {J.~P.}\ \bibnamefont {Sethna}},\
  }\bibfield  {title} {\bibinfo {title} {Low-temperature properties of a model
  glass. II. Specific heat and thermal transport},\ }\href
  {https://doi.org/10.1103/PhysRevB.41.7799} {\bibfield  {journal} {\bibinfo
  {journal} {Phys. Rev. B}\ }\textbf {\bibinfo {volume} {41}},\ \bibinfo
  {pages} {7799} (\bibinfo {year} {1990}{\natexlab{b}})}\BibitemShut {NoStop}%
\bibitem [{\citenamefont {K\"uhn}\ and\ \citenamefont
  {Horstmann}(1997)}]{Kuhn_Horstmann_prl_1997}%
  \BibitemOpen
  \bibfield  {author} {\bibinfo {author} {\bibfnamefont {R.}~\bibnamefont
  {K\"uhn}}\ and\ \bibinfo {author} {\bibfnamefont {U.}~\bibnamefont
  {Horstmann}},\ }\bibfield  {title} {\bibinfo {title} {Random matrix approach
  to glassy physics: Low temperatures and beyond},\ }\href
  {https://doi.org/10.1103/PhysRevLett.78.4067} {\bibfield  {journal} {\bibinfo
   {journal} {Phys. Rev. Lett.}\ }\textbf {\bibinfo {volume} {78}},\ \bibinfo
  {pages} {4067} (\bibinfo {year} {1997})}\BibitemShut {NoStop}%
\bibitem [{\citenamefont {Schirmacher}(2011)}]{schirmacher2011comments}%
  \BibitemOpen
  \bibfield  {author} {\bibinfo {author} {\bibfnamefont {W.}~\bibnamefont
  {Schirmacher}},\ }\bibfield  {title} {\bibinfo {title} {Some comments on
  fluctuating-elasticity and local oscillator models for anomalous vibrational
  excitations in glasses},\ }\href
  {https://doi.org/https://doi.org/10.1016/j.jnoncrysol.2010.07.052} {\bibfield
   {journal} {\bibinfo  {journal} {J. Non-Cryst. Solids}\ }\textbf {\bibinfo
  {volume} {357}},\ \bibinfo {pages} {518 } (\bibinfo {year}
  {2011})}\BibitemShut {NoStop}%
\bibitem [{\citenamefont {Das}\ \emph {et~al.}(2020)\citenamefont {Das},
  \citenamefont {Hentschel}, \citenamefont {Lerner},\ and\ \citenamefont
  {Procaccia}}]{itamar_gps_prb2020}%
  \BibitemOpen
  \bibfield  {author} {\bibinfo {author} {\bibfnamefont {P.}~\bibnamefont
  {Das}}, \bibinfo {author} {\bibfnamefont {H.~G.~E.}\ \bibnamefont
  {Hentschel}}, \bibinfo {author} {\bibfnamefont {E.}~\bibnamefont {Lerner}},\
  and\ \bibinfo {author} {\bibfnamefont {I.}~\bibnamefont {Procaccia}},\
  }\bibfield  {title} {\bibinfo {title} {Robustness of density of low-frequency
  states in amorphous solids},\ }\href
  {https://doi.org/10.1103/PhysRevB.102.014202} {\bibfield  {journal} {\bibinfo
   {journal} {Phys. Rev. B}\ }\textbf {\bibinfo {volume} {102}},\ \bibinfo
  {pages} {014202} (\bibinfo {year} {2020})}\BibitemShut {NoStop}%
\bibitem [{\citenamefont {Schober}\ and\ \citenamefont
  {Laird}(1991)}]{Schober_Laird_numerics_PRB}%
  \BibitemOpen
  \bibfield  {author} {\bibinfo {author} {\bibfnamefont {H.~R.}\ \bibnamefont
  {Schober}}\ and\ \bibinfo {author} {\bibfnamefont {B.~B.}\ \bibnamefont
  {Laird}},\ }\bibfield  {title} {\bibinfo {title} {Localized low-frequency
  vibrational modes in glasses},\ }\href
  {https://doi.org/10.1103/PhysRevB.44.6746} {\bibfield  {journal} {\bibinfo
  {journal} {Phys. Rev. B}\ }\textbf {\bibinfo {volume} {44}},\ \bibinfo
  {pages} {6746} (\bibinfo {year} {1991})}\BibitemShut {NoStop}%
\bibitem [{\citenamefont {Schober}\ and\ \citenamefont
  {Oligschleger}(1996)}]{Schober_Oligschleger_numerics_PRB}%
  \BibitemOpen
  \bibfield  {author} {\bibinfo {author} {\bibfnamefont {H.~R.}\ \bibnamefont
  {Schober}}\ and\ \bibinfo {author} {\bibfnamefont {C.}~\bibnamefont
  {Oligschleger}},\ }\bibfield  {title} {\bibinfo {title} {Low-frequency
  vibrations in a model glass},\ }\href
  {https://doi.org/10.1103/PhysRevB.53.11469} {\bibfield  {journal} {\bibinfo
  {journal} {Phys. Rev. B}\ }\textbf {\bibinfo {volume} {53}},\ \bibinfo
  {pages} {11469} (\bibinfo {year} {1996})}\BibitemShut {NoStop}%
\bibitem [{\citenamefont {Buchenau}\ \emph {et~al.}(1988)\citenamefont
  {Buchenau}, \citenamefont {Zhou}, \citenamefont {Nucker}, \citenamefont
  {Gilroy},\ and\ \citenamefont {Phillips}}]{Buchenau_1988}%
  \BibitemOpen
  \bibfield  {author} {\bibinfo {author} {\bibfnamefont {U.}~\bibnamefont
  {Buchenau}}, \bibinfo {author} {\bibfnamefont {H.~M.}\ \bibnamefont {Zhou}},
  \bibinfo {author} {\bibfnamefont {N.}~\bibnamefont {Nucker}}, \bibinfo
  {author} {\bibfnamefont {K.~S.}\ \bibnamefont {Gilroy}},\ and\ \bibinfo
  {author} {\bibfnamefont {W.~A.}\ \bibnamefont {Phillips}},\ }\bibfield
  {title} {\bibinfo {title} {Structural relaxation in vitreous silica},\ }\href
  {https://doi.org/10.1103/PhysRevLett.60.1318} {\bibfield  {journal} {\bibinfo
   {journal} {Phys. Rev. Lett.}\ }\textbf {\bibinfo {volume} {60}},\ \bibinfo
  {pages} {1318} (\bibinfo {year} {1988})}\BibitemShut {NoStop}%
\bibitem [{\citenamefont {Schober}\ \emph {et~al.}(1993)\citenamefont
  {Schober}, \citenamefont {Oligschleger},\ and\ \citenamefont
  {Laird}}]{schober1993_numerics}%
  \BibitemOpen
  \bibfield  {author} {\bibinfo {author} {\bibfnamefont {H.}~\bibnamefont
  {Schober}}, \bibinfo {author} {\bibfnamefont {C.}~\bibnamefont
  {Oligschleger}},\ and\ \bibinfo {author} {\bibfnamefont {B.~B.}~\bibnamefont
  {Laird}},\ }\bibfield  {title} {\bibinfo {title} {Low-frequency vibrations
  and relaxations in glasses},\ }\href
  {https://doi.org/https://doi.org/10.1016/0022-3093(93)90106-8} {\bibfield
  {journal} {\bibinfo  {journal} {J. Non-Cryst. Solids}\ }\textbf {\bibinfo
  {volume} {156}},\ \bibinfo {pages} {965 } (\bibinfo {year}
  {1993})}\BibitemShut {NoStop}%
\bibitem [{\citenamefont {Oligschleger}\ and\ \citenamefont
  {Schober}(1993)}]{Schober_Oligschleger_se_1993_numerics}%
  \BibitemOpen
  \bibfield  {author} {\bibinfo {author} {\bibfnamefont {C.}~\bibnamefont
  {Oligschleger}}\ and\ \bibinfo {author} {\bibfnamefont {H.}~\bibnamefont
  {Schober}},\ }\bibfield  {title} {\bibinfo {title} {Dynamics of Se glasses},\
  }\href {https://doi.org/https://doi.org/10.1016/0378-4371(93)90438-A}
  {\bibfield  {journal} {\bibinfo  {journal} {Phys. A: Stat. Mech. Appl.}\
  }\textbf {\bibinfo {volume} {201}},\ \bibinfo {pages} {391} (\bibinfo {year}
  {1993})}\BibitemShut {NoStop}%
\bibitem [{\citenamefont {Luchnikov}\ \emph {et~al.}(2000)\citenamefont
  {Luchnikov}, \citenamefont {Medvedev}, \citenamefont {Naberukhin},\ and\
  \citenamefont {Schober}}]{Schober_numerics_prb_2000}%
  \BibitemOpen
  \bibfield  {author} {\bibinfo {author} {\bibfnamefont {V.~A.}\ \bibnamefont
  {Luchnikov}}, \bibinfo {author} {\bibfnamefont {N.~N.}\ \bibnamefont
  {Medvedev}}, \bibinfo {author} {\bibfnamefont {Y.~I.}\ \bibnamefont
  {Naberukhin}},\ and\ \bibinfo {author} {\bibfnamefont {H.~R.}\ \bibnamefont
  {Schober}},\ }\bibfield  {title} {\bibinfo {title} {Voronoi-Delaunay analysis
  of normal modes in a simple model glass},\ }\href
  {https://doi.org/10.1103/PhysRevB.62.3181} {\bibfield  {journal} {\bibinfo
  {journal} {Phys. Rev. B}\ }\textbf {\bibinfo {volume} {62}},\ \bibinfo
  {pages} {3181} (\bibinfo {year} {2000})}\BibitemShut {NoStop}%
\bibitem [{\citenamefont {Schober}\ and\ \citenamefont
  {Ruocco}(2004)}]{schober_and_ruocco_2004}%
  \BibitemOpen
  \bibfield  {author} {\bibinfo {author} {\bibfnamefont {H.~R.}\ \bibnamefont
  {Schober}}\ and\ \bibinfo {author} {\bibfnamefont {G.}~\bibnamefont
  {Ruocco}},\ }\bibfield  {title} {\bibinfo {title} {Size effects and
  quasilocalized vibrations},\ }\href
  {https://doi.org/10.1080/14786430310001644107} {\bibfield  {journal}
  {\bibinfo  {journal} {Philos. Mag.}\ }\textbf {\bibinfo {volume} {84}},\
  \bibinfo {pages} {1361} (\bibinfo {year} {2004})}\BibitemShut {NoStop}%
\bibitem [{\citenamefont {Schober}(2004)}]{Schober_2004}%
  \BibitemOpen
  \bibfield  {author} {\bibinfo {author} {\bibfnamefont {H.~R.}\ \bibnamefont
  {Schober}},\ }\bibfield  {title} {\bibinfo {title} {Vibrations and
  relaxations in a soft sphere glass: boson peak and structure factors},\
  }\href {https://doi.org/10.1088/0953-8984/16/27/005} {\bibfield  {journal}
  {\bibinfo  {journal} {J. Condens. Matter Phys.}\ }\textbf {\bibinfo {volume}
  {16}},\ \bibinfo {pages} {S2659} (\bibinfo {year} {2004})}\BibitemShut
  {NoStop}%
\bibitem [{foo({\natexlab{a}})}]{footnote5}%
  \BibitemOpen
  \bibinfo {note} {Earlier numerical work on models of amorphous silicon had
  identified localized low-frequency vibrational modes emanating from
  coordination defects~\cite{Biswas_1988}, see also comments
  in~\cite{Schober_Laird_numerics_PRL}}\BibitemShut {NoStop}%
\bibitem [{\citenamefont {Klinger}\ and\ \citenamefont
  {Kosevich}(2002)}]{klinger2002_hybridizations}%
  \BibitemOpen
  \bibfield  {author} {\bibinfo {author} {\bibfnamefont {M.}~\bibnamefont
  {Klinger}}\ and\ \bibinfo {author} {\bibfnamefont {A.}~\bibnamefont
  {Kosevich}},\ }\bibfield  {title} {\bibinfo {title} {Soft-mode dynamics model
  of boson peak and high frequency sound in glasses: ``inelastic'' Ioffe-Regel
  crossover and strong hybridization of excitations},\ }\href
  {https://doi.org/https://doi.org/10.1016/S0375-9601(02)00167-6} {\bibfield
  {journal} {\bibinfo  {journal} {Phys. Lett. A}\ }\textbf {\bibinfo {volume}
  {295}},\ \bibinfo {pages} {311} (\bibinfo {year} {2002})}\BibitemShut
  {NoStop}%
\bibitem [{\citenamefont {Baity-Jesi}\ \emph {et~al.}(2015)\citenamefont
  {Baity-Jesi}, \citenamefont {Mart\'{\i}n-Mayor}, \citenamefont {Parisi},\
  and\ \citenamefont {Perez-Gaviro}}]{parisi_spin_glass}%
  \BibitemOpen
  \bibfield  {author} {\bibinfo {author} {\bibfnamefont {M.}~\bibnamefont
  {Baity-Jesi}}, \bibinfo {author} {\bibfnamefont {V.}~\bibnamefont
  {Mart\'{\i}n-Mayor}}, \bibinfo {author} {\bibfnamefont {G.}~\bibnamefont
  {Parisi}},\ and\ \bibinfo {author} {\bibfnamefont {S.}~\bibnamefont
  {Perez-Gaviro}},\ }\bibfield  {title} {\bibinfo {title} {Soft modes,
  localization, and two-level systems in spin glasses},\ }\href
  {https://doi.org/10.1103/PhysRevLett.115.267205} {\bibfield  {journal}
  {\bibinfo  {journal} {Phys. Rev. Lett.}\ }\textbf {\bibinfo {volume} {115}},\
  \bibinfo {pages} {267205} (\bibinfo {year} {2015})}\BibitemShut {NoStop}%
\bibitem [{\citenamefont {Gartner}\ and\ \citenamefont
  {Lerner}(2016{\natexlab{a}})}]{SciPost2016}%
  \BibitemOpen
  \bibfield  {author} {\bibinfo {author} {\bibfnamefont {L.}~\bibnamefont
  {Gartner}}\ and\ \bibinfo {author} {\bibfnamefont {E.}~\bibnamefont
  {Lerner}},\ }\bibfield  {title} {\bibinfo {title} {{Nonlinear modes
  disentangle glassy and Goldstone modes in structural glasses}},\ }\href
  {https://doi.org/10.21468/SciPostPhys.1.2.016} {\bibfield  {journal}
  {\bibinfo  {journal} {SciPost Phys.}\ }\textbf {\bibinfo {volume} {1}},\
  \bibinfo {pages} {016} (\bibinfo {year} {2016}{\natexlab{a}})}\BibitemShut
  {NoStop}%
\bibitem [{\citenamefont {Wijtmans}\ and\ \citenamefont
  {Manning}(2017)}]{manning_defects}%
  \BibitemOpen
  \bibfield  {author} {\bibinfo {author} {\bibfnamefont {S.}~\bibnamefont
  {Wijtmans}}\ and\ \bibinfo {author} {\bibfnamefont {M.~L.}\ \bibnamefont
  {Manning}},\ }\bibfield  {title} {\bibinfo {title} {Disentangling defects and
  sound modes in disordered solids},\ }\href
  {https://doi.org/10.1039/C7SM00792B} {\bibfield  {journal} {\bibinfo
  {journal} {Soft Matter}\ }\textbf {\bibinfo {volume} {13}},\ \bibinfo {pages}
  {5649} (\bibinfo {year} {2017})}\BibitemShut {NoStop}%
\bibitem [{\citenamefont {Angelani}\ \emph {et~al.}(2018)\citenamefont
  {Angelani}, \citenamefont {Paoluzzi}, \citenamefont {Parisi},\ and\
  \citenamefont {Ruocco}}]{paoluzzi_pnas_2018_pinning}%
  \BibitemOpen
  \bibfield  {author} {\bibinfo {author} {\bibfnamefont {L.}~\bibnamefont
  {Angelani}}, \bibinfo {author} {\bibfnamefont {M.}~\bibnamefont {Paoluzzi}},
  \bibinfo {author} {\bibfnamefont {G.}~\bibnamefont {Parisi}},\ and\ \bibinfo
  {author} {\bibfnamefont {G.}~\bibnamefont {Ruocco}},\ }\bibfield  {title}
  {\bibinfo {title} {Probing the non-Debye low-frequency excitations in glasses
  through random pinning},\ }\href {https://doi.org/10.1073/pnas.1805024115}
  {\bibfield  {journal} {\bibinfo  {journal} {Proc. Natl. Acad. Sci. U.S.A.}\
  }\textbf {\bibinfo {volume} {115}},\ \bibinfo {pages} {8700} (\bibinfo {year}
  {2018})}\BibitemShut {NoStop}%
\bibitem [{\citenamefont {Kapteijns}\ \emph {et~al.}(2020)\citenamefont
  {Kapteijns}, \citenamefont {Richard},\ and\ \citenamefont
  {Lerner}}]{episode_1_2020}%
  \BibitemOpen
  \bibfield  {author} {\bibinfo {author} {\bibfnamefont {G.}~\bibnamefont
  {Kapteijns}}, \bibinfo {author} {\bibfnamefont {D.}~\bibnamefont {Richard}},\
  and\ \bibinfo {author} {\bibfnamefont {E.}~\bibnamefont {Lerner}},\
  }\bibfield  {title} {\bibinfo {title} {Nonlinear quasilocalized excitations
  in glasses: True representatives of soft spots},\ }\href
  {https://doi.org/10.1103/PhysRevE.101.032130} {\bibfield  {journal} {\bibinfo
   {journal} {Phys. Rev. E}\ }\textbf {\bibinfo {volume} {101}},\ \bibinfo
  {pages} {032130} (\bibinfo {year} {2020})}\BibitemShut {NoStop}%
\bibitem [{\citenamefont {Giannini}\ \emph {et~al.}(2021)\citenamefont
  {Giannini}, \citenamefont {Richard}, \citenamefont {Manning},\ and\
  \citenamefont {Lerner}}]{julia_arXiv}%
  \BibitemOpen
  \bibfield  {author} {\bibinfo {author} {\bibfnamefont {J.~A.}\ \bibnamefont
  {Giannini}}, \bibinfo {author} {\bibfnamefont {D.}~\bibnamefont {Richard}},
  \bibinfo {author} {\bibfnamefont {M.~L.}\ \bibnamefont {Manning}},\ and\
  \bibinfo {author} {\bibfnamefont {E.}~\bibnamefont {Lerner}},\ }\bibfield
  {title} {\bibinfo {title} {Bond-space operator disentangles quasi-localized
  and phononic modes in structural glasses},\ }\href
  {https://arxiv.org/abs/2106.16231} {\bibfield  {journal} {\bibinfo  {journal}
  {arXiv preprint arXiv:2106.16231}\ } (\bibinfo {year} {2021})}\BibitemShut
  {NoStop}%
\bibitem [{\citenamefont {Kapteijns}\ \emph
  {et~al.}(2021{\natexlab{a}})\citenamefont {Kapteijns}, \citenamefont
  {Bouchbinder},\ and\ \citenamefont {Lerner}}]{phonon_widths2}%
  \BibitemOpen
  \bibfield  {author} {\bibinfo {author} {\bibfnamefont {G.}~\bibnamefont
  {Kapteijns}}, \bibinfo {author} {\bibfnamefont {E.}~\bibnamefont
  {Bouchbinder}},\ and\ \bibinfo {author} {\bibfnamefont {E.}~\bibnamefont
  {Lerner}},\ }\bibfield  {title} {\bibinfo {title} {A unified quantifier of
  mechanical disorder in solids},\ }\href {https://arxiv.org/abs/2106.12613}
  {\bibfield  {journal} {\bibinfo  {journal} {arXiv preprint arXiv:2106.12613}\
  } (\bibinfo {year} {2021}{\natexlab{a}})}\BibitemShut {NoStop}%
\bibitem [{\citenamefont {Bouchbinder}\ and\ \citenamefont
  {Lerner}(2018)}]{phonon_widths}%
  \BibitemOpen
  \bibfield  {author} {\bibinfo {author} {\bibfnamefont {E.}~\bibnamefont
  {Bouchbinder}}\ and\ \bibinfo {author} {\bibfnamefont {E.}~\bibnamefont
  {Lerner}},\ }\bibfield  {title} {\bibinfo {title} {Universal disorder-induced
  broadening of phonon bands: from disordered lattices to glasses},\ }\href
  {http://stacks.iop.org/1367-2630/20/i=7/a=073022} {\bibfield  {journal}
  {\bibinfo  {journal} {New J. Phys.}\ }\textbf {\bibinfo {volume} {20}},\
  \bibinfo {pages} {073022} (\bibinfo {year} {2018})}\BibitemShut {NoStop}%
\bibitem [{\citenamefont {Ninarello}\ \emph {et~al.}(2017)\citenamefont
  {Ninarello}, \citenamefont {Berthier},\ and\ \citenamefont
  {Coslovich}}]{LB_swap_prx}%
  \BibitemOpen
  \bibfield  {author} {\bibinfo {author} {\bibfnamefont {A.}~\bibnamefont
  {Ninarello}}, \bibinfo {author} {\bibfnamefont {L.}~\bibnamefont
  {Berthier}},\ and\ \bibinfo {author} {\bibfnamefont {D.}~\bibnamefont
  {Coslovich}},\ }\bibfield  {title} {\bibinfo {title} {Models and algorithms
  for the next generation of glass transition studies},\ }\href
  {https://doi.org/10.1103/PhysRevX.7.021039} {\bibfield  {journal} {\bibinfo
  {journal} {Phys. Rev. X}\ }\textbf {\bibinfo {volume} {7}},\ \bibinfo {pages}
  {021039} (\bibinfo {year} {2017})}\BibitemShut {NoStop}%
\bibitem [{\citenamefont {Wang}\ \emph
  {et~al.}(2021{\natexlab{a}})\citenamefont {Wang}, \citenamefont {Szamel},\
  and\ \citenamefont {Flenner}}]{grzegorz_arXiv_2021_2D_modes}%
  \BibitemOpen
  \bibfield  {author} {\bibinfo {author} {\bibfnamefont {L.}~\bibnamefont
  {Wang}}, \bibinfo {author} {\bibfnamefont {G.}~\bibnamefont {Szamel}},\ and\
  \bibinfo {author} {\bibfnamefont {E.}~\bibnamefont {Flenner}},\ }\bibfield
  {title} {\bibinfo {title} {Low-frequency excess vibrational modes in
  two-dimensional glasses},\ }\href {https://arxiv.org/abs/2107.01505}
  {\bibfield  {journal} {\bibinfo  {journal} {arXiv preprint arXiv:2107.01505}\
  } (\bibinfo {year} {2021}{\natexlab{a}})}\BibitemShut {NoStop}%
\bibitem [{\citenamefont {O'Hern}\ \emph {et~al.}(2003)\citenamefont {O'Hern},
  \citenamefont {Silbert}, \citenamefont {Liu},\ and\ \citenamefont
  {Nagel}}]{ohern2003}%
  \BibitemOpen
  \bibfield  {author} {\bibinfo {author} {\bibfnamefont {C.~S.}\ \bibnamefont
  {O'Hern}}, \bibinfo {author} {\bibfnamefont {L.~E.}\ \bibnamefont {Silbert}},
  \bibinfo {author} {\bibfnamefont {A.~J.}\ \bibnamefont {Liu}},\ and\ \bibinfo
  {author} {\bibfnamefont {S.~R.}\ \bibnamefont {Nagel}},\ }\bibfield  {title}
  {\bibinfo {title} {Jamming at zero temperature and zero applied stress: The
  epitome of disorder},\ }\href {https://doi.org/10.1103/PhysRevE.68.011306}
  {\bibfield  {journal} {\bibinfo  {journal} {Phys. Rev. E}\ }\textbf {\bibinfo
  {volume} {68}},\ \bibinfo {pages} {011306} (\bibinfo {year}
  {2003})}\BibitemShut {NoStop}%
\bibitem [{\citenamefont {Silbert}\ \emph {et~al.}(2005)\citenamefont
  {Silbert}, \citenamefont {Liu},\ and\ \citenamefont
  {Nagel}}]{Silbert_prl_2005}%
  \BibitemOpen
  \bibfield  {author} {\bibinfo {author} {\bibfnamefont {L.~E.}\ \bibnamefont
  {Silbert}}, \bibinfo {author} {\bibfnamefont {A.~J.}\ \bibnamefont {Liu}},\
  and\ \bibinfo {author} {\bibfnamefont {S.~R.}\ \bibnamefont {Nagel}},\
  }\bibfield  {title} {\bibinfo {title} {Vibrations and diverging length scales
  near the unjamming transition},\ }\href
  {https://doi.org/10.1103/PhysRevLett.95.098301} {\bibfield  {journal}
  {\bibinfo  {journal} {Phys. Rev. Lett.}\ }\textbf {\bibinfo {volume} {95}},\
  \bibinfo {pages} {098301} (\bibinfo {year} {2005})}\BibitemShut {NoStop}%
\bibitem [{\citenamefont {Monaco}\ and\ \citenamefont
  {Giordano}(2009)}]{Monaco_pnas_2009}%
  \BibitemOpen
  \bibfield  {author} {\bibinfo {author} {\bibfnamefont {G.}~\bibnamefont
  {Monaco}}\ and\ \bibinfo {author} {\bibfnamefont {V.~M.}\ \bibnamefont
  {Giordano}},\ }\bibfield  {title} {\bibinfo {title} {Breakdown of the debye
  approximation for the acoustic modes with nanometric wavelengths in
  glasses},\ }\href {https://doi.org/10.1073/pnas.0808965106} {\bibfield
  {journal} {\bibinfo  {journal} {Proc. Natl. Acad. Sci. U.S.A.}\ }\textbf
  {\bibinfo {volume} {106}},\ \bibinfo {pages} {3659} (\bibinfo {year}
  {2009})}\BibitemShut {NoStop}%
\bibitem [{\citenamefont {Chumakov}\ \emph
  {et~al.}(2011{\natexlab{b}})\citenamefont {Chumakov}, \citenamefont {Monaco},
  \citenamefont {Monaco}, \citenamefont {Crichton}, \citenamefont {Bosak},
  \citenamefont {R\"uffer}, \citenamefont {Meyer}, \citenamefont {Kargl},
  \citenamefont {Comez}, \citenamefont {Fioretto}, \citenamefont {Giefers},
  \citenamefont {Roitsch}, \citenamefont {Wortmann}, \citenamefont {Manghnani},
  \citenamefont {Hushur}, \citenamefont {Williams}, \citenamefont {Balogh},
  \citenamefont {Parli\ifmmode~\acute{n}\else \'{n}\fi{}ski}, \citenamefont
  {Jochym},\ and\ \citenamefont {Piekarz}}]{Monaco_prl_2011}%
  \BibitemOpen
  \bibfield  {author} {\bibinfo {author} {\bibfnamefont {A.~I.}\ \bibnamefont
  {Chumakov}}, \bibinfo {author} {\bibfnamefont {G.}~\bibnamefont {Monaco}},
  \bibinfo {author} {\bibfnamefont {A.}~\bibnamefont {Monaco}}, \bibinfo
  {author} {\bibfnamefont {W.~A.}\ \bibnamefont {Crichton}}, \bibinfo {author}
  {\bibfnamefont {A.}~\bibnamefont {Bosak}}, \bibinfo {author} {\bibfnamefont
  {R.}~\bibnamefont {R\"uffer}}, \bibinfo {author} {\bibfnamefont
  {A.}~\bibnamefont {Meyer}}, \bibinfo {author} {\bibfnamefont
  {F.}~\bibnamefont {Kargl}}, \bibinfo {author} {\bibfnamefont
  {L.}~\bibnamefont {Comez}}, \bibinfo {author} {\bibfnamefont
  {D.}~\bibnamefont {Fioretto}}, \bibinfo {author} {\bibfnamefont
  {H.}~\bibnamefont {Giefers}}, \bibinfo {author} {\bibfnamefont
  {S.}~\bibnamefont {Roitsch}}, \bibinfo {author} {\bibfnamefont
  {G.}~\bibnamefont {Wortmann}}, \bibinfo {author} {\bibfnamefont {M.~H.}\
  \bibnamefont {Manghnani}}, \bibinfo {author} {\bibfnamefont {A.}~\bibnamefont
  {Hushur}}, \bibinfo {author} {\bibfnamefont {Q.}~\bibnamefont {Williams}},
  \bibinfo {author} {\bibfnamefont {J.}~\bibnamefont {Balogh}}, \bibinfo
  {author} {\bibfnamefont {K.}~\bibnamefont {Parli\ifmmode~\acute{n}\else
  \'{n}\fi{}ski}}, \bibinfo {author} {\bibfnamefont {P.}~\bibnamefont
  {Jochym}},\ and\ \bibinfo {author} {\bibfnamefont {P.}~\bibnamefont
  {Piekarz}},\ }\bibfield  {title} {\bibinfo {title} {Equivalence of the boson
  peak in glasses to the transverse acoustic van hove singularity in
  crystals},\ }\href {https://doi.org/10.1103/PhysRevLett.106.225501}
  {\bibfield  {journal} {\bibinfo  {journal} {Phys. Rev. Lett.}\ }\textbf
  {\bibinfo {volume} {106}},\ \bibinfo {pages} {225501} (\bibinfo {year}
  {2011}{\natexlab{b}})}\BibitemShut {NoStop}%
\bibitem [{\citenamefont {Leonforte}\ \emph {et~al.}(2005)\citenamefont
  {Leonforte}, \citenamefont {Boissi\`ere}, \citenamefont {Tanguy},
  \citenamefont {Wittmer},\ and\ \citenamefont {Barrat}}]{barrat_3d}%
  \BibitemOpen
  \bibfield  {author} {\bibinfo {author} {\bibfnamefont {F.}~\bibnamefont
  {Leonforte}}, \bibinfo {author} {\bibfnamefont {R.}~\bibnamefont
  {Boissi\`ere}}, \bibinfo {author} {\bibfnamefont {A.}~\bibnamefont {Tanguy}},
  \bibinfo {author} {\bibfnamefont {J.~P.}\ \bibnamefont {Wittmer}},\ and\
  \bibinfo {author} {\bibfnamefont {J.-L.}\ \bibnamefont {Barrat}},\ }\bibfield
   {title} {\bibinfo {title} {Continuum limit of amorphous elastic bodies. III.
  Three-dimensional systems},\ }\href
  {https://doi.org/10.1103/PhysRevB.72.224206} {\bibfield  {journal} {\bibinfo
  {journal} {Phys. Rev. B}\ }\textbf {\bibinfo {volume} {72}},\ \bibinfo
  {pages} {224206} (\bibinfo {year} {2005})}\BibitemShut {NoStop}%
\bibitem [{\citenamefont {Tanguy}(2015)}]{Tanguy2015}%
  \BibitemOpen
  \bibfield  {author} {\bibinfo {author} {\bibfnamefont {A.}~\bibnamefont
  {Tanguy}},\ }\bibfield  {title} {\bibinfo {title} {Vibration modes and
  characteristic length scales in amorphous materials},\ }\href
  {https://doi.org/10.1007/s11837-015-1480-y} {\bibfield  {journal} {\bibinfo
  {journal} {JOM}\ }\textbf {\bibinfo {volume} {67}},\ \bibinfo {pages} {1832}
  (\bibinfo {year} {2015})}\BibitemShut {NoStop}%
\bibitem [{\citenamefont {DeGiuli}\ \emph
  {et~al.}(2014{\natexlab{b}})\citenamefont {DeGiuli}, \citenamefont {Lerner},
  \citenamefont {Brito},\ and\ \citenamefont
  {Wyart}}]{eric_hard_sphere_vdos_pnas2014}%
  \BibitemOpen
  \bibfield  {author} {\bibinfo {author} {\bibfnamefont {E.}~\bibnamefont
  {DeGiuli}}, \bibinfo {author} {\bibfnamefont {E.}~\bibnamefont {Lerner}},
  \bibinfo {author} {\bibfnamefont {C.}~\bibnamefont {Brito}},\ and\ \bibinfo
  {author} {\bibfnamefont {M.}~\bibnamefont {Wyart}},\ }\bibfield  {title}
  {\bibinfo {title} {Force distribution affects vibrational properties in
  hard-sphere glasses},\ }\href {https://doi.org/10.1073/pnas.1415298111}
  {\bibfield  {journal} {\bibinfo  {journal} {Proc. Natl. Acad. Sci. U.S.A.}\
  }\textbf {\bibinfo {volume} {111}},\ \bibinfo {pages} {17054} (\bibinfo
  {year} {2014}{\natexlab{b}})}\BibitemShut {NoStop}%
\bibitem [{\citenamefont {Franz}\ \emph {et~al.}(2015)\citenamefont {Franz},
  \citenamefont {Parisi}, \citenamefont {Urbani},\ and\ \citenamefont
  {Zamponi}}]{silvio}%
  \BibitemOpen
  \bibfield  {author} {\bibinfo {author} {\bibfnamefont {S.}~\bibnamefont
  {Franz}}, \bibinfo {author} {\bibfnamefont {G.}~\bibnamefont {Parisi}},
  \bibinfo {author} {\bibfnamefont {P.}~\bibnamefont {Urbani}},\ and\ \bibinfo
  {author} {\bibfnamefont {F.}~\bibnamefont {Zamponi}},\ }\bibfield  {title}
  {\bibinfo {title} {Universal spectrum of normal modes in low-temperature
  glasses},\ }\href {https://doi.org/10.1073/pnas.1511134112} {\bibfield
  {journal} {\bibinfo  {journal} {Proc. Natl. Acad. Sci. U.S.A.}\ }\textbf
  {\bibinfo {volume} {112}},\ \bibinfo {pages} {14539} (\bibinfo {year}
  {2015})}\BibitemShut {NoStop}%
\bibitem [{\citenamefont {Karmakar}\ \emph {et~al.}(2012)\citenamefont
  {Karmakar}, \citenamefont {Lerner},\ and\ \citenamefont
  {Procaccia}}]{karmakar_lengthscale}%
  \BibitemOpen
  \bibfield  {author} {\bibinfo {author} {\bibfnamefont {S.}~\bibnamefont
  {Karmakar}}, \bibinfo {author} {\bibfnamefont {E.}~\bibnamefont {Lerner}},\
  and\ \bibinfo {author} {\bibfnamefont {I.}~\bibnamefont {Procaccia}},\
  }\bibfield  {title} {\bibinfo {title} {Direct estimate of the static
  length-scale accompanying the glass transition},\ }\href
  {https://doi.org/http://dx.doi.org/10.1016/j.physa.2011.11.020} {\bibfield
  {journal} {\bibinfo  {journal} {Physica A}\ }\textbf {\bibinfo {volume}
  {391}},\ \bibinfo {pages} {1001 } (\bibinfo {year} {2012})}\BibitemShut
  {NoStop}%
\bibitem [{\citenamefont {Biswas}\ \emph {et~al.}(1988)\citenamefont {Biswas},
  \citenamefont {Bouchard}, \citenamefont {Kamitakahara}, \citenamefont
  {Grest},\ and\ \citenamefont {Soukoulis}}]{Biswas_1988}%
  \BibitemOpen
  \bibfield  {author} {\bibinfo {author} {\bibfnamefont {R.}~\bibnamefont
  {Biswas}}, \bibinfo {author} {\bibfnamefont {A.~M.}\ \bibnamefont
  {Bouchard}}, \bibinfo {author} {\bibfnamefont {W.~A.}\ \bibnamefont
  {Kamitakahara}}, \bibinfo {author} {\bibfnamefont {G.~S.}\ \bibnamefont
  {Grest}},\ and\ \bibinfo {author} {\bibfnamefont {C.~M.}\ \bibnamefont
  {Soukoulis}},\ }\bibfield  {title} {\bibinfo {title} {Vibrational
  localization in amorphous silicon},\ }\href
  {https://doi.org/10.1103/PhysRevLett.60.2280} {\bibfield  {journal} {\bibinfo
   {journal} {Phys. Rev. Lett.}\ }\textbf {\bibinfo {volume} {60}},\ \bibinfo
  {pages} {2280} (\bibinfo {year} {1988})}\BibitemShut {NoStop}%
\bibitem [{\citenamefont {Lerner}(2016)}]{micromechanics2016}%
  \BibitemOpen
  \bibfield  {author} {\bibinfo {author} {\bibfnamefont {E.}~\bibnamefont
  {Lerner}},\ }\bibfield  {title} {\bibinfo {title} {Micromechanics of
  nonlinear plastic modes},\ }\href
  {https://doi.org/10.1103/PhysRevE.93.053004} {\bibfield  {journal} {\bibinfo
  {journal} {Phys. Rev. E}\ }\textbf {\bibinfo {volume} {93}},\ \bibinfo
  {pages} {053004} (\bibinfo {year} {2016})}\BibitemShut {NoStop}%
\bibitem [{\citenamefont {Shimada}\ \emph
  {et~al.}(2020{\natexlab{a}})\citenamefont {Shimada}, \citenamefont {Mizuno},
  \citenamefont {Berthier},\ and\ \citenamefont
  {Ikeda}}]{Atsushi_high_d_pre_2020}%
  \BibitemOpen
  \bibfield  {author} {\bibinfo {author} {\bibfnamefont {M.}~\bibnamefont
  {Shimada}}, \bibinfo {author} {\bibfnamefont {H.}~\bibnamefont {Mizuno}},
  \bibinfo {author} {\bibfnamefont {L.}~\bibnamefont {Berthier}},\ and\
  \bibinfo {author} {\bibfnamefont {A.}~\bibnamefont {Ikeda}},\ }\bibfield
  {title} {\bibinfo {title} {Low-frequency vibrations of jammed packings in
  large spatial dimensions},\ }\href
  {https://doi.org/10.1103/PhysRevE.101.052906} {\bibfield  {journal} {\bibinfo
   {journal} {Phys. Rev. E}\ }\textbf {\bibinfo {volume} {101}},\ \bibinfo
  {pages} {052906} (\bibinfo {year} {2020}{\natexlab{a}})}\BibitemShut
  {NoStop}%
\bibitem [{\citenamefont {Tsai}\ \emph {et~al.}(1978)\citenamefont {Tsai},
  \citenamefont {Abraham},\ and\ \citenamefont {Pound}}]{tsai_swap}%
  \BibitemOpen
  \bibfield  {author} {\bibinfo {author} {\bibfnamefont {N.-H.}\ \bibnamefont
  {Tsai}}, \bibinfo {author} {\bibfnamefont {F.~F.}\ \bibnamefont {Abraham}},\
  and\ \bibinfo {author} {\bibfnamefont {G.}~\bibnamefont {Pound}},\ }\bibfield
   {title} {\bibinfo {title} {The structure and thermodynamics of binary
  microclusters: A monte carlo simulation},\ }\href
  {https://doi.org/https://doi.org/10.1016/0039-6028(78)90134-6} {\bibfield
  {journal} {\bibinfo  {journal} {Surf. Sci.}\ }\textbf {\bibinfo {volume}
  {77}},\ \bibinfo {pages} {465 } (\bibinfo {year} {1978})}\BibitemShut
  {NoStop}%
\bibitem [{\citenamefont {Gazzillo}\ and\ \citenamefont
  {Pastore}(1989)}]{gazzillo_swap}%
  \BibitemOpen
  \bibfield  {author} {\bibinfo {author} {\bibfnamefont {D.}~\bibnamefont
  {Gazzillo}}\ and\ \bibinfo {author} {\bibfnamefont {G.}~\bibnamefont
  {Pastore}},\ }\bibfield  {title} {\bibinfo {title} {Equation of state for
  symmetric non-additive hard-sphere fluids: An approximate analytic expression
  and new Monte Carlo results},\ }\href
  {https://doi.org/https://doi.org/10.1016/0009-2614(89)87505-0} {\bibfield
  {journal} {\bibinfo  {journal} {Chem. Phys. Lett.}\ }\textbf {\bibinfo
  {volume} {159}},\ \bibinfo {pages} {388 } (\bibinfo {year}
  {1989})}\BibitemShut {NoStop}%
\bibitem [{\citenamefont {Grigera}\ and\ \citenamefont
  {Parisi}(2001)}]{grigera_swap}%
  \BibitemOpen
  \bibfield  {author} {\bibinfo {author} {\bibfnamefont {T.~S.}\ \bibnamefont
  {Grigera}}\ and\ \bibinfo {author} {\bibfnamefont {G.}~\bibnamefont
  {Parisi}},\ }\bibfield  {title} {\bibinfo {title} {Fast Monte Carlo algorithm
  for supercooled soft spheres},\ }\href
  {https://doi.org/10.1103/PhysRevE.63.045102} {\bibfield  {journal} {\bibinfo
  {journal} {Phys. Rev. E}\ }\textbf {\bibinfo {volume} {63}},\ \bibinfo
  {pages} {045102} (\bibinfo {year} {2001})}\BibitemShut {NoStop}%
\bibitem [{\citenamefont {Bonfanti}\ \emph {et~al.}(2020)\citenamefont
  {Bonfanti}, \citenamefont {Guerra}, \citenamefont {Mondal}, \citenamefont
  {Procaccia},\ and\ \citenamefont {Zapperi}}]{universal_VDoS_ip}%
  \BibitemOpen
  \bibfield  {author} {\bibinfo {author} {\bibfnamefont {S.}~\bibnamefont
  {Bonfanti}}, \bibinfo {author} {\bibfnamefont {R.}~\bibnamefont {Guerra}},
  \bibinfo {author} {\bibfnamefont {C.}~\bibnamefont {Mondal}}, \bibinfo
  {author} {\bibfnamefont {I.}~\bibnamefont {Procaccia}},\ and\ \bibinfo
  {author} {\bibfnamefont {S.}~\bibnamefont {Zapperi}},\ }\bibfield  {title}
  {\bibinfo {title} {Universal low-frequency vibrational modes in silica
  glasses},\ }\href {https://doi.org/10.1103/PhysRevLett.125.085501} {\bibfield
   {journal} {\bibinfo  {journal} {Phys. Rev. Lett.}\ }\textbf {\bibinfo
  {volume} {125}},\ \bibinfo {pages} {085501} (\bibinfo {year}
  {2020})}\BibitemShut {NoStop}%
\bibitem [{\citenamefont {Shimada}\ \emph {et~al.}(2018)\citenamefont
  {Shimada}, \citenamefont {Mizuno}, \citenamefont {Wyart},\ and\ \citenamefont
  {Ikeda}}]{atsushi_core_size_pre}%
  \BibitemOpen
  \bibfield  {author} {\bibinfo {author} {\bibfnamefont {M.}~\bibnamefont
  {Shimada}}, \bibinfo {author} {\bibfnamefont {H.}~\bibnamefont {Mizuno}},
  \bibinfo {author} {\bibfnamefont {M.}~\bibnamefont {Wyart}},\ and\ \bibinfo
  {author} {\bibfnamefont {A.}~\bibnamefont {Ikeda}},\ }\bibfield  {title}
  {\bibinfo {title} {Spatial structure of quasilocalized vibrations in nearly
  jammed amorphous solids},\ }\href
  {https://doi.org/10.1103/PhysRevE.98.060901} {\bibfield  {journal} {\bibinfo
  {journal} {Phys. Rev. E}\ }\textbf {\bibinfo {volume} {98}},\ \bibinfo
  {pages} {060901} (\bibinfo {year} {2018})}\BibitemShut {NoStop}%
\bibitem [{\citenamefont {Richard}\ \emph
  {et~al.}(2021{\natexlab{b}})\citenamefont {Richard}, \citenamefont {Lerner},\
  and\ \citenamefont {Bouchbinder}}]{david_fracture_2021}%
  \BibitemOpen
  \bibfield  {author} {\bibinfo {author} {\bibfnamefont {D.}~\bibnamefont
  {Richard}}, \bibinfo {author} {\bibfnamefont {E.}~\bibnamefont {Lerner}},\
  and\ \bibinfo {author} {\bibfnamefont {E.}~\bibnamefont {Bouchbinder}},\
  }\bibfield  {title} {\bibinfo {title} {Brittle to ductile transitions in
  glasses: Roles of soft defects and loading geometry},\ }\href
  {https://arxiv.org/abs/2103.05258} {\bibfield  {journal} {\bibinfo  {journal}
  {arXiv preprint arXiv:2103.05258}\ } (\bibinfo {year}
  {2021}{\natexlab{b}})}\BibitemShut {NoStop}%
\bibitem [{\citenamefont {Rainone}\ \emph
  {et~al.}(2020{\natexlab{b}})\citenamefont {Rainone}, \citenamefont
  {Bouchbinder},\ and\ \citenamefont {Lerner}}]{cge2_jcp2020}%
  \BibitemOpen
  \bibfield  {author} {\bibinfo {author} {\bibfnamefont {C.}~\bibnamefont
  {Rainone}}, \bibinfo {author} {\bibfnamefont {E.}~\bibnamefont
  {Bouchbinder}},\ and\ \bibinfo {author} {\bibfnamefont {E.}~\bibnamefont
  {Lerner}},\ }\bibfield  {title} {\bibinfo {title} {Statistical mechanics of
  local force dipole responses in computer glasses},\ }\href
  {https://doi.org/10.1063/5.0005655} {\bibfield  {journal} {\bibinfo
  {journal} {J. Chem. Phys.}\ }\textbf {\bibinfo {volume} {152}},\ \bibinfo
  {pages} {194503} (\bibinfo {year} {2020}{\natexlab{b}})}\BibitemShut
  {NoStop}%
\bibitem [{\citenamefont {Sun}\ \emph {et~al.}(2016)\citenamefont {Sun},
  \citenamefont {Hu}, \citenamefont {Wang}, \citenamefont {Zhu}, \citenamefont
  {Wen}, \citenamefont {Wang}, \citenamefont {Liu},\ and\ \citenamefont
  {Yang}}]{experimental_inannealability_AM_2016}%
  \BibitemOpen
  \bibfield  {author} {\bibinfo {author} {\bibfnamefont {B.}~\bibnamefont
  {Sun}}, \bibinfo {author} {\bibfnamefont {Y.}~\bibnamefont {Hu}}, \bibinfo
  {author} {\bibfnamefont {D.}~\bibnamefont {Wang}}, \bibinfo {author}
  {\bibfnamefont {Z.}~\bibnamefont {Zhu}}, \bibinfo {author} {\bibfnamefont
  {P.}~\bibnamefont {Wen}}, \bibinfo {author} {\bibfnamefont {W.}~\bibnamefont
  {Wang}}, \bibinfo {author} {\bibfnamefont {C.}~\bibnamefont {Liu}},\ and\
  \bibinfo {author} {\bibfnamefont {Y.}~\bibnamefont {Yang}},\ }\bibfield
  {title} {\bibinfo {title} {Correlation between local elastic heterogeneities
  and overall elastic properties in metallic glasses},\ }\href
  {https://doi.org/https://doi.org/10.1016/j.actamat.2016.09.014} {\bibfield
  {journal} {\bibinfo  {journal} {Acta Mater.}\ }\textbf {\bibinfo {volume}
  {121}},\ \bibinfo {pages} {266 } (\bibinfo {year} {2016})}\BibitemShut
  {NoStop}%
\bibitem [{\citenamefont {Khomenko}\ \emph {et~al.}(2020)\citenamefont
  {Khomenko}, \citenamefont {Scalliet}, \citenamefont {Berthier}, \citenamefont
  {Reichman},\ and\ \citenamefont {Zamponi}}]{zamponi_tls_prl_2020}%
  \BibitemOpen
  \bibfield  {author} {\bibinfo {author} {\bibfnamefont {D.}~\bibnamefont
  {Khomenko}}, \bibinfo {author} {\bibfnamefont {C.}~\bibnamefont {Scalliet}},
  \bibinfo {author} {\bibfnamefont {L.}~\bibnamefont {Berthier}}, \bibinfo
  {author} {\bibfnamefont {D.~R.}\ \bibnamefont {Reichman}},\ and\ \bibinfo
  {author} {\bibfnamefont {F.}~\bibnamefont {Zamponi}},\ }\bibfield  {title}
  {\bibinfo {title} {Depletion of two-level systems in ultrastable
  computer-generated glasses},\ }\href
  {https://doi.org/10.1103/PhysRevLett.124.225901} {\bibfield  {journal}
  {\bibinfo  {journal} {Phys. Rev. Lett.}\ }\textbf {\bibinfo {volume} {124}},\
  \bibinfo {pages} {225901} (\bibinfo {year} {2020})}\BibitemShut {NoStop}%
\bibitem [{\citenamefont {Gonz\'alez-L\'opez}\ \emph
  {et~al.}(2021)\citenamefont {Gonz\'alez-L\'opez}, \citenamefont {Shivam},
  \citenamefont {Zheng}, \citenamefont {Ciamarra},\ and\ \citenamefont
  {Lerner}}]{sticky_spheres1_karina_pre2021}%
  \BibitemOpen
  \bibfield  {author} {\bibinfo {author} {\bibfnamefont {K.}~\bibnamefont
  {Gonz\'alez-L\'opez}}, \bibinfo {author} {\bibfnamefont {M.}~\bibnamefont
  {Shivam}}, \bibinfo {author} {\bibfnamefont {Y.}~\bibnamefont {Zheng}},
  \bibinfo {author} {\bibfnamefont {M.~P.}\ \bibnamefont {Ciamarra}},\ and\
  \bibinfo {author} {\bibfnamefont {E.}~\bibnamefont {Lerner}},\ }\bibfield
  {title} {\bibinfo {title} {Mechanical disorder of sticky-sphere glasses. I.
  Effect of attractive interactions},\ }\href
  {https://doi.org/10.1103/PhysRevE.103.022605} {\bibfield  {journal} {\bibinfo
   {journal} {Phys. Rev. E}\ }\textbf {\bibinfo {volume} {103}},\ \bibinfo
  {pages} {022605} (\bibinfo {year} {2021})}\BibitemShut {NoStop}%
\bibitem [{\citenamefont {Lerner}\ \emph {et~al.}(2014)\citenamefont {Lerner},
  \citenamefont {DeGiuli}, \citenamefont {During},\ and\ \citenamefont
  {Wyart}}]{breakdown}%
  \BibitemOpen
  \bibfield  {author} {\bibinfo {author} {\bibfnamefont {E.}~\bibnamefont
  {Lerner}}, \bibinfo {author} {\bibfnamefont {E.}~\bibnamefont {DeGiuli}},
  \bibinfo {author} {\bibfnamefont {G.}~\bibnamefont {During}},\ and\ \bibinfo
  {author} {\bibfnamefont {M.}~\bibnamefont {Wyart}},\ }\bibfield  {title}
  {\bibinfo {title} {Breakdown of continuum elasticity in amorphous solids},\
  }\href {https://doi.org/10.1039/C4SM00311J} {\bibfield  {journal} {\bibinfo
  {journal} {Soft Matter}\ }\textbf {\bibinfo {volume} {10}},\ \bibinfo {pages}
  {5085} (\bibinfo {year} {2014})}\BibitemShut {NoStop}%
\bibitem [{\citenamefont {Lerner}(2020)}]{lerner2019finite}%
  \BibitemOpen
  \bibfield  {author} {\bibinfo {author} {\bibfnamefont {E.}~\bibnamefont
  {Lerner}},\ }\bibfield  {title} {\bibinfo {title} {Finite-size effects in the
  nonphononic density of states in computer glasses},\ }\href
  {https://doi.org/10.1103/PhysRevE.101.032120} {\bibfield  {journal} {\bibinfo
   {journal} {Phys. Rev. E}\ }\textbf {\bibinfo {volume} {101}},\ \bibinfo
  {pages} {032120} (\bibinfo {year} {2020})}\BibitemShut {NoStop}%
\bibitem [{\citenamefont {González-López}\ and\ \citenamefont
  {Lerner}(2020)}]{karina_Tx_jcp_2020}%
  \BibitemOpen
  \bibfield  {author} {\bibinfo {author} {\bibfnamefont {K.}~\bibnamefont
  {González-López}}\ and\ \bibinfo {author} {\bibfnamefont {E.}~\bibnamefont
  {Lerner}},\ }\bibfield  {title} {\bibinfo {title} {An energy-landscape-based
  crossover temperature in glass-forming liquids},\ }\href
  {https://doi.org/10.1063/5.0034719} {\bibfield  {journal} {\bibinfo
  {journal} {J. Chem. Phys.}\ }\textbf {\bibinfo {volume} {153}},\ \bibinfo
  {pages} {241101} (\bibinfo {year} {2020})}\BibitemShut {NoStop}%
\bibitem [{\citenamefont {Buchenau}(2020)}]{uli_arXiv_stz_creation}%
  \BibitemOpen
  \bibfield  {author} {\bibinfo {author} {\bibfnamefont {U.}~\bibnamefont
  {Buchenau}},\ }\bibfield  {title} {\bibinfo {title} {Strain field of soft
  modes in glasses},\ }\href {https://arxiv.org/abs/2010.10870} {\bibfield
  {journal} {\bibinfo  {journal} {arXiv preprint arXiv:2010.10870}\ } (\bibinfo
  {year} {2020})}\BibitemShut {NoStop}%
\bibitem [{\citenamefont {Ji}\ \emph {et~al.}(2020)\citenamefont {Ji},
  \citenamefont {de~Geus}, \citenamefont {Popovi\ifmmode~\acute{c}\else
  \'{c}\fi{}}, \citenamefont {Agoritsas},\ and\ \citenamefont
  {Wyart}}]{mw_thermal_origin_of_qle_pre2020}%
  \BibitemOpen
  \bibfield  {author} {\bibinfo {author} {\bibfnamefont {W.}~\bibnamefont
  {Ji}}, \bibinfo {author} {\bibfnamefont {T.~W.~J.}\ \bibnamefont {de~Geus}},
  \bibinfo {author} {\bibfnamefont {M.}~\bibnamefont
  {Popovi\ifmmode~\acute{c}\else \'{c}\fi{}}}, \bibinfo {author} {\bibfnamefont
  {E.}~\bibnamefont {Agoritsas}},\ and\ \bibinfo {author} {\bibfnamefont
  {M.}~\bibnamefont {Wyart}},\ }\bibfield  {title} {\bibinfo {title} {Thermal
  origin of quasilocalized excitations in glasses},\ }\href
  {https://doi.org/10.1103/PhysRevE.102.062110} {\bibfield  {journal} {\bibinfo
   {journal} {Phys. Rev. E}\ }\textbf {\bibinfo {volume} {102}},\ \bibinfo
  {pages} {062110} (\bibinfo {year} {2020})}\BibitemShut {NoStop}%
\bibitem [{\citenamefont {Bouchbinder}\ and\ \citenamefont
  {Langer}(2009{\natexlab{a}})}]{Bouchbinder2009b}%
  \BibitemOpen
  \bibfield  {author} {\bibinfo {author} {\bibfnamefont {E.}~\bibnamefont
  {Bouchbinder}}\ and\ \bibinfo {author} {\bibfnamefont {J.}~\bibnamefont
  {Langer}},\ }\bibfield  {title} {\bibinfo {title} {Nonequilibrium
  thermodynamics of driven amorphous materials. II. Effective-temperature
  theory},\ }\href {https://doi.org/10.1103/PhysRevE.80.031132} {\bibfield
  {journal} {\bibinfo  {journal} {Phys. Rev. E}\ }\textbf {\bibinfo {volume}
  {80}},\ \bibinfo {pages} {031132} (\bibinfo {year}
  {2009}{\natexlab{a}})}\BibitemShut {NoStop}%
\bibitem [{\citenamefont {Bouchbinder}\ and\ \citenamefont
  {Langer}(2009{\natexlab{b}})}]{Bouchbinder2009c}%
  \BibitemOpen
  \bibfield  {author} {\bibinfo {author} {\bibfnamefont {E.}~\bibnamefont
  {Bouchbinder}}\ and\ \bibinfo {author} {\bibfnamefont {J.}~\bibnamefont
  {Langer}},\ }\bibfield  {title} {\bibinfo {title} {Nonequilibrium
  thermodynamics of driven amorphous materials. III. Shear-transformation-zone
  plasticity},\ }\href {https://doi.org/10.1103/PhysRevE.80.031133} {\bibfield
  {journal} {\bibinfo  {journal} {Phys. Rev. E}\ }\textbf {\bibinfo {volume}
  {80}},\ \bibinfo {pages} {031133} (\bibinfo {year}
  {2009}{\natexlab{b}})}\BibitemShut {NoStop}%
\bibitem [{\citenamefont {Rycroft}\ and\ \citenamefont
  {Bouchbinder}(2012)}]{Rycroft2012}%
  \BibitemOpen
  \bibfield  {author} {\bibinfo {author} {\bibfnamefont {C.~H.}\ \bibnamefont
  {Rycroft}}\ and\ \bibinfo {author} {\bibfnamefont {E.}~\bibnamefont
  {Bouchbinder}},\ }\bibfield  {title} {\bibinfo {title} {Fracture toughness of
  metallic glasses: Annealing-induced embrittlement},\ }\href
  {https://doi.org/10.1103/PhysRevLett.109.194301} {\bibfield  {journal}
  {\bibinfo  {journal} {Phys. Rev. Lett.}\ }\textbf {\bibinfo {volume} {109}},\
  \bibinfo {pages} {194301} (\bibinfo {year} {2012})}\BibitemShut {NoStop}%
\bibitem [{\citenamefont {Vasoya}\ \emph {et~al.}(2016)\citenamefont {Vasoya},
  \citenamefont {Rycroft},\ and\ \citenamefont {Bouchbinder}}]{Vasoya2016}%
  \BibitemOpen
  \bibfield  {author} {\bibinfo {author} {\bibfnamefont {M.}~\bibnamefont
  {Vasoya}}, \bibinfo {author} {\bibfnamefont {C.~H.}\ \bibnamefont
  {Rycroft}},\ and\ \bibinfo {author} {\bibfnamefont {E.}~\bibnamefont
  {Bouchbinder}},\ }\bibfield  {title} {\bibinfo {title} {Notch fracture
  toughness of glasses: Dependence on rate, age, and geometry},\ }\href
  {https://doi.org/10.1103/PhysRevApplied.6.024008} {\bibfield  {journal}
  {\bibinfo  {journal} {Phys. Rev. Appl.}\ }\textbf {\bibinfo {volume} {6}},\
  \bibinfo {pages} {024008} (\bibinfo {year} {2016})}\BibitemShut {NoStop}%
\bibitem [{\citenamefont {Ketkaew}\ \emph {et~al.}(2018)\citenamefont
  {Ketkaew}, \citenamefont {Chen}, \citenamefont {Wang}, \citenamefont {Datye},
  \citenamefont {Fan}, \citenamefont {Pereira}, \citenamefont {Schwarz},
  \citenamefont {Liu}, \citenamefont {Yamada}, \citenamefont {Dmowski},
  \citenamefont {Shattuck}, \citenamefont {O'Hern}, \citenamefont {Egami},
  \citenamefont {Bouchbinder},\ and\ \citenamefont
  {Schroers}}]{Eran_mechanical_glass_transition}%
  \BibitemOpen
  \bibfield  {author} {\bibinfo {author} {\bibfnamefont {J.}~\bibnamefont
  {Ketkaew}}, \bibinfo {author} {\bibfnamefont {W.}~\bibnamefont {Chen}},
  \bibinfo {author} {\bibfnamefont {H.}~\bibnamefont {Wang}}, \bibinfo {author}
  {\bibfnamefont {A.}~\bibnamefont {Datye}}, \bibinfo {author} {\bibfnamefont
  {M.}~\bibnamefont {Fan}}, \bibinfo {author} {\bibfnamefont {G.}~\bibnamefont
  {Pereira}}, \bibinfo {author} {\bibfnamefont {U.~D.}\ \bibnamefont
  {Schwarz}}, \bibinfo {author} {\bibfnamefont {Z.}~\bibnamefont {Liu}},
  \bibinfo {author} {\bibfnamefont {R.}~\bibnamefont {Yamada}}, \bibinfo
  {author} {\bibfnamefont {W.}~\bibnamefont {Dmowski}}, \bibinfo {author}
  {\bibfnamefont {M.~D.}\ \bibnamefont {Shattuck}}, \bibinfo {author}
  {\bibfnamefont {C.~S.}\ \bibnamefont {O'Hern}}, \bibinfo {author}
  {\bibfnamefont {T.}~\bibnamefont {Egami}}, \bibinfo {author} {\bibfnamefont
  {E.}~\bibnamefont {Bouchbinder}},\ and\ \bibinfo {author} {\bibfnamefont
  {J.}~\bibnamefont {Schroers}},\ }\bibfield  {title} {\bibinfo {title}
  {Mechanical glass transition revealed by the fracture toughness of metallic
  glasses},\ }\href {https://doi.org/10.1038/s41467-018-05682-8} {\bibfield
  {journal} {\bibinfo  {journal} {Nat. Commun.}\ }\textbf {\bibinfo {volume}
  {9}},\ \bibinfo {pages} {3271} (\bibinfo {year} {2018})}\BibitemShut
  {NoStop}%
\bibitem [{\citenamefont {Karmakar}\ \emph {et~al.}(2011)\citenamefont
  {Karmakar}, \citenamefont {Lerner}, \citenamefont {Procaccia},\ and\
  \citenamefont {Zylberg}}]{itamar_sticky_spheres_potential_pre_2011}%
  \BibitemOpen
  \bibfield  {author} {\bibinfo {author} {\bibfnamefont {S.}~\bibnamefont
  {Karmakar}}, \bibinfo {author} {\bibfnamefont {E.}~\bibnamefont {Lerner}},
  \bibinfo {author} {\bibfnamefont {I.}~\bibnamefont {Procaccia}},\ and\
  \bibinfo {author} {\bibfnamefont {J.}~\bibnamefont {Zylberg}},\ }\bibfield
  {title} {\bibinfo {title} {Effect of the interparticle potential on the yield
  stress of amorphous solids},\ }\href
  {https://doi.org/10.1103/PhysRevE.83.046106} {\bibfield  {journal} {\bibinfo
  {journal} {Phys. Rev. E}\ }\textbf {\bibinfo {volume} {83}},\ \bibinfo
  {pages} {046106} (\bibinfo {year} {2011})}\BibitemShut {NoStop}%
\bibitem [{\citenamefont {Gonz\'alez-L\'opez}\ \emph
  {et~al.}(2020)\citenamefont {Gonz\'alez-L\'opez}, \citenamefont
  {Bouchbinder},\ and\ \citenamefont {Lerner}}]{karina_minimal_disorder}%
  \BibitemOpen
  \bibfield  {author} {\bibinfo {author} {\bibfnamefont {K.}~\bibnamefont
  {Gonz\'alez-L\'opez}}, \bibinfo {author} {\bibfnamefont {E.}~\bibnamefont
  {Bouchbinder}},\ and\ \bibinfo {author} {\bibfnamefont {E.}~\bibnamefont
  {Lerner}},\ }\bibfield  {title} {\bibinfo {title} {Does a universal lower
  bound on glassy mechanical disorder exist?},\ }\href
  {https://arxiv.org/abs/2012.03634} {\bibfield  {journal} {\bibinfo  {journal}
  {arXiv preprint arXiv:2012.03634}\ } (\bibinfo {year} {2020})}\BibitemShut
  {NoStop}%
\bibitem [{\citenamefont {Landau}\ and\ \citenamefont
  {Lifshitz}(1995)}]{landau_lifshitz_elasticity}%
  \BibitemOpen
  \bibfield  {author} {\bibinfo {author} {\bibfnamefont {L.~D.}\ \bibnamefont
  {Landau}}\ and\ \bibinfo {author} {\bibfnamefont {E.~M.}\ \bibnamefont
  {Lifshitz}},\ }\href@noop {} {\emph {\bibinfo {title} {Theory of
  elasticity}}}\ (\bibinfo  {publisher} {Butterworth-Heinemann, London},\
  \bibinfo {year} {1995})\BibitemShut {NoStop}%
\bibitem [{\citenamefont {Albaret}\ \emph {et~al.}(2016)\citenamefont
  {Albaret}, \citenamefont {Tanguy}, \citenamefont {Boioli},\ and\
  \citenamefont {Rodney}}]{Rodney_pre_2016}%
  \BibitemOpen
  \bibfield  {author} {\bibinfo {author} {\bibfnamefont {T.}~\bibnamefont
  {Albaret}}, \bibinfo {author} {\bibfnamefont {A.}~\bibnamefont {Tanguy}},
  \bibinfo {author} {\bibfnamefont {F.}~\bibnamefont {Boioli}},\ and\ \bibinfo
  {author} {\bibfnamefont {D.}~\bibnamefont {Rodney}},\ }\bibfield  {title}
  {\bibinfo {title} {Mapping between atomistic simulations and eshelby
  inclusions in the shear deformation of an amorphous silicon model},\ }\href
  {https://doi.org/10.1103/PhysRevE.93.053002} {\bibfield  {journal} {\bibinfo
  {journal} {Phys. Rev. E}\ }\textbf {\bibinfo {volume} {93}},\ \bibinfo
  {pages} {053002} (\bibinfo {year} {2016})}\BibitemShut {NoStop}%
\bibitem [{\citenamefont {Nicolas}\ and\ \citenamefont
  {Rottler}(2018)}]{Rottler_pre_2018}%
  \BibitemOpen
  \bibfield  {author} {\bibinfo {author} {\bibfnamefont {A.}~\bibnamefont
  {Nicolas}}\ and\ \bibinfo {author} {\bibfnamefont {J.}~\bibnamefont
  {Rottler}},\ }\bibfield  {title} {\bibinfo {title} {Orientation of plastic
  rearrangements in two-dimensional model glasses under shear},\ }\href
  {https://doi.org/10.1103/PhysRevE.97.063002} {\bibfield  {journal} {\bibinfo
  {journal} {Phys. Rev. E}\ }\textbf {\bibinfo {volume} {97}},\ \bibinfo
  {pages} {063002} (\bibinfo {year} {2018})}\BibitemShut {NoStop}%
\bibitem [{\citenamefont {Liu}\ \emph {et~al.}(2011)\citenamefont {Liu},
  \citenamefont {Nagel}, \citenamefont {Van~Saarloos},\ and\ \citenamefont
  {Wyart}}]{liu2011jamming}%
  \BibitemOpen
  \bibfield  {author} {\bibinfo {author} {\bibfnamefont {A.~J.}\ \bibnamefont
  {Liu}}, \bibinfo {author} {\bibfnamefont {S.~R.}\ \bibnamefont {Nagel}},
  \bibinfo {author} {\bibfnamefont {W.}~\bibnamefont {Van~Saarloos}},\ and\
  \bibinfo {author} {\bibfnamefont {M.}~\bibnamefont {Wyart}},\ }\bibfield
  {title} {\bibinfo {title} {The jamming scenario-an introduction and
  outlook},\ }in\ \href@noop {} {\emph {\bibinfo {booktitle} {Dynamical
  heterogeneities in glasses, colloids, and granular media}}}\ (\bibinfo
  {publisher} {Oxford University Press},\ \bibinfo {year} {2011})\BibitemShut
  {NoStop}%
\bibitem [{\citenamefont {Liu}\ and\ \citenamefont {Nagel}(2010)}]{liu_review}%
  \BibitemOpen
  \bibfield  {author} {\bibinfo {author} {\bibfnamefont {A.~J.}\ \bibnamefont
  {Liu}}\ and\ \bibinfo {author} {\bibfnamefont {S.~R.}\ \bibnamefont
  {Nagel}},\ }\bibfield  {title} {\bibinfo {title} {The jamming transition and
  the marginally jammed solid},\ }\href
  {https://doi.org/10.1146/annurev-conmatphys-070909-104045} {\bibfield
  {journal} {\bibinfo  {journal} {Annu. Rev. Condens. Matter Phys.}\ }\textbf
  {\bibinfo {volume} {1}},\ \bibinfo {pages} {347} (\bibinfo {year}
  {2010})}\BibitemShut {NoStop}%
\bibitem [{\citenamefont {van Hecke}(2010)}]{van_hecke_review}%
  \BibitemOpen
  \bibfield  {author} {\bibinfo {author} {\bibfnamefont {M.}~\bibnamefont {van
  Hecke}},\ }\bibfield  {title} {\bibinfo {title} {Jamming of soft particles:
  geometry, mechanics, scaling and isostaticity},\ }\href
  {http://stacks.iop.org/0953-8984/22/i=3/a=033101} {\bibfield  {journal}
  {\bibinfo  {journal} {J. Phys.: Condens. Matter}\ }\textbf {\bibinfo {volume}
  {22}},\ \bibinfo {pages} {033101} (\bibinfo {year} {2010})}\BibitemShut
  {NoStop}%
\bibitem [{\citenamefont {Schoenholz}\ \emph {et~al.}(2013)\citenamefont
  {Schoenholz}, \citenamefont {Goodrich}, \citenamefont {Kogan}, \citenamefont
  {Liu},\ and\ \citenamefont {Nagel}}]{liu_transverse_length_2013}%
  \BibitemOpen
  \bibfield  {author} {\bibinfo {author} {\bibfnamefont {S.~S.}\ \bibnamefont
  {Schoenholz}}, \bibinfo {author} {\bibfnamefont {C.~P.}\ \bibnamefont
  {Goodrich}}, \bibinfo {author} {\bibfnamefont {O.}~\bibnamefont {Kogan}},
  \bibinfo {author} {\bibfnamefont {A.~J.}\ \bibnamefont {Liu}},\ and\ \bibinfo
  {author} {\bibfnamefont {S.~R.}\ \bibnamefont {Nagel}},\ }\bibfield  {title}
  {\bibinfo {title} {Stability of jammed packings ii: the transverse length
  scale},\ }\href {https://doi.org/10.1039/C3SM51096D} {\bibfield  {journal}
  {\bibinfo  {journal} {Soft Matter}\ }\textbf {\bibinfo {volume} {9}},\
  \bibinfo {pages} {11000} (\bibinfo {year} {2013})}\BibitemShut {NoStop}%
\bibitem [{\citenamefont {Baumgarten}\ \emph {et~al.}(2017)\citenamefont
  {Baumgarten}, \citenamefont {V\aa{}gberg},\ and\ \citenamefont
  {Tighe}}]{brian_prl_2017}%
  \BibitemOpen
  \bibfield  {author} {\bibinfo {author} {\bibfnamefont {K.}~\bibnamefont
  {Baumgarten}}, \bibinfo {author} {\bibfnamefont {D.}~\bibnamefont
  {V\aa{}gberg}},\ and\ \bibinfo {author} {\bibfnamefont {B.~P.}\ \bibnamefont
  {Tighe}},\ }\bibfield  {title} {\bibinfo {title} {Nonlocal elasticity near
  jamming in frictionless soft spheres},\ }\href
  {https://doi.org/10.1103/PhysRevLett.118.098001} {\bibfield  {journal}
  {\bibinfo  {journal} {Phys. Rev. Lett.}\ }\textbf {\bibinfo {volume} {118}},\
  \bibinfo {pages} {098001} (\bibinfo {year} {2017})}\BibitemShut {NoStop}%
\bibitem [{\citenamefont
  {Lerner}(2018)}]{quasilocalized_states_of_self_stress}%
  \BibitemOpen
  \bibfield  {author} {\bibinfo {author} {\bibfnamefont {E.}~\bibnamefont
  {Lerner}},\ }\bibfield  {title} {\bibinfo {title} {Quasilocalized states of
  self stress in packing-derived networks},\ }\href
  {https://doi.org/10.1140/epje/i2018-11705-9} {\bibfield  {journal} {\bibinfo
  {journal} {Eur. Phys. J. E}\ }\textbf {\bibinfo {volume} {41}},\ \bibinfo
  {pages} {93} (\bibinfo {year} {2018})}\BibitemShut {NoStop}%
\bibitem [{\citenamefont {Ji}\ \emph {et~al.}(2021)\citenamefont {Ji},
  \citenamefont {de~Geus}, \citenamefont {Agoritsas},\ and\ \citenamefont
  {Wyart}}]{mw_hopping_qle_arXiv_2021}%
  \BibitemOpen
  \bibfield  {author} {\bibinfo {author} {\bibfnamefont {W.}~\bibnamefont
  {Ji}}, \bibinfo {author} {\bibfnamefont {T.~W.}\ \bibnamefont {de~Geus}},
  \bibinfo {author} {\bibfnamefont {E.}~\bibnamefont {Agoritsas}},\ and\
  \bibinfo {author} {\bibfnamefont {M.}~\bibnamefont {Wyart}},\ }\bibfield
  {title} {\bibinfo {title} {Geometry of hopping processes and local
  excitations in glasses},\ }\href {https://arxiv.org/abs/2106.13153}
  {\bibfield  {journal} {\bibinfo  {journal} {arXiv preprint arXiv:2106.13153}\
  } (\bibinfo {year} {2021})}\BibitemShut {NoStop}%
\bibitem [{\citenamefont {Bouchbinder}\ \emph {et~al.}(2021)\citenamefont
  {Bouchbinder}, \citenamefont {Lerner}, \citenamefont {Rainone}, \citenamefont
  {Urbani},\ and\ \citenamefont
  {Zamponi}}]{meanfield_qle_pierfrancesco_prb_2021}%
  \BibitemOpen
  \bibfield  {author} {\bibinfo {author} {\bibfnamefont {E.}~\bibnamefont
  {Bouchbinder}}, \bibinfo {author} {\bibfnamefont {E.}~\bibnamefont {Lerner}},
  \bibinfo {author} {\bibfnamefont {C.}~\bibnamefont {Rainone}}, \bibinfo
  {author} {\bibfnamefont {P.}~\bibnamefont {Urbani}},\ and\ \bibinfo {author}
  {\bibfnamefont {F.}~\bibnamefont {Zamponi}},\ }\bibfield  {title} {\bibinfo
  {title} {Low-frequency vibrational spectrum of mean-field disordered
  systems},\ }\href {https://doi.org/10.1103/PhysRevB.103.174202} {\bibfield
  {journal} {\bibinfo  {journal} {Phys. Rev. B}\ }\textbf {\bibinfo {volume}
  {103}},\ \bibinfo {pages} {174202} (\bibinfo {year} {2021})}\BibitemShut
  {NoStop}%
\bibitem [{\citenamefont {Schirmacher}(2006)}]{Schirmacher_2006}%
  \BibitemOpen
  \bibfield  {author} {\bibinfo {author} {\bibfnamefont {W.}~\bibnamefont
  {Schirmacher}},\ }\bibfield  {title} {\bibinfo {title} {Thermal conductivity
  of glassy materials and the boson peak},\ }\href
  {https://doi.org/10.1209/epl/i2005-10471-9} {\bibfield  {journal} {\bibinfo
  {journal} {Europhys. Lett.}\ }\textbf {\bibinfo {volume} {73}},\ \bibinfo
  {pages} {892} (\bibinfo {year} {2006})}\BibitemShut {NoStop}%
\bibitem [{\citenamefont {Charbonneau}\ \emph
  {et~al.}(2014{\natexlab{a}})\citenamefont {Charbonneau}, \citenamefont
  {Kurchan}, \citenamefont {Parisi}, \citenamefont {Urbani},\ and\
  \citenamefont {Zamponi}}]{Zamponi_2014}%
  \BibitemOpen
  \bibfield  {author} {\bibinfo {author} {\bibfnamefont {P.}~\bibnamefont
  {Charbonneau}}, \bibinfo {author} {\bibfnamefont {J.}~\bibnamefont
  {Kurchan}}, \bibinfo {author} {\bibfnamefont {G.}~\bibnamefont {Parisi}},
  \bibinfo {author} {\bibfnamefont {P.}~\bibnamefont {Urbani}},\ and\ \bibinfo
  {author} {\bibfnamefont {F.}~\bibnamefont {Zamponi}},\ }\bibfield  {title}
  {\bibinfo {title} {Exact theory of dense amorphous hard spheres in high
  dimension. III. The full replica symmetry breaking solution},\ }\href
  {http://stacks.iop.org/1742-5468/2014/i=10/a=P10009} {\bibfield  {journal}
  {\bibinfo  {journal} {J. Stat. Mech. Theor. Exp.}\ }\textbf {\bibinfo
  {volume} {2014}},\ \bibinfo {pages} {P10009} (\bibinfo {year}
  {2014}{\natexlab{a}})}\BibitemShut {NoStop}%
\bibitem [{\citenamefont {Charbonneau}\ \emph
  {et~al.}(2014{\natexlab{b}})\citenamefont {Charbonneau}, \citenamefont
  {Kurchan}, \citenamefont {Parisi}, \citenamefont {Urbani},\ and\
  \citenamefont {Zamponi}}]{parisi_fractal}%
  \BibitemOpen
  \bibfield  {author} {\bibinfo {author} {\bibfnamefont {P.}~\bibnamefont
  {Charbonneau}}, \bibinfo {author} {\bibfnamefont {J.}~\bibnamefont
  {Kurchan}}, \bibinfo {author} {\bibfnamefont {G.}~\bibnamefont {Parisi}},
  \bibinfo {author} {\bibfnamefont {P.}~\bibnamefont {Urbani}},\ and\ \bibinfo
  {author} {\bibfnamefont {F.}~\bibnamefont {Zamponi}},\ }\bibfield  {title}
  {\bibinfo {title} {Fractal free energy landscapes in structural glasses},\
  }\href {https://doi.org/10.1038/ncomms4725} {\bibfield  {journal} {\bibinfo
  {journal} {Nat. Commun.}\ }\textbf {\bibinfo {volume} {5}},\ \bibinfo {pages}
  {3725} (\bibinfo {year} {2014}{\natexlab{b}})}\BibitemShut {NoStop}%
\bibitem [{\citenamefont {Charbonneau}\ \emph {et~al.}(2017)\citenamefont
  {Charbonneau}, \citenamefont {Kurchan}, \citenamefont {Parisi}, \citenamefont
  {Urbani},\ and\ \citenamefont {Zamponi}}]{zamponi_hard_spheres_review_2017}%
  \BibitemOpen
  \bibfield  {author} {\bibinfo {author} {\bibfnamefont {P.}~\bibnamefont
  {Charbonneau}}, \bibinfo {author} {\bibfnamefont {J.}~\bibnamefont
  {Kurchan}}, \bibinfo {author} {\bibfnamefont {G.}~\bibnamefont {Parisi}},
  \bibinfo {author} {\bibfnamefont {P.}~\bibnamefont {Urbani}},\ and\ \bibinfo
  {author} {\bibfnamefont {F.}~\bibnamefont {Zamponi}},\ }\bibfield  {title}
  {\bibinfo {title} {Glass and jamming transitions: From exact results to
  finite-dimensional descriptions},\ }\href
  {https://doi.org/10.1146/annurev-conmatphys-031016-025334} {\bibfield
  {journal} {\bibinfo  {journal} {Annu. Rev. Condens. Matter Phys.}\ }\textbf
  {\bibinfo {volume} {8}},\ \bibinfo {pages} {265} (\bibinfo {year}
  {2017})}\BibitemShut {NoStop}%
\bibitem [{\citenamefont {Benetti}\ \emph {et~al.}(2018)\citenamefont
  {Benetti}, \citenamefont {Parisi}, \citenamefont {Pietracaprina},\ and\
  \citenamefont {Sicuro}}]{parisi_mean_field_w4}%
  \BibitemOpen
  \bibfield  {author} {\bibinfo {author} {\bibfnamefont {F.~P.~C.}\
  \bibnamefont {Benetti}}, \bibinfo {author} {\bibfnamefont {G.}~\bibnamefont
  {Parisi}}, \bibinfo {author} {\bibfnamefont {F.}~\bibnamefont
  {Pietracaprina}},\ and\ \bibinfo {author} {\bibfnamefont {G.}~\bibnamefont
  {Sicuro}},\ }\bibfield  {title} {\bibinfo {title} {Mean-field model for the
  density of states of jammed soft spheres},\ }\href
  {https://doi.org/10.1103/PhysRevE.97.062157} {\bibfield  {journal} {\bibinfo
  {journal} {Phys. Rev. E}\ }\textbf {\bibinfo {volume} {97}},\ \bibinfo
  {pages} {062157} (\bibinfo {year} {2018})}\BibitemShut {NoStop}%
\bibitem [{\citenamefont {Sharma}\ \emph {et~al.}(2016)\citenamefont {Sharma},
  \citenamefont {Yeo},\ and\ \citenamefont {Moore}}]{Sharma_pre_2016}%
  \BibitemOpen
  \bibfield  {author} {\bibinfo {author} {\bibfnamefont {A.}~\bibnamefont
  {Sharma}}, \bibinfo {author} {\bibfnamefont {J.}~\bibnamefont {Yeo}},\ and\
  \bibinfo {author} {\bibfnamefont {M.~A.}\ \bibnamefont {Moore}},\ }\bibfield
  {title} {\bibinfo {title} {Metastable minima of the heisenberg spin glass in
  a random magnetic field},\ }\href
  {https://doi.org/10.1103/PhysRevE.94.052143} {\bibfield  {journal} {\bibinfo
  {journal} {Phys. Rev. E}\ }\textbf {\bibinfo {volume} {94}},\ \bibinfo
  {pages} {052143} (\bibinfo {year} {2016})}\BibitemShut {NoStop}%
\bibitem [{\citenamefont {Rainone}\ \emph {et~al.}(2021)\citenamefont
  {Rainone}, \citenamefont {Urbani}, \citenamefont {Zamponi}, \citenamefont
  {Lerner},\ and\ \citenamefont {Bouchbinder}}]{scipost_mean_field_qles_2021}%
  \BibitemOpen
  \bibfield  {author} {\bibinfo {author} {\bibfnamefont {C.}~\bibnamefont
  {Rainone}}, \bibinfo {author} {\bibfnamefont {P.}~\bibnamefont {Urbani}},
  \bibinfo {author} {\bibfnamefont {F.}~\bibnamefont {Zamponi}}, \bibinfo
  {author} {\bibfnamefont {E.}~\bibnamefont {Lerner}},\ and\ \bibinfo {author}
  {\bibfnamefont {E.}~\bibnamefont {Bouchbinder}},\ }\bibfield  {title}
  {\bibinfo {title} {{Mean-field model of interacting quasilocalized
  excitations in glasses}},\ }\href
  {https://doi.org/10.21468/SciPostPhysCore.4.2.008} {\bibfield  {journal}
  {\bibinfo  {journal} {SciPost Phys. Core}\ }\textbf {\bibinfo {volume} {4}},\
  \bibinfo {pages} {8} (\bibinfo {year} {2021})}\BibitemShut {NoStop}%
\bibitem [{\citenamefont {Sompolinsky}\ and\ \citenamefont
  {Zippelius}(1982)}]{SK_model_1982}%
  \BibitemOpen
  \bibfield  {author} {\bibinfo {author} {\bibfnamefont {H.}~\bibnamefont
  {Sompolinsky}}\ and\ \bibinfo {author} {\bibfnamefont {A.}~\bibnamefont
  {Zippelius}},\ }\bibfield  {title} {\bibinfo {title} {Relaxational dynamics
  of the edwards-anderson model and the mean-field theory of spin-glasses},\
  }\href {https://doi.org/10.1103/PhysRevB.25.6860} {\bibfield  {journal}
  {\bibinfo  {journal} {Phys. Rev. B}\ }\textbf {\bibinfo {volume} {25}},\
  \bibinfo {pages} {6860} (\bibinfo {year} {1982})}\BibitemShut {NoStop}%
\bibitem [{\citenamefont {Folena}\ and\ \citenamefont
  {Urbani}(2021)}]{pierfrancesco_arXiv_2021}%
  \BibitemOpen
  \bibfield  {author} {\bibinfo {author} {\bibfnamefont {G.}~\bibnamefont
  {Folena}}\ and\ \bibinfo {author} {\bibfnamefont {P.}~\bibnamefont
  {Urbani}},\ }\bibfield  {title} {\bibinfo {title} {Marginal stability of
  local energy minima in soft anharmonic mean field spin glasses},\ }\href
  {https://arxiv.org/abs/2106.16221} {\bibfield  {journal} {\bibinfo  {journal}
  {arXiv preprint arXiv:2106.16221}\ } (\bibinfo {year} {2021})}\BibitemShut
  {NoStop}%
\bibitem [{\citenamefont {Gartner}\ and\ \citenamefont
  {Lerner}(2016{\natexlab{b}})}]{plastic_modes_prerc}%
  \BibitemOpen
  \bibfield  {author} {\bibinfo {author} {\bibfnamefont {L.}~\bibnamefont
  {Gartner}}\ and\ \bibinfo {author} {\bibfnamefont {E.}~\bibnamefont
  {Lerner}},\ }\bibfield  {title} {\bibinfo {title} {Nonlinear plastic modes in
  disordered solids},\ }\href {https://doi.org/10.1103/PhysRevE.93.011001}
  {\bibfield  {journal} {\bibinfo  {journal} {Phys. Rev. E}\ }\textbf {\bibinfo
  {volume} {93}},\ \bibinfo {pages} {011001} (\bibinfo {year}
  {2016}{\natexlab{b}})}\BibitemShut {NoStop}%
\bibitem [{foo({\natexlab{b}})}]{footnote2}%
  \BibitemOpen
  \bibinfo {note} {Both schemes for obtaining nonlinear quasilocalized
  excitations $\piv$ require choosing some meaningful initial guess, see
  discussions in~\cite{SciPost2016,episode_1_2020}}\BibitemShut {NoStop}%
\bibitem [{\citenamefont {Richard}\ \emph
  {et~al.}(2021{\natexlab{c}})\citenamefont {Richard}, \citenamefont
  {Kapteijns}, \citenamefont {Giannini}, \citenamefont {Manning},\ and\
  \citenamefont {Lerner}}]{pseudo_harmonic_prl}%
  \BibitemOpen
  \bibfield  {author} {\bibinfo {author} {\bibfnamefont {D.}~\bibnamefont
  {Richard}}, \bibinfo {author} {\bibfnamefont {G.}~\bibnamefont {Kapteijns}},
  \bibinfo {author} {\bibfnamefont {J.~A.}\ \bibnamefont {Giannini}}, \bibinfo
  {author} {\bibfnamefont {M.~L.}\ \bibnamefont {Manning}},\ and\ \bibinfo
  {author} {\bibfnamefont {E.}~\bibnamefont {Lerner}},\ }\bibfield  {title}
  {\bibinfo {title} {Simple and broadly applicable definition of shear
  transformation zones},\ }\href
  {https://doi.org/10.1103/PhysRevLett.126.015501} {\bibfield  {journal}
  {\bibinfo  {journal} {Phys. Rev. Lett.}\ }\textbf {\bibinfo {volume} {126}},\
  \bibinfo {pages} {015501} (\bibinfo {year} {2021}{\natexlab{c}})}\BibitemShut
  {NoStop}%
\bibitem [{\citenamefont {Gendelman}\ \emph {et~al.}(2015)\citenamefont
  {Gendelman}, \citenamefont {Jaiswal}, \citenamefont {Procaccia},
  \citenamefont {Gupta},\ and\ \citenamefont {Zylberg}}]{Gendelman_2015}%
  \BibitemOpen
  \bibfield  {author} {\bibinfo {author} {\bibfnamefont {O.}~\bibnamefont
  {Gendelman}}, \bibinfo {author} {\bibfnamefont {P.~K.}\ \bibnamefont
  {Jaiswal}}, \bibinfo {author} {\bibfnamefont {I.}~\bibnamefont {Procaccia}},
  \bibinfo {author} {\bibfnamefont {B.~S.}\ \bibnamefont {Gupta}},\ and\
  \bibinfo {author} {\bibfnamefont {J.}~\bibnamefont {Zylberg}},\ }\bibfield
  {title} {\bibinfo {title} {Shear transformation zones: State determined or
  protocol dependent?},\ }\href {https://doi.org/10.1209/0295-5075/109/16002}
  {\bibfield  {journal} {\bibinfo  {journal} {Europhys. Lett.}\ }\textbf
  {\bibinfo {volume} {109}},\ \bibinfo {pages} {16002} (\bibinfo {year}
  {2015})}\BibitemShut {NoStop}%
\bibitem [{\citenamefont {Zylberg}\ \emph {et~al.}(2017)\citenamefont
  {Zylberg}, \citenamefont {Lerner}, \citenamefont {Bar-Sinai},\ and\
  \citenamefont {Bouchbinder}}]{lte_pnas}%
  \BibitemOpen
  \bibfield  {author} {\bibinfo {author} {\bibfnamefont {J.}~\bibnamefont
  {Zylberg}}, \bibinfo {author} {\bibfnamefont {E.}~\bibnamefont {Lerner}},
  \bibinfo {author} {\bibfnamefont {Y.}~\bibnamefont {Bar-Sinai}},\ and\
  \bibinfo {author} {\bibfnamefont {E.}~\bibnamefont {Bouchbinder}},\
  }\bibfield  {title} {\bibinfo {title} {Local thermal energy as a structural
  indicator in glasses},\ }\href {https://doi.org/10.1073/pnas.1704403114}
  {\bibfield  {journal} {\bibinfo  {journal} {Proc. Natl. Acad. Sci. U.S.A.}\
  }\textbf {\bibinfo {volume} {114}},\ \bibinfo {pages} {7289} (\bibinfo {year}
  {2017})}\BibitemShut {NoStop}%
\bibitem [{\citenamefont {Schwartzman-Nowik}\ \emph {et~al.}(2019)\citenamefont
  {Schwartzman-Nowik}, \citenamefont {Lerner},\ and\ \citenamefont
  {Bouchbinder}}]{zohar_prerc}%
  \BibitemOpen
  \bibfield  {author} {\bibinfo {author} {\bibfnamefont {Z.}~\bibnamefont
  {Schwartzman-Nowik}}, \bibinfo {author} {\bibfnamefont {E.}~\bibnamefont
  {Lerner}},\ and\ \bibinfo {author} {\bibfnamefont {E.}~\bibnamefont
  {Bouchbinder}},\ }\bibfield  {title} {\bibinfo {title} {Anisotropic
  structural predictor in glassy materials},\ }\href
  {https://doi.org/10.1103/PhysRevE.99.060601} {\bibfield  {journal} {\bibinfo
  {journal} {Phys. Rev. E}\ }\textbf {\bibinfo {volume} {99}},\ \bibinfo
  {pages} {060601} (\bibinfo {year} {2019})}\BibitemShut {NoStop}%
\bibitem [{\citenamefont {Moriel}\ \emph {et~al.}(2019)\citenamefont {Moriel},
  \citenamefont {Kapteijns}, \citenamefont {Rainone}, \citenamefont {Zylberg},
  \citenamefont {Lerner},\ and\ \citenamefont {Bouchbinder}}]{scattering_jcp}%
  \BibitemOpen
  \bibfield  {author} {\bibinfo {author} {\bibfnamefont {A.}~\bibnamefont
  {Moriel}}, \bibinfo {author} {\bibfnamefont {G.}~\bibnamefont {Kapteijns}},
  \bibinfo {author} {\bibfnamefont {C.}~\bibnamefont {Rainone}}, \bibinfo
  {author} {\bibfnamefont {J.}~\bibnamefont {Zylberg}}, \bibinfo {author}
  {\bibfnamefont {E.}~\bibnamefont {Lerner}},\ and\ \bibinfo {author}
  {\bibfnamefont {E.}~\bibnamefont {Bouchbinder}},\ }\bibfield  {title}
  {\bibinfo {title} {Wave attenuation in glasses: Rayleigh and
  generalized-rayleigh scattering scaling},\ }\href
  {https://doi.org/10.1063/1.5111192} {\bibfield  {journal} {\bibinfo
  {journal} {J. Chem. Phys.}\ }\textbf {\bibinfo {volume} {151}},\ \bibinfo
  {pages} {104503} (\bibinfo {year} {2019})}\BibitemShut {NoStop}%
\bibitem [{\citenamefont {Grace}\ and\ \citenamefont
  {Anderson}(1989)}]{Grace_prb_1989}%
  \BibitemOpen
  \bibfield  {author} {\bibinfo {author} {\bibfnamefont {J.~M.}\ \bibnamefont
  {Grace}}\ and\ \bibinfo {author} {\bibfnamefont {A.~C.}\ \bibnamefont
  {Anderson}},\ }\bibfield  {title} {\bibinfo {title} {Low-temperature specific
  heat and thermal conductivity of a glassy polymer under applied pressure},\
  }\href {https://doi.org/10.1103/PhysRevB.40.1901} {\bibfield  {journal}
  {\bibinfo  {journal} {Phys. Rev. B}\ }\textbf {\bibinfo {volume} {40}},\
  \bibinfo {pages} {1901} (\bibinfo {year} {1989})}\BibitemShut {NoStop}%
\bibitem [{\citenamefont {Brand}\ and\ \citenamefont
  {L\"{o}hneysen}(1991)}]{Brand_1991}%
  \BibitemOpen
  \bibfield  {author} {\bibinfo {author} {\bibfnamefont {O.}~\bibnamefont
  {Brand}}\ and\ \bibinfo {author} {\bibfnamefont {H.~V.}\ \bibnamefont
  {L\"{o}hneysen}},\ }\bibfield  {title} {\bibinfo {title} {Rigidity
  percolation and low-energy excitations in amorphous As$_{x}$Se$_{1-x}$},\
  }\href {https://doi.org/10.1209/0295-5075/16/5/008} {\bibfield  {journal}
  {\bibinfo  {journal} {Europhys. Lett.}\ }\textbf {\bibinfo {volume} {16}},\
  \bibinfo {pages} {455} (\bibinfo {year} {1991})}\BibitemShut {NoStop}%
\bibitem [{\citenamefont {Khomenko}\ \emph {et~al.}(2021)\citenamefont
  {Khomenko}, \citenamefont {Reichman},\ and\ \citenamefont
  {Zamponi}}]{zamponi_qle_tls_prm2021}%
  \BibitemOpen
  \bibfield  {author} {\bibinfo {author} {\bibfnamefont {D.}~\bibnamefont
  {Khomenko}}, \bibinfo {author} {\bibfnamefont {D.~R.}\ \bibnamefont
  {Reichman}},\ and\ \bibinfo {author} {\bibfnamefont {F.}~\bibnamefont
  {Zamponi}},\ }\bibfield  {title} {\bibinfo {title} {Relationship between
  two-level systems and quasilocalized normal modes in glasses},\ }\href
  {https://doi.org/10.1103/PhysRevMaterials.5.055602} {\bibfield  {journal}
  {\bibinfo  {journal} {Phys. Rev. Materials}\ }\textbf {\bibinfo {volume}
  {5}},\ \bibinfo {pages} {055602} (\bibinfo {year} {2021})}\BibitemShut
  {NoStop}%
\bibitem [{\citenamefont {Kumar}\ \emph {et~al.}(2021)\citenamefont {Kumar},
  \citenamefont {Procaccia},\ and\ \citenamefont
  {Singh}}]{itamar_where_are_tls_2021}%
  \BibitemOpen
  \bibfield  {author} {\bibinfo {author} {\bibfnamefont {A.}~\bibnamefont
  {Kumar}}, \bibinfo {author} {\bibfnamefont {I.}~\bibnamefont {Procaccia}},\
  and\ \bibinfo {author} {\bibfnamefont {M.}~\bibnamefont {Singh}},\ }\bibfield
   {title} {\bibinfo {title} {Density of quasi-localized modes in glasses:
  where are the two-level systems?},\ }\href {https://arxiv.org/abs/2102.12368}
  {\bibfield  {journal} {\bibinfo  {journal} {arXiv preprint arXiv:2102.12368}\
  } (\bibinfo {year} {2021})}\BibitemShut {NoStop}%
\bibitem [{\citenamefont {Lerner}(2019)}]{boring_paper}%
  \BibitemOpen
  \bibfield  {author} {\bibinfo {author} {\bibfnamefont {E.}~\bibnamefont
  {Lerner}},\ }\bibfield  {title} {\bibinfo {title} {Mechanical properties of
  simple computer glasses},\ }\href
  {https://doi.org/https://doi.org/10.1016/j.jnoncrysol.2019.119570} {\bibfield
   {journal} {\bibinfo  {journal} {J. Non-Cryst. Solids}\ }\textbf {\bibinfo
  {volume} {522}},\ \bibinfo {pages} {119570} (\bibinfo {year}
  {2019})}\BibitemShut {NoStop}%
\bibitem [{\citenamefont {Kapteijns}\ \emph
  {et~al.}(2021{\natexlab{b}})\citenamefont {Kapteijns}, \citenamefont
  {Richard}, \citenamefont {Bouchbinder}, \citenamefont {Schrøder},
  \citenamefont {Dyre},\ and\ \citenamefont {Lerner}}]{jeppe_project_jcp}%
  \BibitemOpen
  \bibfield  {author} {\bibinfo {author} {\bibfnamefont {G.}~\bibnamefont
  {Kapteijns}}, \bibinfo {author} {\bibfnamefont {D.}~\bibnamefont {Richard}},
  \bibinfo {author} {\bibfnamefont {E.}~\bibnamefont {Bouchbinder}}, \bibinfo
  {author} {\bibfnamefont {T.~B.}\ \bibnamefont {Schrøder}}, \bibinfo {author}
  {\bibfnamefont {J.~C.}\ \bibnamefont {Dyre}},\ and\ \bibinfo {author}
  {\bibfnamefont {E.}~\bibnamefont {Lerner}},\ }\bibfield  {title} {\bibinfo
  {title} {Does mesoscopic elasticity control viscous slowing down in
  glassforming liquids?},\ }\href {https://doi.org/10.1063/5.0051193}
  {\bibfield  {journal} {\bibinfo  {journal} {J. Chem. Phys.}\ }\textbf
  {\bibinfo {volume} {155}},\ \bibinfo {pages} {074502} (\bibinfo {year}
  {2021}{\natexlab{b}})}\BibitemShut {NoStop}%
\bibitem [{\citenamefont {Ji}\ \emph {et~al.}(2019)\citenamefont {Ji},
  \citenamefont {Popovi\ifmmode~\acute{c}\else \'{c}\fi{}}, \citenamefont
  {de~Geus}, \citenamefont {Lerner},\ and\ \citenamefont
  {Wyart}}]{MW_theta_and_omega}%
  \BibitemOpen
  \bibfield  {author} {\bibinfo {author} {\bibfnamefont {W.}~\bibnamefont
  {Ji}}, \bibinfo {author} {\bibfnamefont {M.}~\bibnamefont
  {Popovi\ifmmode~\acute{c}\else \'{c}\fi{}}}, \bibinfo {author} {\bibfnamefont
  {T.~W.~J.}\ \bibnamefont {de~Geus}}, \bibinfo {author} {\bibfnamefont
  {E.}~\bibnamefont {Lerner}},\ and\ \bibinfo {author} {\bibfnamefont
  {M.}~\bibnamefont {Wyart}},\ }\bibfield  {title} {\bibinfo {title} {Theory
  for the density of interacting quasilocalized modes in amorphous solids},\
  }\href {https://doi.org/10.1103/PhysRevE.99.023003} {\bibfield  {journal}
  {\bibinfo  {journal} {Phys. Rev. E}\ }\textbf {\bibinfo {volume} {99}},\
  \bibinfo {pages} {023003} (\bibinfo {year} {2019})}\BibitemShut {NoStop}%
\bibitem [{\citenamefont {Ikeda}(2019)}]{Harukuni_pre_2019}%
  \BibitemOpen
  \bibfield  {author} {\bibinfo {author} {\bibfnamefont {H.}~\bibnamefont
  {Ikeda}},\ }\bibfield  {title} {\bibinfo {title} {Universal non-mean-field
  scaling in the density of states of amorphous solids},\ }\href
  {https://doi.org/10.1103/PhysRevE.99.050901} {\bibfield  {journal} {\bibinfo
  {journal} {Phys. Rev. E}\ }\textbf {\bibinfo {volume} {99}},\ \bibinfo
  {pages} {050901} (\bibinfo {year} {2019})}\BibitemShut {NoStop}%
\bibitem [{\citenamefont {Shimada}\ \emph
  {et~al.}(2020{\natexlab{b}})\citenamefont {Shimada}, \citenamefont {Mizuno},\
  and\ \citenamefont {Ikeda}}]{atsushi_dipole_instability_soft_matter2020}%
  \BibitemOpen
  \bibfield  {author} {\bibinfo {author} {\bibfnamefont {M.}~\bibnamefont
  {Shimada}}, \bibinfo {author} {\bibfnamefont {H.}~\bibnamefont {Mizuno}},\
  and\ \bibinfo {author} {\bibfnamefont {A.}~\bibnamefont {Ikeda}},\ }\bibfield
   {title} {\bibinfo {title} {Vibrational spectrum derived from local
  mechanical response in disordered solids},\ }\href
  {https://doi.org/10.1039/D0SM00376J} {\bibfield  {journal} {\bibinfo
  {journal} {Soft Matter}\ }\textbf {\bibinfo {volume} {16}},\ \bibinfo {pages}
  {7279} (\bibinfo {year} {2020}{\natexlab{b}})}\BibitemShut {NoStop}%
\bibitem [{\citenamefont {Shimada}\ \emph {et~al.}(2021)\citenamefont
  {Shimada}, \citenamefont {Mizuno},\ and\ \citenamefont
  {Ikeda}}]{novel_instability_atsushi_2021}%
  \BibitemOpen
  \bibfield  {author} {\bibinfo {author} {\bibfnamefont {M.}~\bibnamefont
  {Shimada}}, \bibinfo {author} {\bibfnamefont {H.}~\bibnamefont {Mizuno}},\
  and\ \bibinfo {author} {\bibfnamefont {A.}~\bibnamefont {Ikeda}},\ }\bibfield
   {title} {\bibinfo {title} {Novel elastic instability of amorphous solids in
  finite spatial dimensions},\ }\href {https://doi.org/10.1039/D0SM01583K}
  {\bibfield  {journal} {\bibinfo  {journal} {Soft Matter}\ }\textbf {\bibinfo
  {volume} {17}},\ \bibinfo {pages} {346} (\bibinfo {year} {2021})}\BibitemShut
  {NoStop}%
\bibitem [{\citenamefont {Shimada}\ and\ \citenamefont
  {De~Giuli}(2020)}]{eric_random_quench_dipoles_arXiv2020}%
  \BibitemOpen
  \bibfield  {author} {\bibinfo {author} {\bibfnamefont {M.}~\bibnamefont
  {Shimada}}\ and\ \bibinfo {author} {\bibfnamefont {E.}~\bibnamefont
  {De~Giuli}},\ }\bibfield  {title} {\bibinfo {title} {Random quench predicts
  universal properties of amorphous solids},\ }\href
  {https://arxiv.org/abs/2008.11896} {\bibfield  {journal} {\bibinfo  {journal}
  {arXiv preprint arXiv:2008.11896}\ } (\bibinfo {year} {2020})}\BibitemShut
  {NoStop}%
\bibitem [{\citenamefont {Lema\^{\i}tre}(2018)}]{lemaitre_stres_correlations}%
  \BibitemOpen
  \bibfield  {author} {\bibinfo {author} {\bibfnamefont {A.}~\bibnamefont
  {Lema\^{\i}tre}},\ }\bibfield  {title} {\bibinfo {title} {Stress correlations
  in glasses},\ }\href {https://doi.org/10.1063/1.5041461} {\bibfield
  {journal} {\bibinfo  {journal} {J. Chem. Phys.}\ }\textbf {\bibinfo {volume}
  {149}},\ \bibinfo {pages} {104107} (\bibinfo {year} {2018})}\BibitemShut
  {NoStop}%
\bibitem [{\citenamefont {DeGiuli}(2018)}]{eric_field_theory_prl_2018}%
  \BibitemOpen
  \bibfield  {author} {\bibinfo {author} {\bibfnamefont {E.}~\bibnamefont
  {DeGiuli}},\ }\bibfield  {title} {\bibinfo {title} {Field theory for
  amorphous solids},\ }\href {https://doi.org/10.1103/PhysRevLett.121.118001}
  {\bibfield  {journal} {\bibinfo  {journal} {Phys. Rev. Lett.}\ }\textbf
  {\bibinfo {volume} {121}},\ \bibinfo {pages} {118001} (\bibinfo {year}
  {2018})}\BibitemShut {NoStop}%
\bibitem [{\citenamefont {Stanifer}\ \emph {et~al.}(2018)\citenamefont
  {Stanifer}, \citenamefont {Morse}, \citenamefont {Middleton},\ and\
  \citenamefont {Manning}}]{lisa_random_matrix_2019}%
  \BibitemOpen
  \bibfield  {author} {\bibinfo {author} {\bibfnamefont {E.}~\bibnamefont
  {Stanifer}}, \bibinfo {author} {\bibfnamefont {P.~K.}\ \bibnamefont {Morse}},
  \bibinfo {author} {\bibfnamefont {A.~A.}\ \bibnamefont {Middleton}},\ and\
  \bibinfo {author} {\bibfnamefont {M.~L.}\ \bibnamefont {Manning}},\
  }\bibfield  {title} {\bibinfo {title} {Simple random matrix model for the
  vibrational spectrum of structural glasses},\ }\href
  {https://doi.org/10.1103/PhysRevE.98.042908} {\bibfield  {journal} {\bibinfo
  {journal} {Phys. Rev. E}\ }\textbf {\bibinfo {volume} {98}},\ \bibinfo
  {pages} {042908} (\bibinfo {year} {2018})}\BibitemShut {NoStop}%
\bibitem [{\citenamefont {Moriel}(2021)}]{avraham_minimal_complexes_2021}%
  \BibitemOpen
  \bibfield  {author} {\bibinfo {author} {\bibfnamefont {A.}~\bibnamefont
  {Moriel}},\ }\bibfield  {title} {\bibinfo {title} {Internally stressed and
  positionally disordered minimal complexes yield glasslike nonphononic
  excitations},\ }\href {https://doi.org/10.1103/PhysRevLett.126.088004}
  {\bibfield  {journal} {\bibinfo  {journal} {Phys. Rev. Lett.}\ }\textbf
  {\bibinfo {volume} {126}},\ \bibinfo {pages} {088004} (\bibinfo {year}
  {2021})}\BibitemShut {NoStop}%
\bibitem [{\citenamefont {Xu}\ \emph {et~al.}(2017)\citenamefont {Xu},
  \citenamefont {Liu},\ and\ \citenamefont {Nagel}}]{ning_xu_prl_2017}%
  \BibitemOpen
  \bibfield  {author} {\bibinfo {author} {\bibfnamefont {N.}~\bibnamefont
  {Xu}}, \bibinfo {author} {\bibfnamefont {A.~J.}\ \bibnamefont {Liu}},\ and\
  \bibinfo {author} {\bibfnamefont {S.~R.}\ \bibnamefont {Nagel}},\ }\bibfield
  {title} {\bibinfo {title} {Instabilities of jammed packings of frictionless
  spheres under load},\ }\href {https://doi.org/10.1103/PhysRevLett.119.215502}
  {\bibfield  {journal} {\bibinfo  {journal} {Phys. Rev. Lett.}\ }\textbf
  {\bibinfo {volume} {119}},\ \bibinfo {pages} {215502} (\bibinfo {year}
  {2017})}\BibitemShut {NoStop}%
\bibitem [{\citenamefont {Paoluzzi}\ \emph {et~al.}(2019)\citenamefont
  {Paoluzzi}, \citenamefont {Angelani}, \citenamefont {Parisi},\ and\
  \citenamefont {Ruocco}}]{paoluzzi_prl_2019}%
  \BibitemOpen
  \bibfield  {author} {\bibinfo {author} {\bibfnamefont {M.}~\bibnamefont
  {Paoluzzi}}, \bibinfo {author} {\bibfnamefont {L.}~\bibnamefont {Angelani}},
  \bibinfo {author} {\bibfnamefont {G.}~\bibnamefont {Parisi}},\ and\ \bibinfo
  {author} {\bibfnamefont {G.}~\bibnamefont {Ruocco}},\ }\bibfield  {title}
  {\bibinfo {title} {Relation between heterogeneous frozen regions in
  supercooled liquids and non-debye spectrum in the corresponding glasses},\
  }\href {https://doi.org/10.1103/PhysRevLett.123.155502} {\bibfield  {journal}
  {\bibinfo  {journal} {Phys. Rev. Lett.}\ }\textbf {\bibinfo {volume} {123}},\
  \bibinfo {pages} {155502} (\bibinfo {year} {2019})}\BibitemShut {NoStop}%
\bibitem [{\citenamefont {Paoluzzi}\ \emph {et~al.}(2020)\citenamefont
  {Paoluzzi}, \citenamefont {Angelani}, \citenamefont {Parisi},\ and\
  \citenamefont {Ruocco}}]{paoluzzi_prr_2020}%
  \BibitemOpen
  \bibfield  {author} {\bibinfo {author} {\bibfnamefont {M.}~\bibnamefont
  {Paoluzzi}}, \bibinfo {author} {\bibfnamefont {L.}~\bibnamefont {Angelani}},
  \bibinfo {author} {\bibfnamefont {G.}~\bibnamefont {Parisi}},\ and\ \bibinfo
  {author} {\bibfnamefont {G.}~\bibnamefont {Ruocco}},\ }\bibfield  {title}
  {\bibinfo {title} {Probing the Debye spectrum in glasses using small system
  sizes},\ }\href {https://doi.org/10.1103/PhysRevResearch.2.043248} {\bibfield
   {journal} {\bibinfo  {journal} {Phys. Rev. Research}\ }\textbf {\bibinfo
  {volume} {2}},\ \bibinfo {pages} {043248} (\bibinfo {year}
  {2020})}\BibitemShut {NoStop}%
\bibitem [{\citenamefont {Das}\ and\ \citenamefont
  {Procaccia}(2021)}]{finite_T_VDoS_IP}%
  \BibitemOpen
  \bibfield  {author} {\bibinfo {author} {\bibfnamefont {P.}~\bibnamefont
  {Das}}\ and\ \bibinfo {author} {\bibfnamefont {I.}~\bibnamefont
  {Procaccia}},\ }\bibfield  {title} {\bibinfo {title} {Universal density of
  low-frequency states in amorphous solids at finite temperatures},\ }\href
  {https://doi.org/10.1103/PhysRevLett.126.085502} {\bibfield  {journal}
  {\bibinfo  {journal} {Phys. Rev. Lett.}\ }\textbf {\bibinfo {volume} {126}},\
  \bibinfo {pages} {085502} (\bibinfo {year} {2021})}\BibitemShut {NoStop}%
\bibitem [{\citenamefont {Krishnan}\ \emph {et~al.}(2021)\citenamefont
  {Krishnan}, \citenamefont {Ramola},\ and\ \citenamefont
  {Karmakar}}]{smarajit_2021_modes_arXiv}%
  \BibitemOpen
  \bibfield  {author} {\bibinfo {author} {\bibfnamefont {V.~V.}\ \bibnamefont
  {Krishnan}}, \bibinfo {author} {\bibfnamefont {K.}~\bibnamefont {Ramola}},\
  and\ \bibinfo {author} {\bibfnamefont {S.}~\bibnamefont {Karmakar}},\
  }\bibfield  {title} {\bibinfo {title} {Universal non-Debye low-frequency
  vibrations in sheared amorphous solids},\ }\href
  {https://arxiv.org/abs/2104.09181} {\bibfield  {journal} {\bibinfo  {journal}
  {arXiv preprint arXiv:2104.09181}\ } (\bibinfo {year} {2021})}\BibitemShut
  {NoStop}%
\bibitem [{\citenamefont {Wang}\ \emph
  {et~al.}(2021{\natexlab{b}})\citenamefont {Wang}, \citenamefont {Szamel},\
  and\ \citenamefont {Flenner}}]{grzegorz_2D_modes_arXiv}%
  \BibitemOpen
  \bibfield  {author} {\bibinfo {author} {\bibfnamefont {L.}~\bibnamefont
  {Wang}}, \bibinfo {author} {\bibfnamefont {G.}~\bibnamefont {Szamel}},\ and\
  \bibinfo {author} {\bibfnamefont {E.}~\bibnamefont {Flenner}},\ }\bibfield
  {title} {\bibinfo {title} {Low-frequency excess vibrational modes in
  two-dimensional glasses},\ }\href {https://arxiv.org/abs/2107.01505}
  {\bibfield  {journal} {\bibinfo  {journal} {arXiv preprint arXiv:2107.01505}\
  } (\bibinfo {year} {2021}{\natexlab{b}})}\BibitemShut {NoStop}%
\bibitem [{\citenamefont {Charbonneau}\ \emph {et~al.}(2016)\citenamefont
  {Charbonneau}, \citenamefont {Corwin}, \citenamefont {Parisi}, \citenamefont
  {Poncet},\ and\ \citenamefont {Zamponi}}]{non_debye_prl_2016}%
  \BibitemOpen
  \bibfield  {author} {\bibinfo {author} {\bibfnamefont {P.}~\bibnamefont
  {Charbonneau}}, \bibinfo {author} {\bibfnamefont {E.~I.}\ \bibnamefont
  {Corwin}}, \bibinfo {author} {\bibfnamefont {G.}~\bibnamefont {Parisi}},
  \bibinfo {author} {\bibfnamefont {A.}~\bibnamefont {Poncet}},\ and\ \bibinfo
  {author} {\bibfnamefont {F.}~\bibnamefont {Zamponi}},\ }\bibfield  {title}
  {\bibinfo {title} {Universal non-debye scaling in the density of states of
  amorphous solids},\ }\href {https://doi.org/10.1103/PhysRevLett.117.045503}
  {\bibfield  {journal} {\bibinfo  {journal} {Phys. Rev. Lett.}\ }\textbf
  {\bibinfo {volume} {117}},\ \bibinfo {pages} {045503} (\bibinfo {year}
  {2016})}\BibitemShut {NoStop}%
\bibitem [{\citenamefont {Buchenau}\ \emph {et~al.}(2021)\citenamefont
  {Buchenau}, \citenamefont {D'Angelo}, \citenamefont {Carini}, \citenamefont
  {Liu},\ and\ \citenamefont {Ramos}}]{buchenau2021sound}%
  \BibitemOpen
  \bibfield  {author} {\bibinfo {author} {\bibfnamefont {U.}~\bibnamefont
  {Buchenau}}, \bibinfo {author} {\bibfnamefont {G.}~\bibnamefont {D'Angelo}},
  \bibinfo {author} {\bibfnamefont {G.}~\bibnamefont {Carini}}, \bibinfo
  {author} {\bibfnamefont {X.}~\bibnamefont {Liu}},\ and\ \bibinfo {author}
  {\bibfnamefont {M.~A.}\ \bibnamefont {Ramos}},\ }\bibfield  {title} {\bibinfo
  {title} {Sound absorption in glasses},\ }\href
  {https://arxiv.org/abs/2012.10139} {\bibfield  {journal} {\bibinfo  {journal}
  {arXiv preprint arXiv:2012.10139}\ } (\bibinfo {year} {2021})}\BibitemShut
  {NoStop}%
\bibitem [{\citenamefont {Ramos}(2004)}]{ramos_2004}%
  \BibitemOpen
  \bibfield  {author} {\bibinfo {author} {\bibfnamefont {M.~A.}\ \bibnamefont
  {Ramos}},\ }\bibfield  {title} {\bibinfo {title} {Are the calorimetric and
  elastic debye temperatures of glasses really different?},\ }\href
  {https://doi.org/10.1080/14786430310001644053} {\bibfield  {journal}
  {\bibinfo  {journal} {Philos. Mag.}\ }\textbf {\bibinfo {volume} {84}},\
  \bibinfo {pages} {1313} (\bibinfo {year} {2004})}\BibitemShut {NoStop}%
\bibitem [{\citenamefont {Schober}\ \emph {et~al.}(2014)\citenamefont
  {Schober}, \citenamefont {Buchenau},\ and\ \citenamefont
  {Gurevich}}]{Schober_PRB_2014}%
  \BibitemOpen
  \bibfield  {author} {\bibinfo {author} {\bibfnamefont {H.~R.}\ \bibnamefont
  {Schober}}, \bibinfo {author} {\bibfnamefont {U.}~\bibnamefont {Buchenau}},\
  and\ \bibinfo {author} {\bibfnamefont {V.~L.}\ \bibnamefont {Gurevich}},\
  }\bibfield  {title} {\bibinfo {title} {Pressure dependence of the boson peak
  in glasses: Correlated and uncorrelated perturbations},\ }\href
  {https://doi.org/10.1103/PhysRevB.89.014204} {\bibfield  {journal} {\bibinfo
  {journal} {Phys. Rev. B}\ }\textbf {\bibinfo {volume} {89}},\ \bibinfo
  {pages} {014204} (\bibinfo {year} {2014})}\BibitemShut {NoStop}%
\bibitem [{\citenamefont {B\"unz}\ \emph {et~al.}(2014)\citenamefont {B\"unz},
  \citenamefont {Brink}, \citenamefont {Tsuchiya}, \citenamefont {Meng},
  \citenamefont {Wilde},\ and\ \citenamefont {Albe}}]{Deformed_BMG_2014}%
  \BibitemOpen
  \bibfield  {author} {\bibinfo {author} {\bibfnamefont {J.}~\bibnamefont
  {B\"unz}}, \bibinfo {author} {\bibfnamefont {T.}~\bibnamefont {Brink}},
  \bibinfo {author} {\bibfnamefont {K.}~\bibnamefont {Tsuchiya}}, \bibinfo
  {author} {\bibfnamefont {F.}~\bibnamefont {Meng}}, \bibinfo {author}
  {\bibfnamefont {G.}~\bibnamefont {Wilde}},\ and\ \bibinfo {author}
  {\bibfnamefont {K.}~\bibnamefont {Albe}},\ }\bibfield  {title} {\bibinfo
  {title} {Low temperature heat capacity of a severely deformed metallic
  glass},\ }\href {https://doi.org/10.1103/PhysRevLett.112.135501} {\bibfield
  {journal} {\bibinfo  {journal} {Phys. Rev. Lett.}\ }\textbf {\bibinfo
  {volume} {112}},\ \bibinfo {pages} {135501} (\bibinfo {year}
  {2014})}\BibitemShut {NoStop}%
\bibitem [{\citenamefont {Widmer-Cooper}\ \emph {et~al.}(2008)\citenamefont
  {Widmer-Cooper}, \citenamefont {Perry}, \citenamefont {Harrowell},\ and\
  \citenamefont {Reichman}}]{widmer2008irreversible}%
  \BibitemOpen
  \bibfield  {author} {\bibinfo {author} {\bibfnamefont {A.}~\bibnamefont
  {Widmer-Cooper}}, \bibinfo {author} {\bibfnamefont {H.}~\bibnamefont
  {Perry}}, \bibinfo {author} {\bibfnamefont {P.}~\bibnamefont {Harrowell}},\
  and\ \bibinfo {author} {\bibfnamefont {D.~R.}\ \bibnamefont {Reichman}},\
  }\bibfield  {title} {\bibinfo {title} {Irreversible reorganization in a
  supercooled liquid originates from localized soft modes},\ }\href
  {https://doi.org/10.1038/nphys1025} {\bibfield  {journal} {\bibinfo
  {journal} {Nature Phys.}\ }\textbf {\bibinfo {volume} {4}},\ \bibinfo {pages}
  {711} (\bibinfo {year} {2008})}\BibitemShut {NoStop}%
\bibitem [{\citenamefont {Widmer-Cooper}\ \emph {et~al.}(2009)\citenamefont
  {Widmer-Cooper}, \citenamefont {Perry}, \citenamefont {Harrowell},\ and\
  \citenamefont {Reichman}}]{harrowell_2009}%
  \BibitemOpen
  \bibfield  {author} {\bibinfo {author} {\bibfnamefont {A.}~\bibnamefont
  {Widmer-Cooper}}, \bibinfo {author} {\bibfnamefont {H.}~\bibnamefont
  {Perry}}, \bibinfo {author} {\bibfnamefont {P.}~\bibnamefont {Harrowell}},\
  and\ \bibinfo {author} {\bibfnamefont {D.~R.}\ \bibnamefont {Reichman}},\
  }\bibfield  {title} {\bibinfo {title} {Localized soft modes and the
  supercooled liquid's irreversible passage through its configuration space},\
  }\href {https://doi.org/10.1063/1.3265983} {\bibfield  {journal} {\bibinfo
  {journal} {J. Chem. Phys.}\ }\textbf {\bibinfo {volume} {131}},\ \bibinfo
  {pages} {194508} (\bibinfo {year} {2009})}\BibitemShut {NoStop}%
\bibitem [{\citenamefont {Oligschleger}\ and\ \citenamefont
  {Schober}(1999)}]{schober_correlate_modes_dynamics_1999}%
  \BibitemOpen
  \bibfield  {author} {\bibinfo {author} {\bibfnamefont {C.}~\bibnamefont
  {Oligschleger}}\ and\ \bibinfo {author} {\bibfnamefont {H.~R.}\ \bibnamefont
  {Schober}},\ }\bibfield  {title} {\bibinfo {title} {Collective jumps in a
  soft-sphere glass},\ }\href {https://doi.org/10.1103/PhysRevB.59.811}
  {\bibfield  {journal} {\bibinfo  {journal} {Phys. Rev. B}\ }\textbf {\bibinfo
  {volume} {59}},\ \bibinfo {pages} {811} (\bibinfo {year} {1999})}\BibitemShut
  {NoStop}%
\end{thebibliography}

%

\end{document}